\documentclass[a4paper,fleqn,usenatbib]{mnras}

\usepackage{booktabs,caption}
\usepackage[flushleft]{threeparttable}
\usepackage{lscape}   

\usepackage{graphicx}	

\usepackage{epsfig}

\title[X/Z Shaped Radio Sources from TGSS ADR 1]{Search for X/Z Shaped Radio Sources from TGSS ADR 1}

\author[Bhukta, Pal and Mondal]{
Netai Bhukta$^{1, 2}$,
Sabyasachi Pal$^{2}$\thanks{E-mail: sabya.pal@gmail.com},
Sushanta K. Mondal$^{1}$
\\
$^{1}$Department of Physics, Sidho Kanho Birsha University, Ranchi Road, Purulia, 723104, India\\
$^{2}$Midnapore City College, Bhadutala, Kuturia, Paschim Midnapur, 721129, India\\
}

\date{Accepted XXX. Received YYY; in original form ZZZ}

\pubyear{2020}

\begin{document}
\label{firstpage}
\pagerange{\pageref{firstpage}--\pageref{lastpage}}
\maketitle

\begin{abstract}
A small sub-class of radio galaxies that exhibit a pair of secondary low surface brightness radio lobes oriented at an angle to the primary high surface brightness lobes are known as X-shaped radio galaxies (XRGs). In cases, it is seen that the less luminous secondary lobes emerge from the edges of the primary high brightened lobes and give a Z-symmetric morphology.  These are known as Z-shaped radio galaxies (ZRGs). From the TIFR GMRT Sky Survey at 150 MHz, we present a systematic search result for XRGs and ZRGs. We identified a total of 58 radio sources, out of which 40 are XRGs and 18 are ZRGs. Taking advantage of the large sample size of the XRGs and ZRGs reported in the current paper, different properties of XRGs and ZRGs are studied. Out of 58 XRGs and ZRGs presented in the current paper, 19 (32 per cent) are FR-I and 33 (57 per cent) are FR-II radio galaxies. For four XRGs and three ZRGs, the morphology is so complex that they could not be classified. We have estimated radio luminosity and spectral index of newly discovered winged radio galaxies and made a comparative study with previously detected XRGs and ZRGs. Most of the XRGs show a steep spectral index between 150 MHz and 1400 MHz and only 14 per cent of the sources show a flat spectrum but for ZRGs, a good fraction of the sources (36 per cent) show a flat spectrum. 
\end{abstract}




\section{Introduction}
\label{sec:intro}
 An average radio galaxy shows a pair of relativistic radio jets coming out from the central black hole or active galactic nuclei (AGN), each one directed in the opposite direction. For a very small sub-class of radio galaxies (3 -- 10 per cent), in addition to the primary jet pair, another extra pair of radio jets or wings are visible \citep{Le84, Le92, Ch07, Ya19}.  These relatively low-luminosity and often symmetric secondary jets extend at an angle to the primary jets with nearly equal or lower than the length of the primary lobes. Winged radio galaxies are classified into two major sub-classes based on the alignment of the secondary lobes with respect to the primary lobes -- `X'-shaped radio galaxies (XRGs) and `Z/S'-shaped radio galaxies (ZRGs). For XRGs, primary and secondary jets pass symmetrically near the centre of the elliptical galaxy which is the source of the jet pairs. The secondary jets for ZRGs are seen to emerge from the edges of the primary jets. 
The primary lobes of the majority of XRGs are Fanaroff-Riley type II (FR-II) \citep{Fa74a} and the remaining are either FR-I or mixed \citep{Le92, Me02, Ch07, Ya19, Be20}. For all known XRGs, the secondary lobes are FR-I type.

3C 272.1 was the first reported radio galaxy with wings \citep{Ri72} which showed a Z like structure. Later NGC 326 \citep{Ek78} and NGC 3309 \citep{Dr88, Ko90} also showed the signature of wings. The sources with X-shape was first classified as XRGs by \citet{Le92} who presented a list of 11 objects with wings. A systematic study to look for radio galaxies with wings was done by \citet{Ch07} (C07 afterwards) who identified 100 XRG candidates using the Very Large Array (VLA) Faint Images of the Radio Sky at Twenty-centimeters (FIRST) survey \citep{Be95}. Recently, \citet{Be20} and \citet{Ya19} also reported 296 and 290 new winged radio galaxies from the FIRST survey. \citet{Pr11} identified 156 XRG candidates by using an automated morphological classification scheme to the FIRST. Out of these 156 sources, 21 sources were already reported in C07. \citet{Ro15, Sa18, Ro18} studied detailed properties of a subset of sources using \citet{Ch07} from follow-up VLA observations. About 30 XRG/ZRG sources were identified using LOFAR Two-metre Sky Survey First Data Release (LoTSS DR1) \citep{Pa21}.  

There is no consensus so far on the origin of the wings in XRGs. Many models are available to explain the mechanism behind the formation of these additional wings \citep{Ro01, De02}.
The prominent models are

(a) Twin AGN Model: Twin AGNs at the core of an elliptical galaxy, each of the AGN is emitting a pair of jets \citep{La07}.

(b) A merger of a pair of black holes \citep{Ro01, Me02, De02, Go11}: The merger of a pair of galaxies each of which contains a super-massive black hole (SMBH) is a possibility for the origin of X-shaped radio morphology. The merger may produce a new jet pair at an axis misaligned from the axis of the jets before the merger.  

(c) Twin jet precession \citep{Re78, Ek78, Pa85, Ma94}: Slow conical precession of twin jets.

Above three models require jets from two black holes.

(d) The plasma backflow: The secondary wings originate from the backflow of synchrotron plasma deflected by the thermal halo surrounding the host \citep{Le84, Wo95} or in non-spherical gas distribution, the cocoon surrounding the radio-jets expands laterally at a high rate producing wings of radio emission along the minor axis \citep{Ca02}.  

Though these models are good enough to explain wing-like behaviour from many galaxies, none of the above models can explain the nature of all the XRGs \citep{Go12}. 

ZRGs are believed to be produced by the merge of two galaxies, each hosting a supermassive black hole in the centre \citep{Go83}. \citet{Zi05} further claimed that the ram pressure of the gas which comes out from the smaller galaxy to the ISM of the larger galaxy in merging galaxies produces ZRGs. 

Though all of the sources presented in the current paper were previously reported in other radio catalogues, and some are well known (e.g. 3C, 4C, NGC etc.), the winged nature of these sources are not reported elsewhere. The current paper is organized as follows. The definition of XRGs and ZRGs, source identification method and identification of optical counterparts of sources are described in section \ref{sec:source-identification}. The different properties of XRGs and ZRGs are described in section \ref{sec:result}. In section \ref{sec:disc}, discussion on different results are made and we summarize the main conclusions in section \ref{sec:conclusion}.

For the entire discussion in this paper, we used the following $\Lambda$CDM cosmology parameters: $H_0= 67.4 $ km s$^{-1}$ Mpc$^{-1}$, $\Omega_m = 0.315$ and $\Omega_{vac} = 0.685$. Results from the final full-mission Planck measurements of the CMB anisotropies are used for estimation of these cosmological constants \citep{Ag20}.

\section{Identifying the XRGs and ZRGs}
\label{sec:source-identification}

\subsection{TGSS ADR 1}
\label{subsec:tgss}
The TIFR GMRT Sky Survey (TGSS) used the Giant Metrewave Radio Telescope (GMRT) at 150 MHz. The survey took place between April 2010 and March 2012. \citet{In17} independently reprocessed the TGSS data using the {\tt SPAM} pipeline. The first alternative data release (ADR 1) of \citet{In17} includes continuum stokes I images of  90 per cent of the full sky or 99.5 per cent of the radio sky north of --53$\degr$ DEC. The median noise of the survey is 3.5 mJy beam$^{-1}$ and the resolution is $25\arcsec \times 25\arcsec$ north of 19$\degr$ DEC and $25\arcsec \times 25\arcsec / \cos(\textrm{DEC}-19\degr)$ south of 19$\degr$. In total, over 2000 hr of observation time was used over about 200 observing sessions. The exposure time on each pointing was about 15 min, split over 3--5 scans to improve uv-coverage.

\subsection{Definition of XRGs and ZRGs}
\label{subsubsec:definition}
The secondary lobes of radio galaxies are known as wings. The winged radio galaxies are classified into two groups depending on the position of the wings - `X'-shaped radio galaxies (XRGs) and `Z'-shaped radio galaxies (ZRGs). In XRGs, it appears that the secondary lobes come from the central spot or near the central region. The near central region is defined as the region within $\sim$25 per cent of the primary jet length from the central core. If the secondary lobes tend to come from the edges of the primary jets or the non-near central area, the sources are classified as ZRGs. Since the convolution size of TGSS images is $\sim25\arcsec$, we have catalogued radio galaxies whose wing size is at least four times the convolution size (i.e. $\sim100\arcsec$). XRGs are further classified into four sub-classes. (1) Small-winged XRGs: In these XRGs, the size of wings are less than primary lobes; (2) Long-winged XRGs: In these XRGs, the size of wings are larger than the primary lobes; (3) Single wing XRGs: In these XRGs, one side of the wing is absent or small ($<25\arcsec$) and (4) Z-symmetry XRGs: both of the jets are distinctly misaligned and forms Z symmetric structure with respect to the central AGN. We need to remember that Z-symmetric XRGs are different from ZRGs where in the case of later, wings appear to come from the edge of the primary jets. 


\subsection{Search Strategy}
\label{subsec:search-strategy}
We have searched for XRGs and ZRGs using TGSS ADR 1 images. TGSS covers 36900 square degrees with 0.62 million radio sources. The good sensitivity and high resolution of TGSS (average rms of 3.5 mJy beam$^{-1}$) helps to study new fainter samples of different types of radio galaxies, such as wide-angled tailed radio galaxies \citep{Bh21} and giant radio galaxies \citep{Bh22}. 

We visually examined all 5536 fields of TGSS images. For the selection of XRG/ZRG source sample from TGSS, unlike checking a small number of sources with different filtering criteria (as was done for C07; \citet{Ya19, Be20}), we tried the rigorous and tedious way and checked each of 5536 fields manually. The earlier method of selection of certain sources with various filtering criteria was prone to miss a number of sources and that is why C07; \citet{Ya19, Be20} discovered a large number of sources from the same FIRST survey and each of them missed many more (which were eventually detected by other surveys). With the resolution of TGSS ($\sim25\arcsec$), we can confidently classify a good fraction of sources in our list to be X-shaped with characteristic wing length $>$80 per cent of the active lobe. Following the precedence of detection of small wing XRGs (14/100 in C07, 32/106 in \citet{Ya19}, and 26/161 in \citet{Be20}, we have also included those sources with the smaller wing size hoping a fraction of these is due to the projection effect. Since the lowest wing size of our catalogue is $\sim$100$''$, wings of all the sources in our catalogue were distinctly identified. It is difficult to measure the extent of secondary jets as often they are diffused and weak and always lack bright peaks (hot-spots) at the end. Images must have enough dynamic range (measured as the ratio of peak flux to rms flux) to detect prolonged low surface brightness wings.

The number of XRGs detected from the FIRST survey (C07; \citet{Ya19, Be20}), which cover 9033 square degrees, is more than the present work as FIRST has better resolution (typically $\sim 5\arcsec$, \citet{Be95}) and sensitivity (typical RMS of 0.13 mJy) compared to TGSS survey in 150 MHz. But, since the FIRST survey was carried out with the VLA B configuration resulting lack of short spacing in uv coverage, sources with large size and diffused emission can not be detected by the FIRST survey. That is why the present paper detect many XRGs and ZRGs in the common coverage region with the FIRST survey when the angular sizes of these sources are large. Here we have excluded all sources from our list which were previously discovered in different surveys (e.g. C07; \citet{Ya19, Be20}) and in individual observations (e.g. \citet{Ri72, Le92}). 

The size of a particular source (or wing) depends on the selection of the lowest contour. We have started contours from 17.5 mJy/beam (5$\sigma$) for images of all XRGs and ZRGs for uniformity. The typical error in the measurement of angular size is one pixel ($\sim$2.5 arcsec) in the image.

\subsection{The Optical/IR Counterparts and Properties}
\label{subsubsec:optical-counterpart}
For each of the newly discovered XRGs and ZRGs, the optical/IR counterpart is searched using the Digital Sky Survey (DSS), the Sloan Digital Sky Survey (SDSS) data catalogue \citep{Gu06, Al15} and NED\footnote{htteps://ned.ipac.caltech.edu}. The identification of the optical/IR counterpart is based on the position of the optical/IR source relative to the morphology of the radio galaxy. DSS2 red images are overlaid with TGSS ADR 1 images. The positions of the optical/IR counterparts of XRGs and ZRGs are used as the positions of these sources and are presented in the 3rd and 4th columns of Table \ref{tab:X-shaped} and Table \ref{tab:Z-shaped}. For XRGs, optical/IR counterparts are found for 26 sources out of a total of 40 sources. For ZRGs, optical/IR counterparts are found for 16 sources out of 18 sources. The location of the core of the radio galaxy or the intersection of both radio lobes is used as the position of the galaxies when no clear optical/IR counterpart is available.   
 
\section{Results}
\label{sec:result}
 
 \begin{figure}
\includegraphics[width=8cm,angle=0,origin=c]{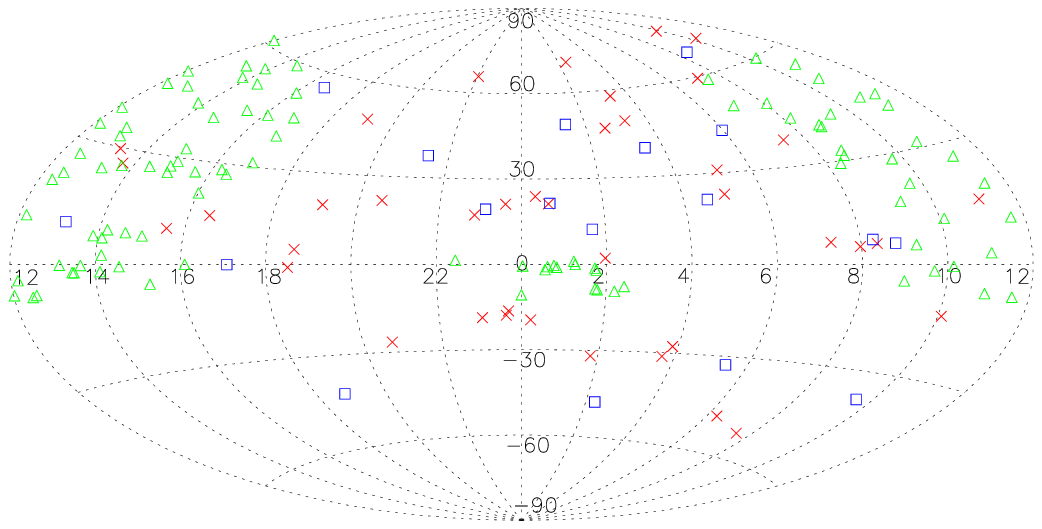}
	\caption{Spatial distribution of X and Z shaped radio galaxies. XRGs and ZRGs presented in the current paper are described by red crosses and blue boxes respectively. XRGs mentioned in \citet{Ch07} are shown by green triangles.}
   
	 \label{fig:source-distribution}
\end{figure}

The discovery of 40 new XRGs and 18 new ZRGs from TGSS ADR 1 is reported. The spatial distribution of XRGs and ZRGs are shown in Fig \ref{fig:source-distribution}. XRGs and ZRGs presented in the current paper are described by red crosses and blue boxes respectively. XRGs mentioned in C07 are shown by green triangles. The sources show a random distribution as expected.

Table \ref{tab:X-shaped} and Table \ref{tab:Z-shaped} list all newly discovered XRGs and ZRGs. The catalogue names from where the optical counterpart is found are listed in the 5th column of those tables. The TGSS flux density at 150 MHz is presented in column 6. Using the NRAO VLA Sky Survey (NVSS; \citet{Co98}), we also tabulate the corresponding flux density for each source at 1400 MHz (column 7).  For most sources, NVSS counterparts are found, but due to less resolution, none of the NVSS images shows the morphological signature of X-shape and Z-shape sources. The two-point spectral index (section \ref{subsec:spectral-index}) between 150 and 1400 MHz $(\alpha_{150}^{1400})$ is calculated when flux density at 1400 MHz is available and is presented in 8th column of Table \ref{tab:X-shaped} and \ref{tab:Z-shaped}. In column 9, the redshifts ($z$) of these sources are mentioned when available. For the sources with known redshift, the luminosities (section \ref{subsec:lum}) of these sources are tabulated in column 10. The names of other catalogues in radio wavelengths are listed in the last column, where these sources were mentioned earlier without identification of them as XRGs or ZRGs. 

Of the 18 ZRGs presented in the current paper, 50 per cent (9) are FR-I radio galaxies and 33 per cent (6) are FR-II radio galaxies. For three ZRGs, the jet morphology is so complex that they could not be classified. The redshifts are known for 13 XRGs and 9 ZRGs out of a total of 40 XRGs and 18 ZRGs respectively. For XRGs, except for one source, all other 12 sources have a redshift less than 0.5. J2337--1752 is an XRG that has the largest redshift ($z$=1.14) amongst all sources presented in the current paper. The nearest XRG is J2336+2108 at $z$=0.05. For ZRGs, redshift $z<0.5$ for all sources. J0817+0709 has the largest redshift among the ZRGs, at $z$ = 0.27 \citep{Ri09}. J0329+3947 is the closest ZRG in our sample with $z$=0.03.

The images of all XRGs are shown in Fig \ref{fig:XRG} and images of all ZRGs are shown in Fig \ref{fig:ZRG}. The optical images from DSS2 (red) are overlaid with the radio images from TGSS ADR 1.

\subsection{Radio Luminosity}
\label{subsec:lum}
We have calculated the radio luminosity ($L_{150}$) of all newly discovered winged sources (when the value of $z$ is available) using standard formula \citep{Do09}

\begin{equation}
L_{150}=4\pi{D_{L}}^{2}S_{0}(1+z)^{\alpha-1}
\end{equation}
where $z$ is the redshift of the radio galaxy, $\alpha$ is the spectral index ($S \propto \nu^{-\alpha}$), $D_{L}$ is the luminosity distance to the source (Mpc), and $S_0$ is the flux density (Jy) at a given frequency. The source radio luminosities at 150 MHz are in the order of $10^{25}$ W Hz$^{-1}$, which is similar to a typical radio galaxy. 
The average of $\log ~L$ [W Hz$^{-1}$] for XRGs is 26.08 (1$\sigma$ standard deviation$=0.76$, median$=25.76$) and that of ZRGs is 25.86 (1$\sigma$ standard deviation$=0.63$, median$=25.90$). For sources found in C07, we calculated the average $\log ~L$ [W Hz$^{-1}$] at 150 MHz to be 25.49 (with 1$\sigma$ standard deviation$=0.71$, the median of 25.31). Here we used the TGSS flux density of these sources at 150 MHz. 
  These Luminosity measurements are expected as XRGs are known to have radio luminosities close to FR-I-FR-II division of $P_{178} \sim 2\times 10^{25}$ W Hz$^{-1}$ sr$^{-1}$\citep{Le92, De02}.

\begin{figure*}
\includegraphics[width=6cm,angle=270,origin=c]{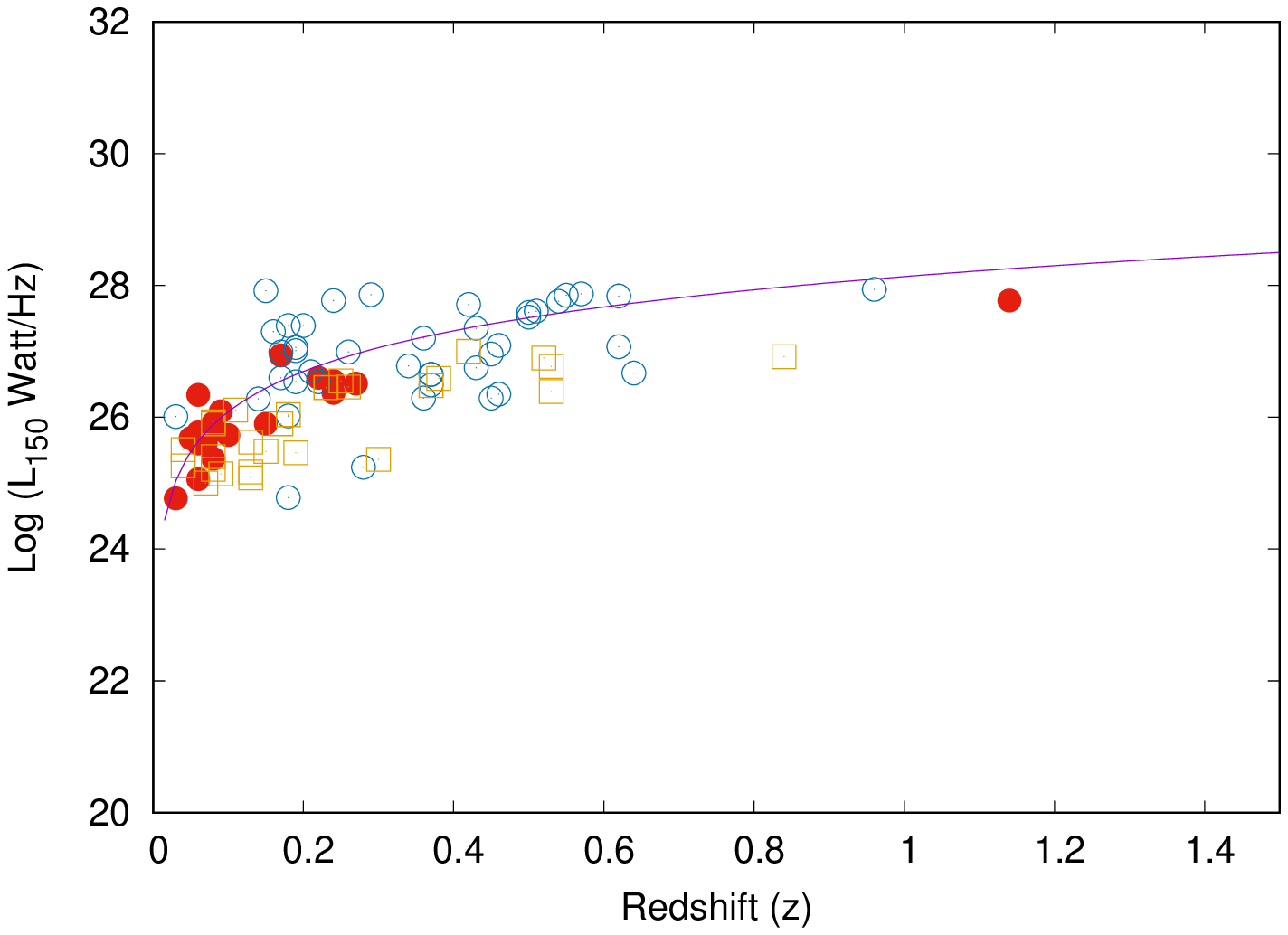}
\includegraphics[width=6cm,angle=270,origin=c]{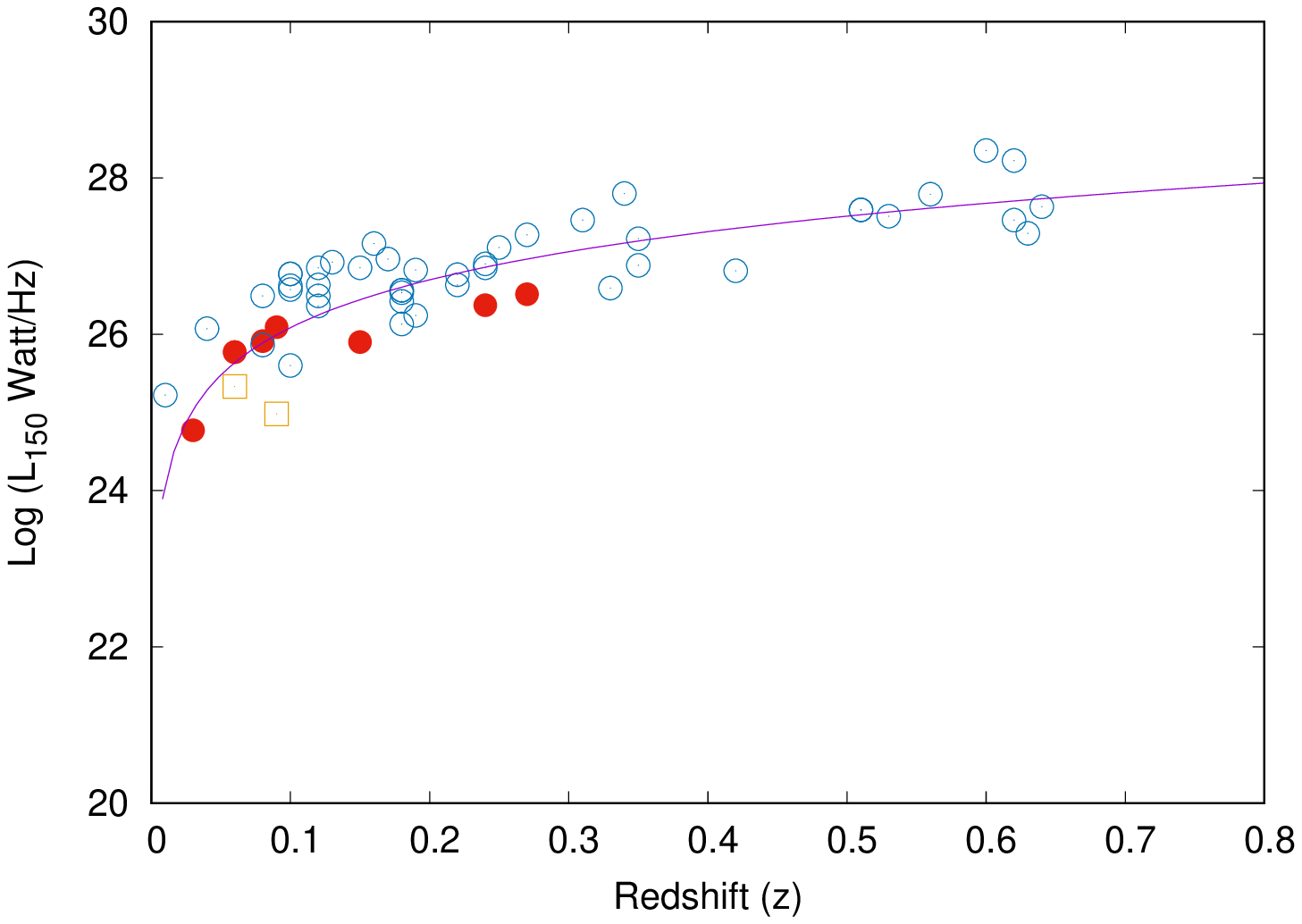}
	\caption{Plot showing distribution of $\log L_{150}$ (W Hz$^{-1}$) with redshift ($z$) for the sources presented in the current paper for XRGs (left) and ZRGs (right). Here, red circles represent TGSS sources. We also include sources presented in \citet{Be20} and \citet{Ch07} (open blue circle and yellow open square). The pink solid line indicates the best-fitted luminosity using points from all surveys in the figure which corresponds to 5.5 Jy flux density for XRGs and 5.0 Jy flux density for ZRGs.}
   \label{fig:lum}
\end{figure*}

Fig \ref{fig:lum} represents the distribution of XRGs and ZRGs in the $\log~L$--$z$ plane.  Most of the sources are within the redshift range of 0.1 to 0.4, and the number of radio sources decreases with increasing redshift beyond $z = 0.5$. There are no available sources in the lower right quadrant of Fig \ref{fig:lum} which is due to the non-identification of low luminosity radio galaxies at high redshift. This sensitivity limit of the survey is known as the Malmquist bias \citep{Ma22, Ma36}. The least luminous XRG in our sample is J0717+0638, with a flux density of $\sim$ 0.43 Jy at 150 MHz. We also include sources presented in \citet{Be20} and C07 using corresponding TGSS Flux at 150 MHz. The pink solid line indicates the best-fitted luminosity using points from all surveys in the figure which corresponds to 5.5 Jy flux density for XRGs and 5.0 Jy flux density for ZRGs. 

\begin{figure*}
\includegraphics[width=6cm,angle=270,origin=c]{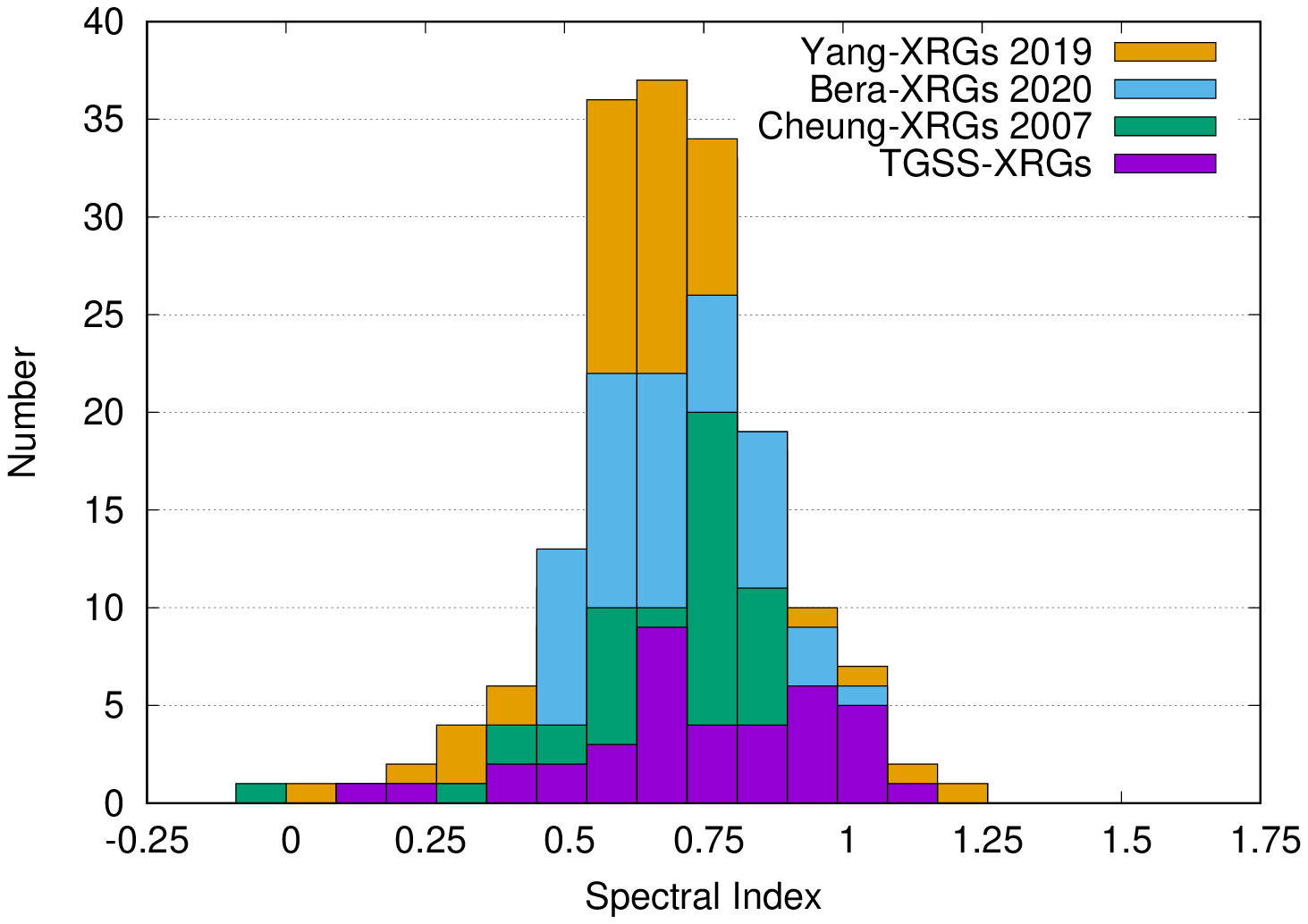}
\includegraphics[width=6cm,angle=270,origin=c]{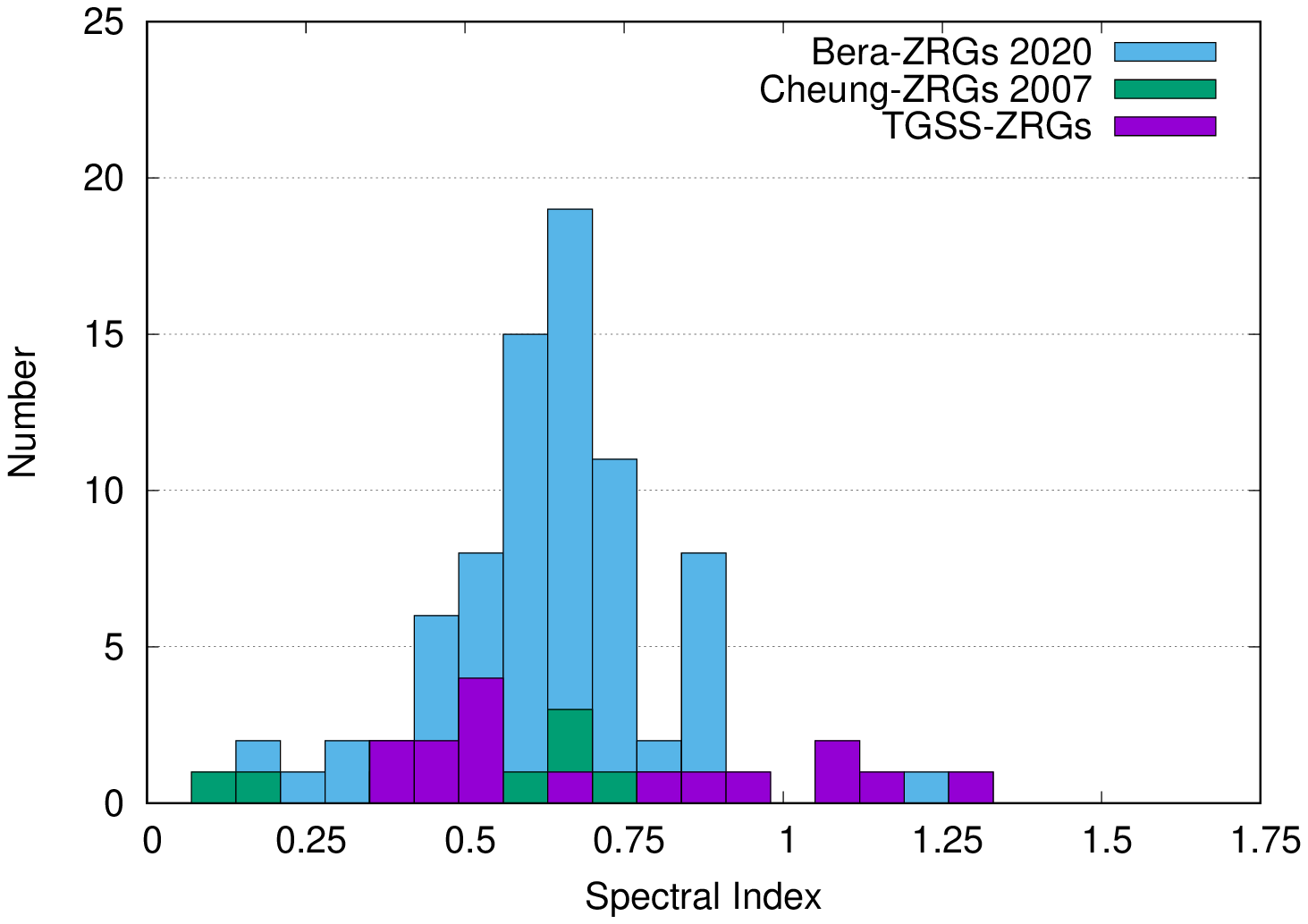}
	\caption{Histogram showing spectral index distribution of radio galaxies presented in the current paper for XRGs (left) and ZRGs (right). We also included sources presented in \citet{Be20}, \citet{Ya19} and \citet{Ch07}. }
   \label{fig:hist-alpha}
\end{figure*}

\subsection{Spectral Index ($\alpha$)}
\label{subsec:spectral-index}

The two-point spectral indices of newly discovered winged radio galaxies between 150 and 1400 MHz are calculated assuming $S \propto \nu^{-\alpha}$, where $\alpha$ is the spectral index and $S_{\nu}$ is the radiative flux density at a given frequency $\nu$. These spectral indices have been determined by integrating over the same aperture at both frequencies. In Table \ref{tab:X-shaped} and Table \ref{tab:Z-shaped}, spectral indices ($\alpha_{150}^{1400}$) are mentioned. For 38 XRGs and 16 ZRGs, spectral index measurements are available. The remaining 2 XRGs and 2 ZRGs are not detectable in NVSS maps since their declination is less than the NVSS coverage. Out of 40 XRGs with spectral index information, 5 (13 per cent) show flat spectrum ($\alpha_{150}^{1400}<0.5$). Out of 18 ZRGs with spectral index information, 4 (22 per cent) are showing flat spectrum ($\alpha_{150}^{1400}<0.5$). Most of the XRGs and ZRGs show a steep radio spectrum ($\alpha_{150}^{1400}>0.5$) which is a common property of lobe dominated radio galaxies.

The uncertainty of spectral index measurements due to flux density uncertainty \citep{Ma16} is
\begin{equation}
	\Delta\alpha=\frac{1}{\ln\frac{\nu_{1}}{\nu_{2}}}\sqrt{\left(\frac{\Delta S_{1}}{S_{1}}\right)^{2}+\left(\frac{\Delta S_{2}}{S_{2}}\right)^{2}}
\end{equation}
where $\nu_{1, 2}$ and $S_{1, 2}$ refer to NVSS and TGSS frequencies and flux densities respectively. The flux density accuracy in TGSS ADR 1 and NVSS is $\sim10$ per cent \citep{In17} and $\sim5$ per cent \citep{Co98}. Using equation 2, the spectral index uncertainty is $\Delta\alpha$=0.05.

Histogram with spectral index distribution for XRGs (left) and ZRGs (right), presented in the current article, is shown in Fig \ref{fig:hist-alpha}. The distribution shows different peaks for XRGs and ZRGs. The histogram peaks near $0.6-0.7$ for XRGs and near $0.4-0.5$ for ZRGs for TGSS sources.  In our catalogue, we have only 15 ZRGs with spectral index information (5 FR-II classifications and 10 FR-I classifications). The majority of the ZRG sources discovered in our sample are FR-I and are showing core-dominated structures which are known to have a less spectral index. In Fig \ref{fig:hist-alpha}, the spectral index distribution of ZRGs in our sample is seen between 0.4 and 0.5. But, \citet{Be20} have a larger sample of 135 ZRGs with spectral index information, 54 per cent of which are FR-II radio galaxies and 27 per cent are FR-I radio galaxies. In this catalogue, the majority are showing steeper lobe-dominated spectral property (which is common for FR-II sources). Since in Fig \ref{fig:hist-alpha}, most of the sources are from \citet{Be20} which is dominated by FR-II sources, the peak for this distribution appears to be near 0.65. For XRG, the total span of $\alpha_{150}^{1400}$ is from $0.40$ to $1.16$ and for ZRGs, the total span of $\alpha^{1400}_{150}$ is from $0.38$ to $1.28$. 

The mean and median values for the spectral index of XRGs are 0.74 and 0.76 respectively. Similarly, for ZRGs, the mean and median values are 0.72 and 0.70 respectively. These mean values of spectral index of XRGs and ZRGs indicate the similar nature of the spectral energy distributions of the XRGs and ZRGs in the present sample to those of winged sources previously discovered in different other surveys (e.g. C07; \citet{Ya19, Be20}). 

\subsection{Angular Size}
{\label{subsubsec:maj-axis}}

The median size of the primary jets for XRGs in our list is 202.25 arcsec. It should be noted that the present study is looking for XRGs with a much higher size than previous systematic studies based on the FIRST survey (C07; \citet{Ya19, Be20}). For example, most of the sources in \citet{Ya19} had $10\arcsec<\theta_{maj}<13\arcsec$.

\begin{figure}
\includegraphics[width=6cm,angle=270,origin=c]{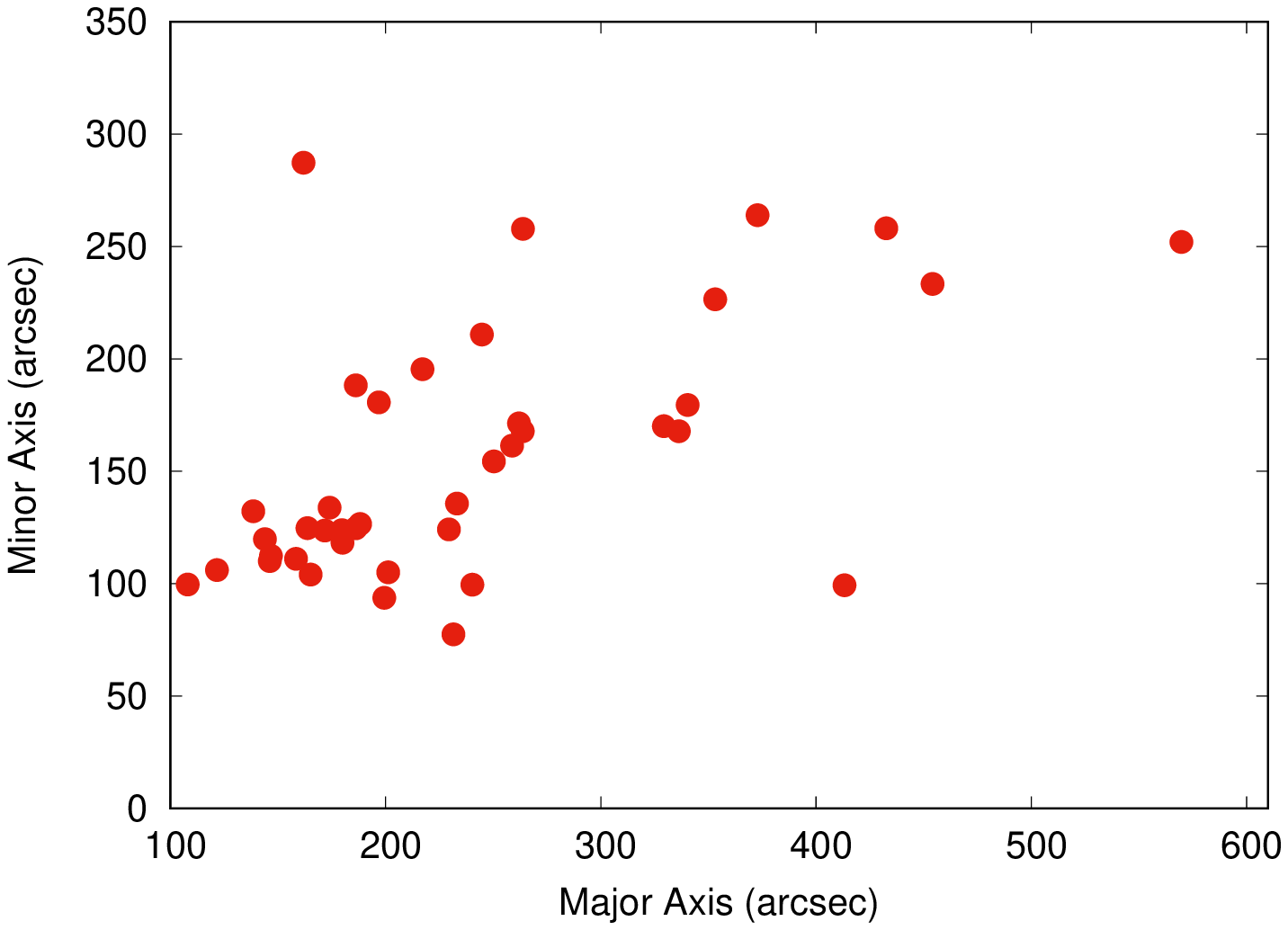}
   \caption{The distribution of size of primary and secondary lobes of X-like shaped radio galaxies in current paper. }
   \label{fig:majorvsminor}
\end{figure}

The variation of the angular size of the primary jets of XRGs with the angular size of the corresponding secondary jets is plotted in Fig \ref{fig:majorvsminor}. It should be remembered that measurement of the angular size of the secondary jet is difficult as most of the time it is diffused and always lack a hotspot at the end. So, the extent of secondary jets will depend on the spatial resolution and depth of the radio map. 

\begin{figure}
\includegraphics[width=6cm,angle=-90,origin=c]{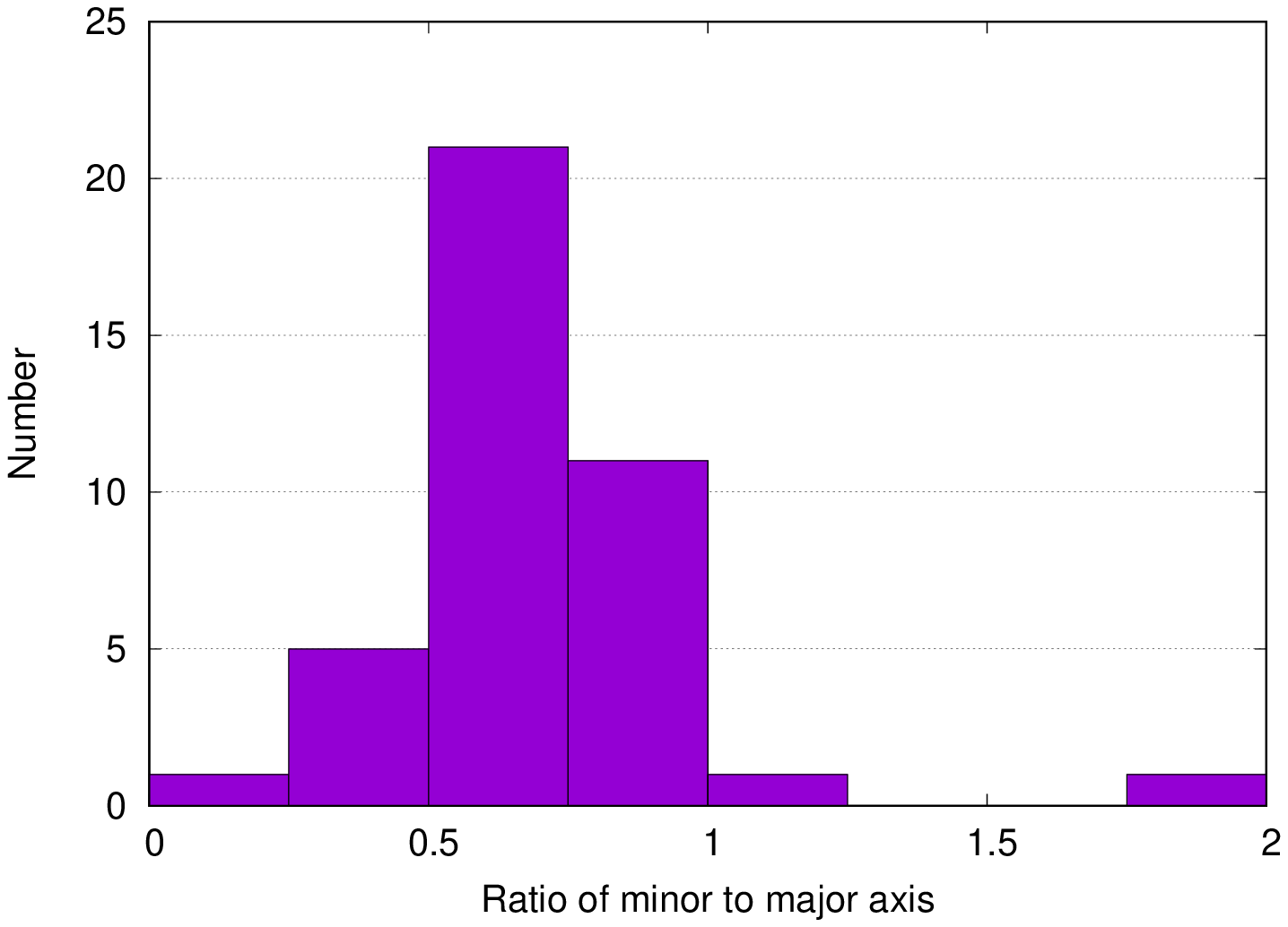}
    \caption{Histogram showing the distribution of ratio of primary and secondary lobe sizes of `X' shaped sources. }
    \label{fig:jetratio}
\end{figure}

For XRGs presented in this paper, a histogram with the distribution of ratio between the secondary and primary axes of radio galaxies is shown in Fig \ref{fig:jetratio}. Amongst the 40 XRGs included in the Figure, 21 XRGs have secondary jets with $50-75$ per cent size of the primary jets.  The secondary jets for 11 XRGs are $75-100$ per cent size of the primary jets. The secondary jets are bigger than the primary jets for 2 XRGs (1319+2938 and J2303–-1841). It should be noted that primary jets also could be affected by the projection effect. Wings with a size more than the primary lobes were found earlier also \citep{Ch07, Be20}. For example, 4C +32.25 shows prominent wings which are more than a factor of 2 extended than its lobes \citep{Kl95}. 
Among the source of XRGs with known redshift, J0157+0209 has the largest linear size of 1.68 Mpc (with $z=0.22$) and J2336+2108 has the smallest linear size 120 kpc (with $z=0.06$). J0157+0209 is a giant size XRG and J2336+2108 is the nearest XRG source within our sample.

In Fig \ref{fig:jetratio}, a histogram with the distribution of ratio between major and minor axis is shown. For most of the winged radio galaxies, the wing size is 50--75 per cent of the primary jet size. Of the sources with known redshift, J0329+3947 is the smallest source with a linear size of 0.23 Mpc.  

\subsection{Angle between the Major and Minor Axes in XRGs}
\label{subsubsec:axis-angle}

\begin{figure}
\includegraphics[width=6cm,angle=270,origin=c]{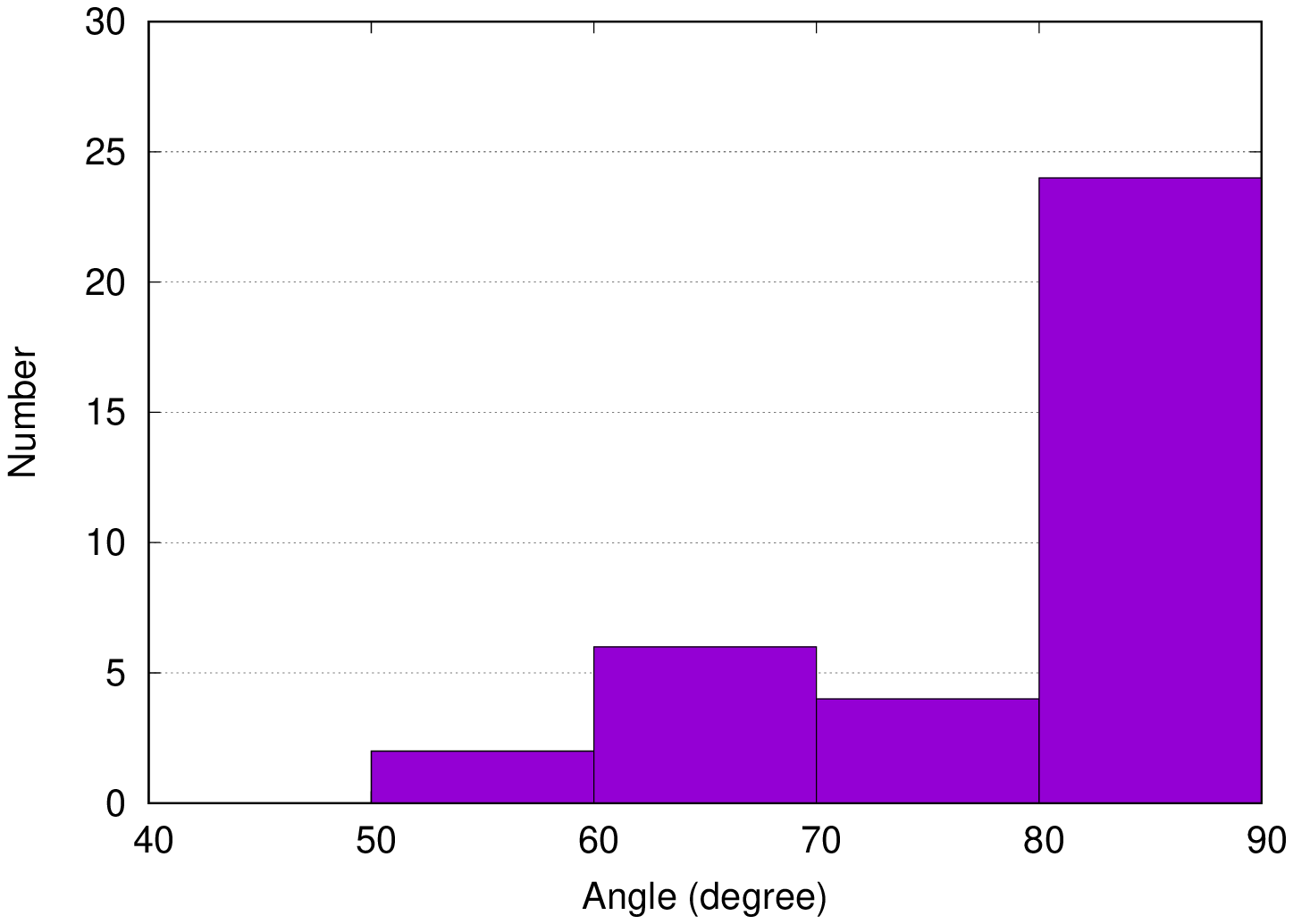}
  \caption{Histogram showing the angle distribution between primary and secondary lobes in XRGs. }
  \label{fig:X-angle}
\end{figure}

The distribution of the angle between primary and secondary axes in XRGs is shown in the histogram of Fig \ref{fig:X-angle}. A good fraction of the sources shows wings in a near perpendicular direction. For 60 per cent of the sources, the secondary wings make an $80-90$ degree angle with the primary lobes. Fig \ref{fig:X-angle} shows that the number of sources with smaller angles between primary and secondary axes gradually decreases. It should be noted that the deficit of wings at a narrow-angle to the primary jets is at least in part due to selection, as they would not be noticeably distinct unless at a substantial angle. 

\section{Discussion}
\label{sec:disc}
It is known that most of the winged radio galaxies are FR-II type \citep{Le92, Ch07, Ya19, Be20}. The present catalogue of XRGs also supports the trend. Twenty-five per cent (10) are FR-I radio galaxies and sixty-five per cent (26) are FR-II radio galaxies within 40 XRGs presented in the current paper. The jet morphology is complex for 4 XRGs and they can not be categorized.

Earlier, systematic searches to look for winged sources were carried out using the FIRST survey (C07; \citet{Be20, Ya19}) at 1400 MHz. Due to high resolution in FIRST image and absence of antennas with shorter baseline, a large number of XRGs and ZRGs with larger size and diffused emission were missed in FIRST based searches many of which is detected in the present study. FIRST also covered a limited portion of the sky compared to TGSS.   

\subsection{Classification of winged radio galaxies}
\label{sec:classification}
In this section, we attempt to classify winged sources presented in the current paper in terms of source morphologies as well as investigate them in terms of the previously established models. In most of the XRGs, the mini orthogonal secondary wings are less than primary jets (known as small-winged XRGs). Few XRGs have a one-sided single wing (known as Single-winged XRGs). In rare cases, wing sizes are greater than the primary jets. These galaxies are known as long-winged XRGs. For a few XRGs, secondary wings are misaligned from each other about the AGN and form Z-symmetry of wings \citep{Le84} (known as Z-symmetry XRGs). For ZRGs, wings originate near the outer point of the primary jet and form an S/Z-shaped radio galaxy. In this section, we will discuss the above-mentioned types of XRGs with their appropriate models. 

\subsubsection{Small-winged XRGs}
In our sample, 22/40 (55 per cent) XRGs have two small size wings which are stretched nearly orthogonally about the host galaxy. The wings of all these sources inherently originated from AGN. A ridge line of the two extended secondary wings is aligned from each other with an AGN. Out of 22 XRGs, 5/22 (23 per cent) sources have a wing size of less than 50 per cent of the primary lobes. The source with the least wing length within our sample is J0507+3057 (99.3 arcsec). 
 
Small-wings in XRGs are believed to grow due to the backflow of plasma in radio jets \citep{Le84, Wo95, Kr05}. In this model, the compact ISM of the highly elliptical galaxy plays a key role in the creation of a secondary lobe pair/wings. The backflowing synchrotron plasma from the two primary lobes experiences buoyancy pressure \citep{Ca02}. As a result, backflowing plasma is deflected away from the axis of the primary lobe, towards the minor axis, where the least resistance of the gas leads to the creation of small wings. The buoyant force effectively controls the prolongation of comparatively small wings \citep{De02}. The backflow model can explain the radio structures associated with the large FR-I \citep{Fa74a} galaxies. But the model is not able to explain why the secondary lobes are larger than the primary lobes in many radio galaxies \citep{Ho10}. 

X-shaped structures are mostly found in edge-brightened radio sources (FR-II) \citep{Ca02, Be20}. Out of 22 small winged XRGs in the present paper, 15 galaxies appear as edge-brightened radio sources and 5 galaxies appear as edge-darkened radio sources. Above mentioned 15 sources strongly support the backflow model for the formation of mini wings. Examples of FR-I type small-winged sources are against models based on backflows for the formation of the secondary radio wings \citep{Sa09}. In such cases, wings may form via flips in the radio ejection axis.      

\subsubsection{Long-winged XRGs}
Only two sources J1319+2938 and J2303--1841 in our sample have secondary wings that are expanded longer than the associated primary lobes. The two wings have nearly the same size. These two candidates are characterized by clear twin wings which create a little angle with the primary radio axis in the reverse direction. Moreover, the wings of the two sources are of FR-I type. The jet axis may have rotated over a small angle to create the pair of long wings with the jet toward the NW to SE for J2303--1841 and along NE to SW for J1319+2938.  

For a few cases, the back-flow diversion model \citep{Ca02} partially explain the above-mentioned properties. According to this model, a high ellipticity galaxy with its asymmetric ISM (excessive pressure gradient in the ISM of host galaxy minor axis), form a cocoon from strong backflowing plasma, the secondary wings may produce along the minor axis of the galaxy due to adjacent prolongation of the excessive-pressured cocoon, thus forming long wings \citep{Ca02}. A 3D numerical simulation indicates a supersonic origin and a subsonic evolution of wings \citep{Ho11}. Long-wing XRGs can be explained by the spin-flip model or rapid jet reorientation model.

\subsubsection{Single wing XRGs}
Some XRG candidates in our catalogue have a wing in only one direction or one of the two wings is very small ($<25\arcsec$). These sources are assigned in our catalogue as ``single-wing XRGs''. In our catalogue, 13 sources are classified as ``single-wing XRGs''. C07 also catalogued some of the XRGs with only one wing but recently, using VLA 1.4 GHz images, \citet{Ro18} found that J0941+2147, J1206+3812, and J1444+4147 have a very faint wing in the other direction.  
 
The identification of clear wings of XRGs depends on several factors: (1) the radio galaxy orientation or evolution, (2) directional offset from the primary lobe pair \citep{Ya19, Be20}. It will also depend on observational limitations connected to angular resolution and the sensitivity to diffuse emission \citep{Wa03, Ch07, Ch09}. The observed faintness and small sizes of wings could often be artefacts arising from energy losses suffered by the relativistic plasma radiating in these older/fossil radio components. To overcome this limitation, low-frequency radio imaging with sufficiently high angular resolution and sensitivity are mandatory \citep{Ya19}.

\subsubsection{Z-symmetry XRGs}

Several X-shaped radio galaxies show Z-symmetric morphologies in their short secondary wings \citep{Go03, Zi05}. The prevalent models (coalescence of two galaxies or backflow diversion of radio jets) can not explain these types of X-shaped radio sources. For these sources, two elongated secondary lobes/wings are not aligned with each other \citep{Mu01} (e.g. NGC 326 \citet{Le84}). The minor wings are offset from each other laterally. There are obvious emission voids in the middle of two wings \citep{Sa18}. Thus, in place of extending in diametrically reverse directions from the galactic nucleus, the lobes/wings exhibit an approximately Z-symmetry about the nucleus. 

In our samples, four XRGs (J0758+0511, J1005--1328, J1908+1838 and J1319+2938) display Z-symmetry of secondary wings. J0758+0511 is a strong Z-symmetry XRG with one edge-brightened lobe (lower end) and short wings. J1005--1328 has dual edge-brightened lobes with its short wings exhibiting an inner Z-shape. J1908+1838 is an inner Z-symmetry type source with a northern edge-brightened lobe and small wings. J1319+2938 is a candidate with one edge-brightened lobe. The northern wing is orthogonal to the primary lobe and another wing is slightly inclined.

The Z-symmetric XRG morphologies with weak secondary wings can not be physically explained by a merger between two galaxies and the backflow of plasma in radio jets. With the help of the spin-flip model \citep{Ro01, De02, Go03}, Z-symmetry of the wings can be easily understood. In this model, the key scenario is a merger between two radio galaxies and each radio galaxy is associated with a supermassive black hole in its centre. As the lighter galaxy spirals close to the common centre, it releases a great amount of gas to the ISM of the heavier active galaxy. Typically, the angular momentum axis of rotating gas is not aligned along the original jets axis \citep{Zie01, Me02}. A meaningful reorientation of the heavier active radio galaxy spin vector along a certain direction at a distance $\sim$10 kpc from the AGN due to adequate ram pressure \citep{Go03} form Z-symmetric secondary wings.  
 
\subsubsection{ZRGs}
Among the 58 winged sources, there are 18 sources with clearly S/Z-shaped structures (5 have FR-II classification and 10 have FR-I classification). Non-collinearity of hotspots or lobe extremities with the core is not uncommon in these sources with 6 of S/Z shaped sources. Remarkably, 3 ZRGs (J0041+2123, J0817+0708, and J1848--4240) have hotspots that are not seen in both lobe pairs. Two ZRGs (J0525--3242 and J0834+6635) are found in the brightest cluster galaxies (BCGs).  

The jet shell interaction model \citep{Go12} explains the S/Z shape of the secondary wings. The coupling between radio galaxy jets and stellar shells provides the basis for this process \citep{Go83}. According to this scenario, two gas-rich disk radio galaxies (different SMBH) form a system of binary black holes that creates a sequence of stellar shells. These shells are also found to contain a significant amount of gas. The rotating arc-shaped sized stellar shells have been seen in $\sim$10 per cent of local low-density environment elliptical galaxies, nicely oriented along the optical major axis \citep{Ma83}. The existent gas in a stellar shell deviates the radio galaxy jets, thus forming the secondary wings \citep{Go12}. This model has the following features: the physical size of the wings depends on the interaction duration between the radio jet and the stellar shells. Physical characteristics such as density and velocity play a role in determining this, as a result, extremely long structures can be formed. Even if this model can recreate the observed ZRG characteristics, only two ZRGs have verified stellar shells thus far: 3C 403 \citep{Ra11} and 4C+00.58 \citep{Ho10}. The S-symmetry of the radio galaxy is the signature of a common clockwise motion of revolution of the stellar shell complex \citep{Gs84, Go10}. This proposed revolution shells can interpret the detected inverse directed geometry of the secondary lobes/wings \citep{Go12}. Lastly, the seen Z-shaped morphology of the secondary wings about the AGN is another inherent result of this model \citep{Go03}. For ZRGs, two radially located partial shells create the Z-deflections of two opposite-sided outer direction radio wings which take place at the extreme position of the radio optical major axis from AGN. The numerical simulations and observation indicate that BCGs are the outcome of various galaxy mergers \citep{De07, Ra10}. Therefore, a system of shells may have been produced after one of these mergers, as found in some other BCGs \citep{Kl19}.

\subsection{Radio properties of XRGs and ZRGs}

All the previous studies \citep{Ma16, Is10, Ka98} found the mean spectral index of radio galaxies in the range of 0.7--0.8 and 0.75 are usually considered to be the typical spectral index for radio galaxies (RGs). For example, \citet{Ma16} found radio galaxies with spectral index in the range --0.5 to 1.5 with median=0.78 and \citet{Is10} found radio galaxies with spectral index in the range --0.5 to 2.5 with median=0.78. The mean and median of the spectral index of winged galaxies presented in the current article is close to the spectral index of a typical radio galaxy which implies that RGs and winged-radio galaxies are similar in terms of the properties of their spectral index. The absence of a radio core, the existence of diffuse amorphous shaped low-surface-brightness emission, and small angular diameters ($<30''$) are the most common morphological characteristics used to identify remnants \citep{Si21}. Most of the XRGs and ZRGs are active RGs and not remnant or dead. Few circumstances cause a small number of RGs to transform into XRGs and ZRGs. The mean and median of the $\alpha^{1400}_{150}$ of XRGs in \citet{Ch07} are 0.70 and 0.73, and for ZRGs are 0.67 and 0.67 respectively. In \citet{Be20}, the mean and median of the $\alpha^{1400}_{150}$ of XRGs are 0.71 and 0.69 respectively. In \citet{Ya19}, the mean and median of the $\alpha^{1400}_{150}$ of XRGs are 0.70 and 0.69 respectively. These show that the mean and median $\alpha^{1400}_{150}$ of winged sources in our sample are nearly the same as other winged sources found from the FIRST survey.

Based on the availability of redshift data, we estimate $L_{150}$ for 12 XRGs and 5 ZRGs from our catalogue. The mean and median values confirm that XRGs and ZRGs have the same distribution of radio luminosity which means that they have a similar order of jet kinetic power. It indicates that the governing conditions in the core engines of XRGs and ZRGs may be the same and as a result, the radio-luminosity of XRGs and ZRGs are in the same order.  The mean and median of $\log ~L_{150}$ [W Hz$^{-1}$] of XRGs  in \citet{Ch07} are 25.50 and 25.32, and in \citet{Be20} are 26.68 and 26.49 respectively. For easy comparison, all the luminosities from different surveys are converted to that of 150 MHz using the TGSS survey.

It should be noted that detected sources in other surveys have been removed from the sample of X and Z shaped sources in the present paper. So, this is not a complete sample. But none of the conclusions in this section should change if those sources were included.

The Malmquist bias means that there will be a tight correlation between luminosity and redshift in a single flux-limited sample. According to \citet{Bl99}, tight correlation depends on (i) the steepness of the distribution in jet kinetic powers, (ii) the energy distribution of the particles injected into the lobes. In our catalogue, XRGs-ZRGs show a strong correlation between projected linear sizes with $z$. This is expected because radio luminosity strongly correlates with two parameters (projected linear size and redshift). But, at high redshift, a negative correlation is established between projected linear size and redshift. The density ($\rho_{\textrm{medium}}$) of the intergalactic medium (IGM) raises with redshift significantly as $(1 + z)^{4}$. The brightness of the radio source surface shrinks as $(1+z)^{-4}$ \citep{Ka98}. 
 
\subsection{Power-linear size diagram}
\begin{figure}
\includegraphics[width=6cm,angle=270,origin=c]{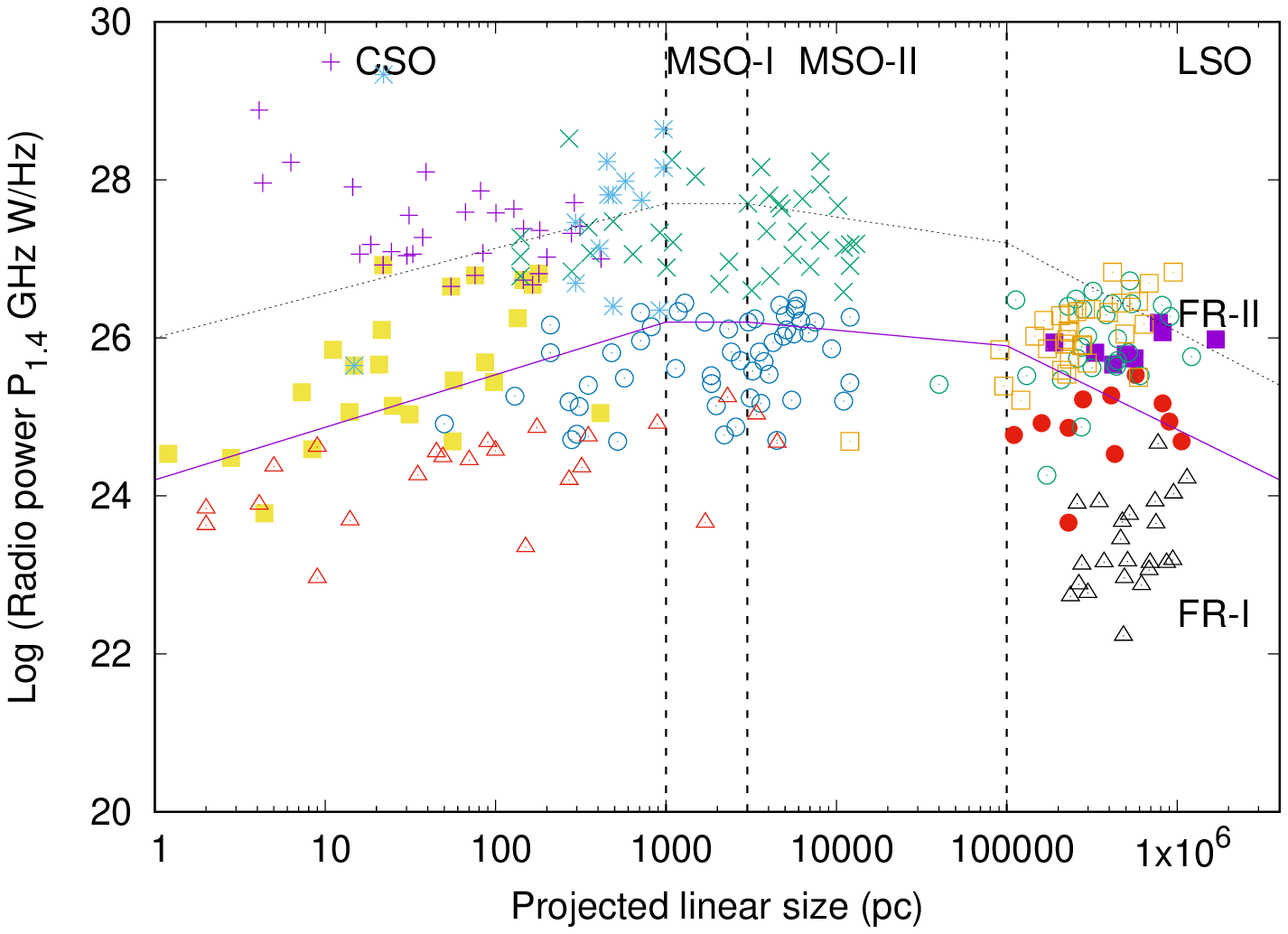}
\caption{Plot showing the variation of the radio power with the linear size of Compact Symmetric Objects (CSOs), Medium-sized Symmetric Objects (MSOs), and Large Symmetric Objects (LSOs) of low and high radio power. The evolutionary paths for the high radio power and low radio power sources are depicted by black and violet dashed lines. The violet filled-squares are TGSS-XRGs (FR-II) and filled red circles are TGSS-ZRGs (FR-I); brown open square and green open circles are FIRST XRGs and ZRGs \citep{Be20}; the yellow squares are low power smallest CSOs \citep{Ku06, Ku10}; the sky-blue circles are low-power Compact Steep Spectrum sources (CSSs) \citep{Da02a, Da02b}; brown open squares are (FR-II); black open-triangles are low power FR-I giant radio sources \citep{La01}, violet cross marks are high power CSOs \citep{Ku06, Ku10}, asterisks are high power GHz Peaked Spectrum (GPS) sources \citep{Xi06, Li07}, green pluses are high power CSS \citep{Fa01}, and red triangles are low power GPS \citep{De09}.}
  \label{fig:X-Z-evolution}
\end{figure}

For the entire family of extra-galactic radio sources, the study of the variation of radio power $P_{rad}$ versus source linear size ($D$) helps to understand their evolutionary properties \citep{Od97, Ku10}. In Fig \ref{fig:X-Z-evolution}, a diagram with the variation of $P_{rad}$ with $D$ is shown which helps to study evolutionary paths for the high radio power and low radio power sources.

Few observables are used to characterise the dynamic evolution of extra-galactic double radio emissions, such as the kinetic power of lobes, the total length of the source, the speed with which the terminal hotspots are advancing, and the density gradients of the ambient medium in the host galaxy along the path of the jets and the lobes \citep{An12}. To retain the overall shape of a symmetric double radio source, modelling of hotspots and lobes as well as the bow shock has been done \citep{Ka07, Ka97}. The power of the source, the local environment of the host galaxy, and the evolutionary age affect the structural and spectral properties of radio sources \citep{Ka07, An12}. Four prominent stages are related to the physical processes of radio sources \citep{An12}. Large-scale Symmetric Object (LSO) is the last evolutionary stage of radio sources which is populated with fully developed FR-I and FR-II sources with a source size of $\ge$ 10 kpc. The radio power decreases sharply as $P_{\textrm{rad}} \propto D^{-1.6}$, where inverse Compton losses originating from the cosmic microwave background (CMB) dominate over the synchrotron losses.

Fig \ref{fig:X-Z-evolution} clearly depicts the above-described evolutionary stages. Earlier, \citet{An12} drawn radio power versus the linear extent of large-scale radio sources of compact, medium-sized, and large symmetric objects for various morphological types (Mtype) and mentioned seven Mtype evolutionary stages. We compare our sample of XRGs and ZRGs with low-power Mtype 1 (FR-I) and Mtype 2 (FR-II) evolutionary pathways of double radio sources and radio powers are continuously growing until a distance of about 100 kpc as visible in Fig \ref{fig:X-Z-evolution}.

The radio power-size (P-D) figure also shows how Mtype 2 sources (FR-II) evolve along evolutionary tracks. Mtype 2 double radio sources have powerful jets with prominent well-confined lobes but the central core region is weak and have no intervening jet structure. They have well-contained lobes with significant terminal hotspots at the leading edge of the lobes, as well as a sharp edge-brightened morphology, which indicates a strong advancing shock. Mtype 2 sources are found in the Compact Symmetric Object (CSO), Medium-sized Symmetric Object (MSO), and Large Symmetric Object (LSO) populations and have sufficient and long-lived jet power. 

Mtype 1 radio sources are characterized by less prominent jets, diffuse, diverging, and less confined lobes or flaring lobes, and hotspots located away from the lobes. Mtype 1 sources are seen in the lower-power CSO, MSO, and LSO sources that have reached the jet instability region in the P–D diagram. Mtype 1 MSO and LSO sources are observed in the FR-I sources with hotspots at $<$50 per cent of the arm lengths.

FR-I sources found in our ZRG sample is consistent with Mtype 1 sources as mentioned in \citet{An12} and FR-II sources in our XRG sample is consistent with Mtype 2 sources as mentioned in \citet{An12}.

\section{Conclusion}
\label{sec:conclusion}
We look for X shaped and Z shaped radio galaxies from TGSS ADR 1 at 150 MHz. Our main findings in this paper are as follows:
\begin{description}

\item[$\bullet$] A total of 58 candidate winged radio sources are discovered, of which 40 are XRGs and 18 are ZRGs. This discovery helps to significantly increase the number of known XRGs and ZRGs and opens up the possibility of follow-up observations of these sources with high-resolution deep radio observations and observations in other wavebands. 
\item[$\bullet$] Out of 58 XRGs and ZRGs presented in the current paper, 19 (32 per cent) are FR-I and 33 (57 per cent) are FR-II radio galaxies. 

\item[$\bullet$] Optical/IR counterparts are identified for 27 out of 40 XRGs (68 per cent) and 16 out of 18 ZRGs (88 per cent). 

\item[$\bullet$] Most of the XRGs show a steep spectral index between 150 MHz and 1400 MHz and only 13 per cent of the sources show a flat spectrum but for ZRGs, a good fraction of the sources (22 per cent) show flat spectrum.

\item[$\bullet$] The XRGs and ZRGs detected in the present study are slightly more luminous compared to those detected from the FIRST survey at 1400 MHz. The average value of luminosities of XRGs and ZRGs are close to the FR-I-FR-II division as found previously (viz. C07). 

\item[$\bullet$] The present sample does not give any conclusive evidence about the origin of X or Z shaped morphologies, though some of the models seem to work well for a group of sources. In section \ref{sec:classification}, we discussed different models for different sub-classes of winged radio sources.
\end{description}
The discovery of a large number of XRGs and ZRGs shows that this type of source is not uncommon and we expect to discover more such objects with a future deeper and high-resolution survey. High-resolution multi-frequency observations of these sources are encouraged to understand the nature of these sources in more detail.
\section*{acknowledgements}
We thank Paddy Leahy and one anonymous reviewer for making helpful suggestions. This research has made use of the NASA/IPAC Extragalactic Database (NED) which is operated by the Jet Propulsion Laboratory, California Institute of Technology, under contract with the National Aeronautics and Space Administration. This publication makes use of data products from the Two Micron All Sky Survey, which is a joint project of the University of Massachusetts and the Infrared Processing and Analysis Center/California Institute of Technology, funded by the National Aeronautics and Space Administration and the National Science Foundation. 
 
\section*{Data Availability Statement}
The data that support the plots within this paper and other findings of this study are available from the corresponding author upon reasonable request. The TGSS ADR 1 images are available at \href{http://tgssadr.strw.leidenuniv.nl/doku.php}{http://tgssadr.strw.leidenuniv.nl/doku.php}.


\begin{landscape}
\begin{table}
    \caption{\bf Candidate X-shaped radio sources}
	\label{tab:X-shaped}
\begin{tabular}{cccccccccccccccccll}
\hline
Cat &Name &R.A.      & Decl.     &Ref. &$F_{150}$&$F_{1400}$ &$\alpha$     &$z$ &$A_{1}$&$A_{2}$&$D_{1}$&$D_{2}$&Luminosity  &Other  &Common &Class\\
No  &     &(J2000.0) &(J2000.0)  &     &(Jy)     &(Jy)       &($\pm$0.05)  &    &($''$) &($''$) &(Mpc)  &(Mpc)  &W Hz$^{-1}$  &Catalog&Name&      \\
    &     &          &           &     &         &           &             &    &       &       &       &       &$\times10^{25}$&    & &     \\
(1) & (2) & (3)      & (4)       & (5) & (6)     & (7)       & (8)         &(9) & (10)  & (11)  & (12)  &(13)   &(14)           &(15)&(16) &(17)\\
\hline
~~1$^{a}$   &J0013--1930&00 13 25.8  &--19 30 05 &WISEA  &1.84 &0.23 &0.93         &0.10&174.0 &133.8 &0.33   &0.25   &5.23        &1, 2    &-- &FR-I \\
~~2$^{a}$   &J0020+2351 &00 20 39.2  &+23 51 32  &WISEA  &1.25 &0.51 &0.40         &--  &144.0&119.8   &--     &--     &--          &1, 9 &--   & FR-II \\
~~3$^{b}$   &J0039+2114 &00 39 42.4  &+21 14 06  &--    &1.27 &0.14 &0.98         &0.06&165.2&104.0  &0.19   &0.13   &1.17        &1 &NGC 193   &-- \\
~~4$^{b}$   &J0148--3156&01 48 15.5  &--31 56 39 &--     &1.76 &1.27 &0.14         &--  &217.2&195.4  &--     &--     &--          &1 &--      &FR-II \\
~~5$^{a}$   &J0157+0209 &01 57 52.5  &+02 09 54  &SDSS   &2.85 &0.68 &0.64         &0.22&454.1&233.3  &1.68   &0.85   &40.5        &1 &--     &FR-II \\
~~6$^{a}$   &J0224+7030 &02 24 19.3  &+70 30 35  &--     &1.43 &0.31 &0.68         &--  &163.7&124.7  &--     &--     &--          &1 &--     &FR-II \\
~~7$^{a}$   &J0234+4712 &02 34 31.7  &+47 11 56  &WISEA  &2.69 &0.54 &0.71         &--  &240.3&99.5   &--     &--     &--          &1, 14&B3 0231+469 &FR-II \\ 
~~8$^{b}$   &J0317+4917 &03 17 36.4  &+49 17 27  &WISEA  &1.06 &0.36 &0.48         &--  &201.2&105.0  &--     &--     &--          &1, 9&--    &FR-II \\
~~9$^{a}$  &J0318+5755 &03 18 58.0  &+57 55 04  &--     &4.27 &0.35 &0.72         &--  &336.2&167.8  &--     &--     &--          &2, 5&--    &FR-II \\ 
~~10$^{a}$  &J0340--3113&03 40 21.7  &--31 13 48 &2MASS  &0.71 &0.09 &0.91         &--  &188.2&126.5  &--     &--     &--          &14, 19&--  &FRII \\
11$^{b}$  &J0351--2744&03 51 35.7  &--27 44 35 &2MASX  &24.0 &5.01 &0.70         &0.06&353.1 &226.5 &0.42   &0.24   &21.9        &1, 14 &--  &FR-II \\
~~12$^{a}$  &J0503+2300 &05 03 03.5  &+23 00 30  &2MASX  &2.22 &0.33 &0.85         &--  &158.4 &111.0 &--     &--     &--          &1 &--     &FR-II \\
~~13$^{a}$  &J0507+3057 &05 07 16.7  &+30 57 58  &WISEA  &2.48 &0.55 &0.67         &--  &413.2 &~99.3  &--     &--     &--          &1, 10, 14&--&FR-II \\
~~14$^{b}$  &J0618--4844&06 18 05.4  &--48 45 00 &MRC    &4.95 &--   &--           &--  &569.7 &252.0 &--     &--     &--          &9, 19&--   &FR-I \\
~~15$^{a}$  &J0707+5929 &07 07 29.3  &+59 29 48  &2MASX  &2.89 &0.69 &0.64         &--  &263.7 &167.8 &--     &--     &--          &1, 5&--    &-- \\
~~16$^{a}$  &J0717+0638 &07 17 36.6  &+06 38 37  &--     &0.43 &0.06 &0.88         &--  &121.7 &106.0 &--     &--     &--          &1  &--    &FR-II \\   
~~17$^{b}$  &J0724+3803 &07 24 06.8  &+38 03 49  &SDSS   &2.02 &0.31 &0.83         &0.24&146.2&110.0  &0.56   &0.38   &36.1        &1, 11, 15&--&FR-II \\
~~18$^{b}$  &J0731--5238&07 31 04.9  &--52 38 08 &2MASX  &4.88 &--   &--           &0.09&372.8&263.9  &0.67   &0.41   &--          &19&--     &FR-II \\
19  &J0758+0511 &07 58 16.3  &+05 11 18  &SDSS   &2.02 &0.19 &1.05         &--  &171.8&123.6  &--     &--     &--          &1, 6&--    &FR-II &\\
~~20$^{b}$  &J0822+0557 &08 22 31.9  &+05 57 07  &--     &1.51 &1.93 &0.20         &0.08&432.6&258.1  &0.41   &0.28   &2.39        &1, 3, 15&3C 198 &FR-I \\
~~21$^{a}$  &J0949+7313 &09 49 16.8  &+73 13 30  &--     &5.96 &1.66 &0.57         &--  &263.8&257.8  &--     &--     &--          &1, 5&--    &FR-II \\
22        &J1005--1328&10 05 02.8  &--13 28 58 &--     &2.66 &0.27 &1.02         &--  &138.6&132.2  &--     &--     &--          &1&--      &FR-II \\
~~23$^{a}$  &J1015+6823 &10 15 18.9  &+68 23 58  &SDSS   &1.81 &0.38 &0.69         &--  &199.4&~93.7   &--     &--     &--          &1, 5&--    &FR-I\\
~~24$^{b}$  &J1107+1551 &11 07 15.3  &+15 51 58  &WISEA  &1.87 &0.21 &0.97         &--  &186.2&124.6  &--     &--     &--          &1, 15&--   &FR-II \\
25        &J1319+2938 &13 19 03.9  &+29 38 35  &2MASX  &3.35 &1.37 &0.40         &0.07&186.2&188.2  &0.28   &0.23   &4.13        &1,18 &--  &FR-I \\
~~26$^{b}$  &J1341+2622 &13 41 54.1  &+26 22 51  &SDSS   &3.74 &0.37 &1.03         &0.08&329.3&170.0  &0.50   &0.25   &6.36        &1, 18&--   &FR-II\\
~~27$^{b}$  &J1534+1016 &15 34 18.0  &+10 16 49  &SDSS   &1.45 &0.37 &0.61         &--  &196.9&180.6  &--     &--     &--          &1, 2&--    &FR-II\\

~~28$^{a}$  &J1630+1435 &16 30 17.6  &+14 35 07  &--     &2.32 &0.48 &0.70         &--  &258.8&161.4  &--     &--     &--          &1, 15&--   &FR-I \\
~~29$^{a}$  &J1830+0059 &18 30 54.2  &+00 59 38  &--     &3.12 &0.37 &0.95         &--  &146.7&112.4  &--     &--     &--          &1, 16&--   &FR-II \\
~~30$^{a}$  &J1839+0452 &18 39 20.7  &+04 52 42  &--     &0.91 &0.11 &0.94         &--  &231.5&~97.4   &--     &--     &--          &1 &--     &FR-II \\
~~31$^{a}$  &J1906+4824 &19 06 56.5  &+48 24 20  &2MASX  &1.50 &0.11 &1.16         &--  &233.2&135.6  &--     &--     &--          &1, 2&--    &FR-II \\
32        &J1908+1938 &19 08 09.8  &+19 38 16  &WISEA  &0.77 &0.07 &1.07         &--  &108.1&~99.6   &--     &--     &--          &1&--      &FR-II \\
~~33$^{b}$  &J2033+2146 &20 33 32.0  &+21 46 22  &WISEA  &10.6 &1.90 &0.76         &0.17&262.0&171.3  &0.78   &0.55   &87.6        &1&4C +21.55&FR-II\\
~~34$^{b}$  &J2043--2633&20 43 42.9  &--26 33 23 &WISEA  &9.54 &2.51 &0.59         &--  &340.3&179.5  &--     &--     &--          &1&--      &-- \\
~~35$^{a}$  &J2202+6538 &22 02 45.4  &+65 38 43  &--     &1.93 &0.29 &0.84         &--  &180.0&118.2  &--     &--     &--          &1&--      &FR-I\\
~~36$^{b}$  &J2251+1717 &22 51 54.8  &+17 17 10  &WISEA  &3.67 &0.66 &0.76         &--  &229.4&124.1  &--     &--     &--          &1, 16&--   &-- \\
37        &J2303--1841&23 03 03.0  &--18 41 26 &WISEA  &4.47 &1.36 &0.53         &0.06&161.9&287.2  &0.23   &0.18   &4.04        &1, 13&PKS 2300-18&FR-II \\
~~38$^{a}$  &J2336+2108 &23 36 30.4  &+21 08 46  &--     &7.72 &1.33 &0.78         &0.05&250.3&154.4  &0.16   &0.12   &4.87        &1&IC 5338 &FR-I \\
~~39$^{b}$  &J2337--1752&23 37 56.6  &--17 52 20 &WISEA  &0.78 &0.08 &1.01         &1.14&179.7&123.8  &1.06   &1.00   &600.4       &17&PKS 2335-18&FR-I \\
~~40$^{a}$  &J2341--1620&23 41 12.8  &--16 20 53 &--     &6.30 &1.40 &0.67         &--  &244.8&210.8  &--     &--     &--          &1, 8&--    &FR-1 \\
\hline
\end{tabular}
\end{table}
\end{landscape}

\begin{landscape}
\begin{table}
    \caption{\bf Candidate Z/S-shaped radio sources}
	\label{tab:Z-shaped}
  \begin{threeparttable}
\begin{tabular}{cccccccccccccll}
\hline
Cat &  Name      & R.A.        & Decl.       & Ref.   &$F_{150}$ &$F_{1400}$ &$\alpha$&$z$  &Luminosity &Z-length  &Linear  &Others&Class \\
No  &            & (J2000.0)   & (J2000.0)   &        &(Jy)      &(Jy)       &($\pm$0.05)   &     &WHz$^{-1}$ &($''$)&Size  &catalog& \\
    &            &             &             &        &          &           &        &     &$\times10^{25}$&  &(Mpc)   &      & \\          
(1) &  (2)       & (3)         & (4)         & (5)    & (6)      & (7)       & (8)    &(9)  & (10)  & (11)      & (12) &(13)   &(14)\\
\hline
1   &J0041+2123  &00 41 47.1   &+21 23 28    &SDSS    &~3.79     &1.09       &0.55     &0.09 &12.47     &272.5 &--  &1, 3&-- & \\
2   &J0123+4858  &01 23 15.5   &+48 58 33    &--      &~1.98     &0.68       &0.47     &--   &--        &278.8 &--  &1, 4, 5&FR-II &\\
3   &J0141+1213  &01 41 09.9   &+12 13 52    &WISEA   &~1.68     &0.49       &0.55     &--   &--        &176.2 &--  &1, 8&FR-I &\\
4   &J0216--4749 &02 16 45.1   &--47 49 09   &ES0     &~3.45     &--         &--       &0.06 &--        &561.4 &--  &9&FR-I &\\
5   &J0329+3947  &03 29 23.9   &+39 47 32    &UGC     &~4.54     &0.35       &1.14     &0.03 &0.60      &371.2 &0.23 &4, 10&FR-I &\\
6   &J0435+2130  &04 35 04.0   &+21 30 25    &--      &~0.48     &1.39       &0.47     &--   &--        &155.3 &--  &1, 6&FR-I &\\
7   &J0525--3242 &05 25 27.2   &--32 42 37   &WISEA   &~4.48     &0.79       &0.78     &--   &--        &437.3 &--  &1, 13&FR-I &\\
8   &J0558+4316  &05 58 06.4   &+43 16 47    &WISEA   &~1.75     &0.15       &1.08     &--   &--        &235.1 &--  &1, 4, 10&FR-II &\\ 
9   &J0817+0708  &08 17 16.1   &+07 08 46    &SDSS    &~1.51     &0.44       &0.55     &0.27 &32.4      &193.7 &0.82&1, 3&--\\
10  &J0834+6635  &08 34 09.0   &+66 35 49    &WISEA   &~2.91     &0.27       &1.06     &--   &--        &165.7 &--  &1, 14&FR-II &\\
11  &J0848+0555  &08 48 37.8   &+05 55 19    &SDSS    &~6.74     &0.94       &0.88     &--   &--        &269.5 &--  &1, 7&FR-II &\\
12  &J0934--3728 &09 34 06.5   &--37 28 10   &WISEA   &~5.96     &0.34       &1.28     &0.06 &5.98      &359.6 &0.43  &1, 11&FR-I &\\
13  &J1309+1029  &13 09 04.6   &+10 29 55    &WISEA   &~3.22     &1.08       &0.51     &0.08 &8.27      &252.4 &--  &1&FR-I &\\
14  &J1639+5546  &16 39 50.3   &+55 46 11    &WISEA   &~1.52     &0.56       &0.64     &0.24 &23.8      &232.3 &0.90&1&FR-I &\\
15  &J1705+0004  &17 05 28.2   &+00 04 24    &WISEA   &~2.18     &0.27       &0.93     &--   &--        &213.1 &--  &1&-- &\\
16  &J1848--4240 &18 48 31.3   &--42 40 50   &WISEA   &~2.90     &--         &--       &0.12 &--        &331.6 &--  &9&FR-II &\\ 
17  &J2125+3738  &21 25 42.6   &+37 38 29    &WISEA   &~3.88     &1.57       &0.40     &--    &--       &257.8 &--  &1, 2, 3&FR-I &\\
18  &J2307+1920  &23 07 26.5   &+19 20 45    &2MASX   &~1.34     &0.57       &0.38     &0.15  &8.02     &212.8 &0.57&1, 11&FR-II &\\
\hline
\end{tabular}
 \begin{tablenotes}
      \small
\item 1: NVSS \citep{Co98}; 2: VLSS \citep{Co07}; 3: 4C \citep{Pi65, Go67, Ca69}; 4: 6C \citep{Ha91}; 5: 87GB \citep{Gr91}; 6: MGC \citep{Vo62}; 7: NVGRC \citep{Pr16}; 8: VFK \citep{Va15}; 9: SUMSS \citep{Ma03}; 10: Cul \citep{Sl95}; 11: MG1 \citep{Be86}; 12: WB \citep{Wh92}; 13: NGC \citep{Dr88}; 14: ABELL \citep{Ab58, Ab89}
 \end{tablenotes}
  \end{threeparttable}
\end{table}
Notes---\\ 
$^a$ indicates small-winged XRGs\\
$^b$ indicates single-wing XRGs\\

\end{landscape}
\begin{figure*}
\psfig{figure=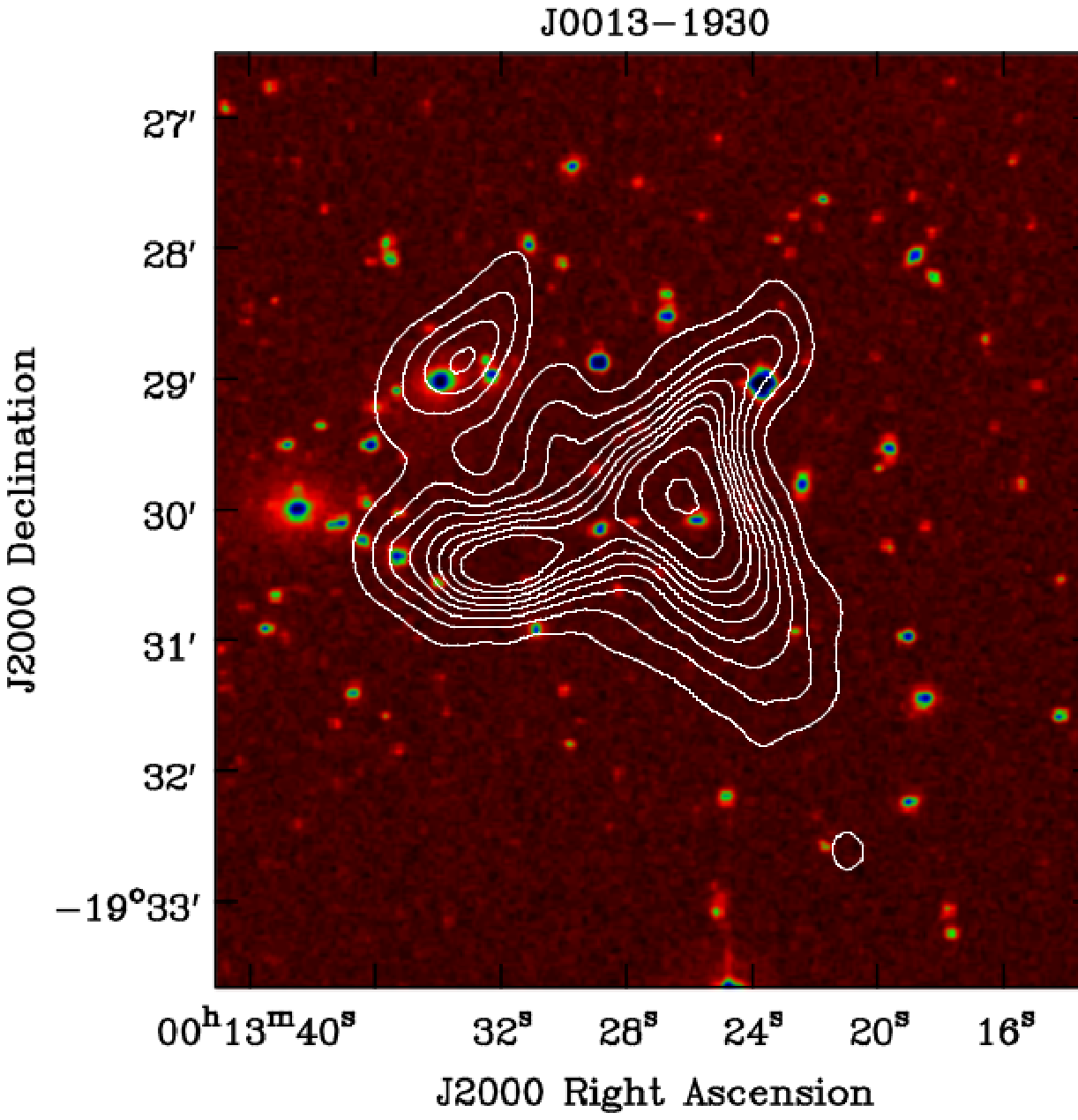,width=5.5cm,height=5cm} 
\psfig{figure=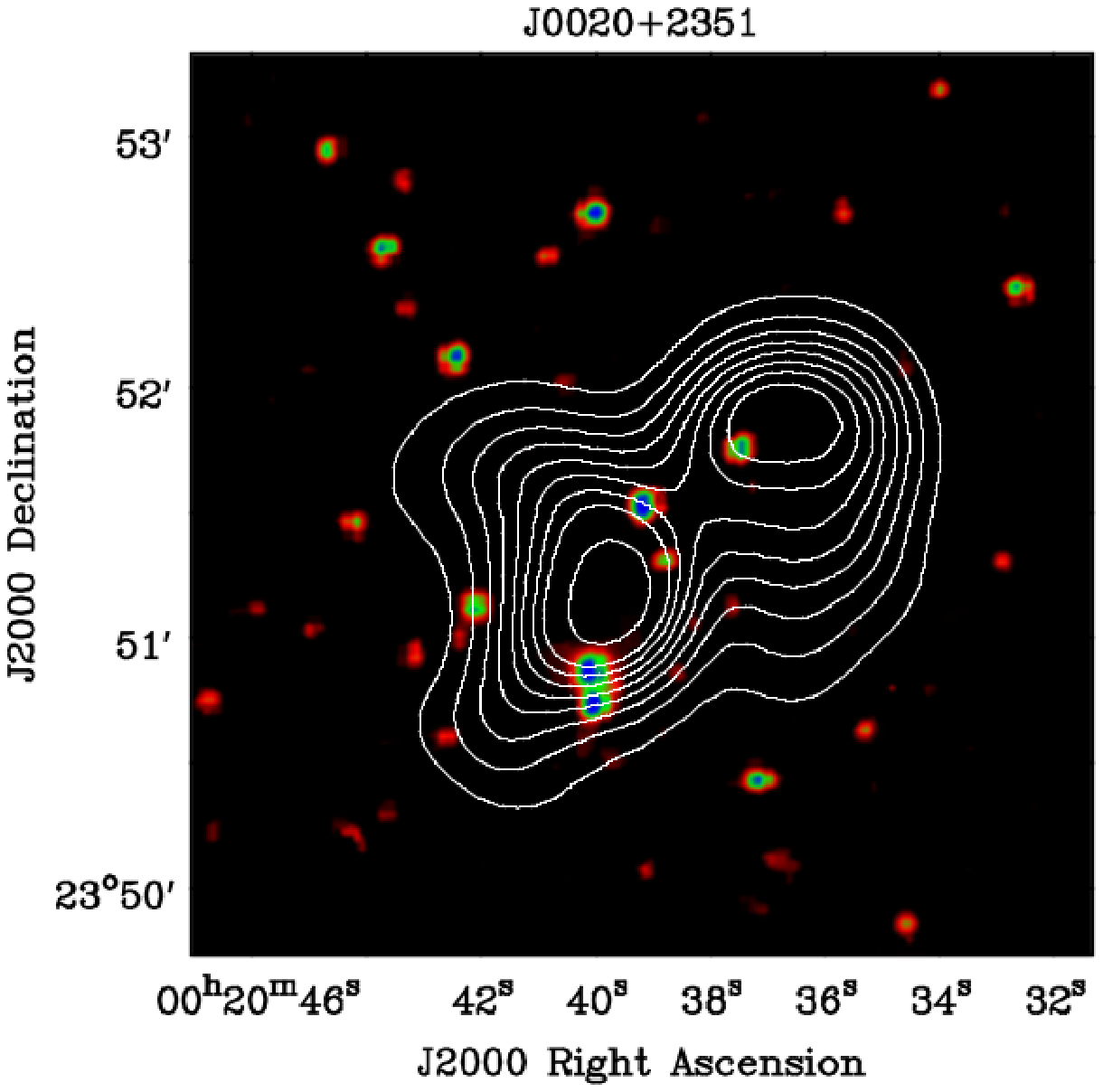,width=5.5cm,height=5cm} 
\psfig{figure=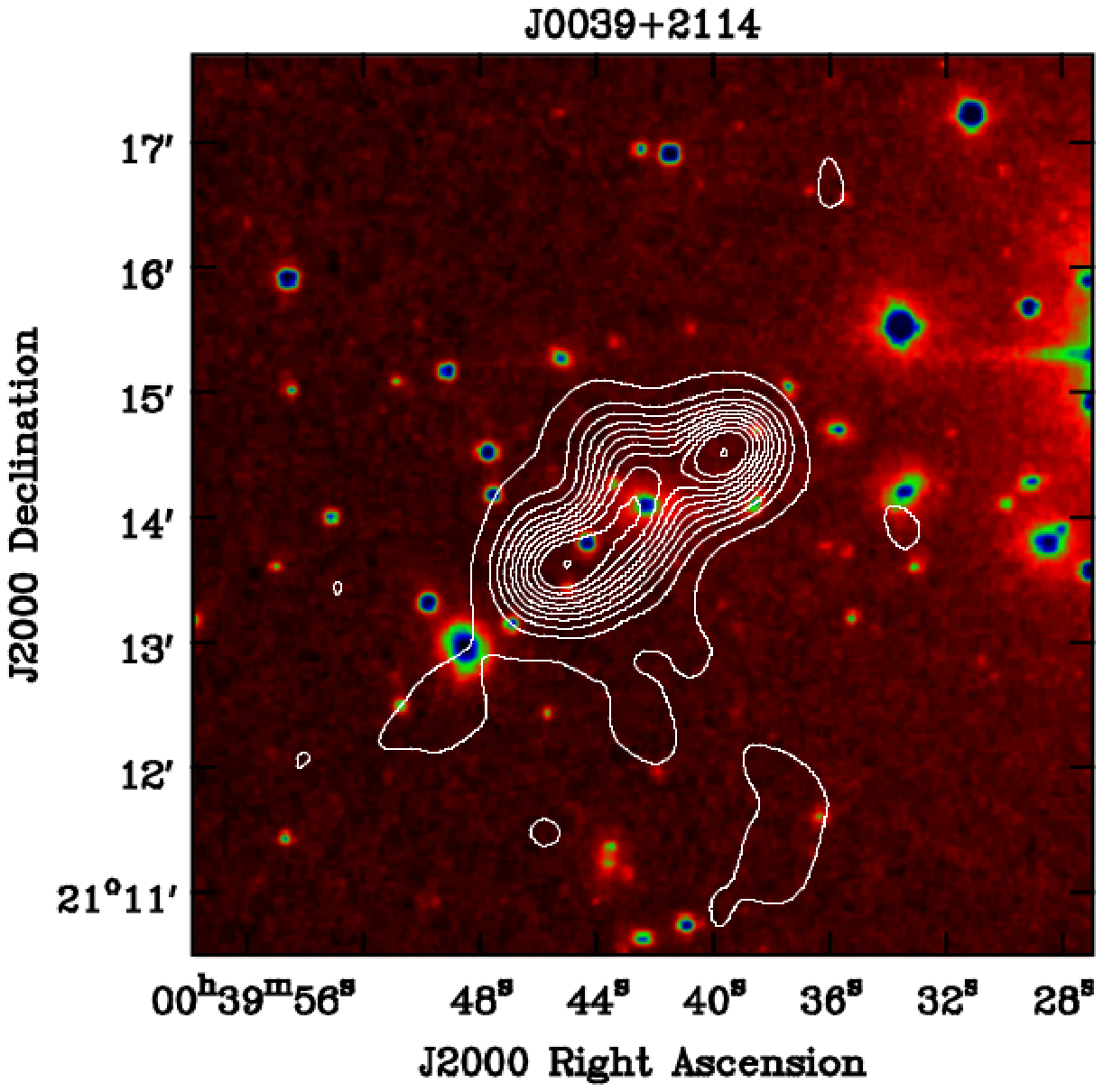,width=5.5cm,height=5cm}
\vskip 0.8cm         
 
\psfig{figure=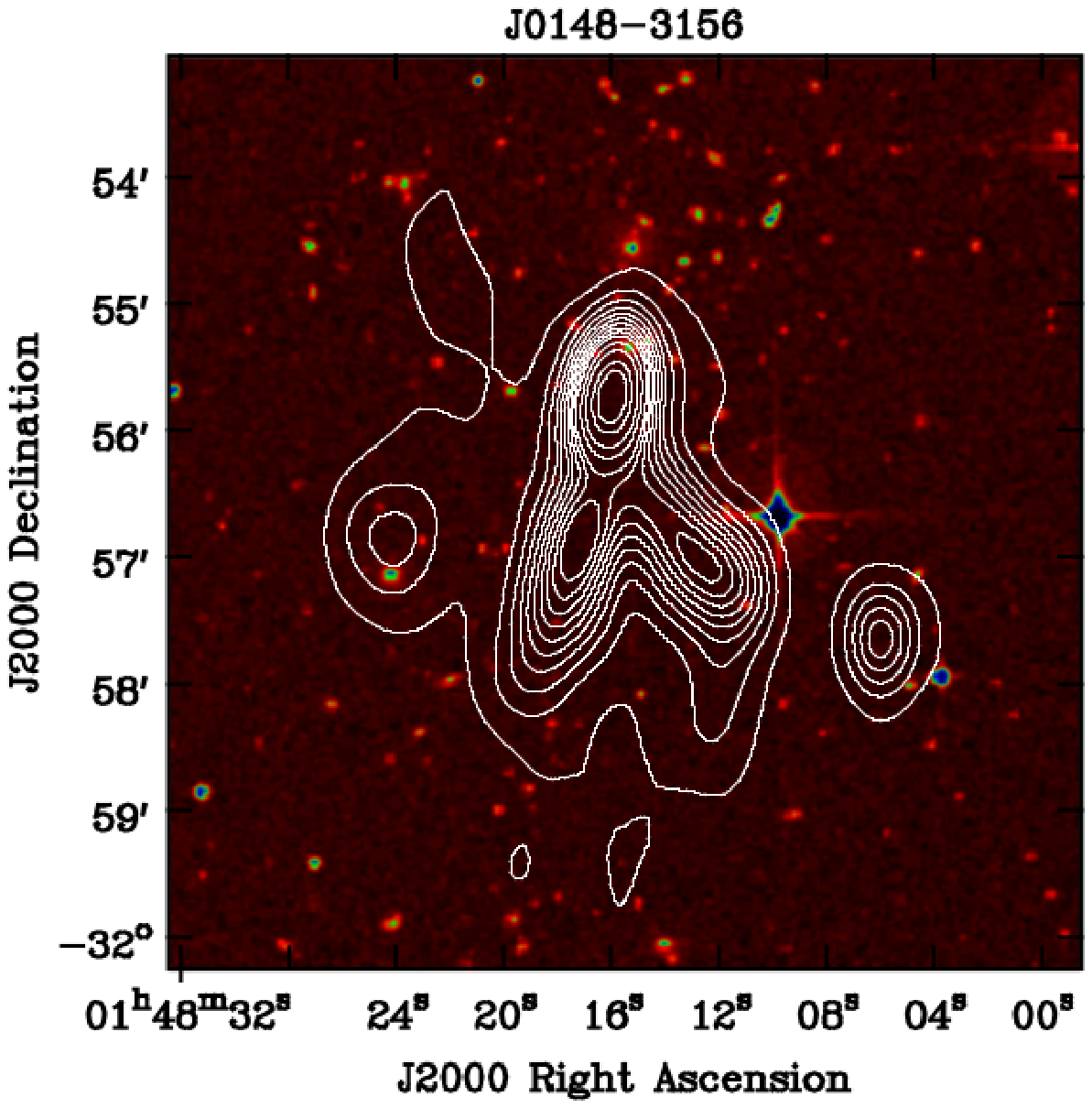,width=5.5cm,height=5cm}
\psfig{figure=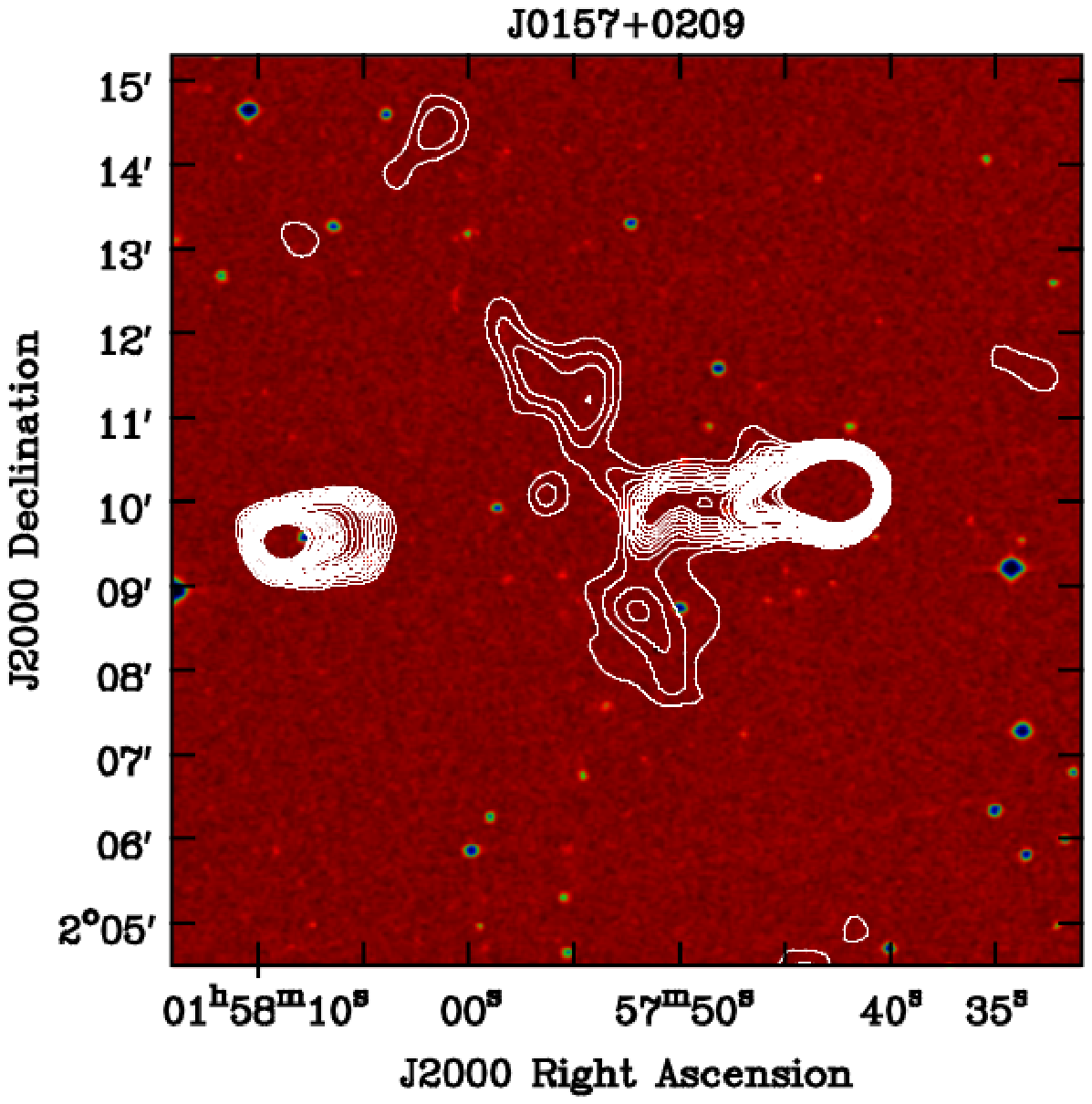,width=5.5cm,height=5cm} 
\psfig{figure=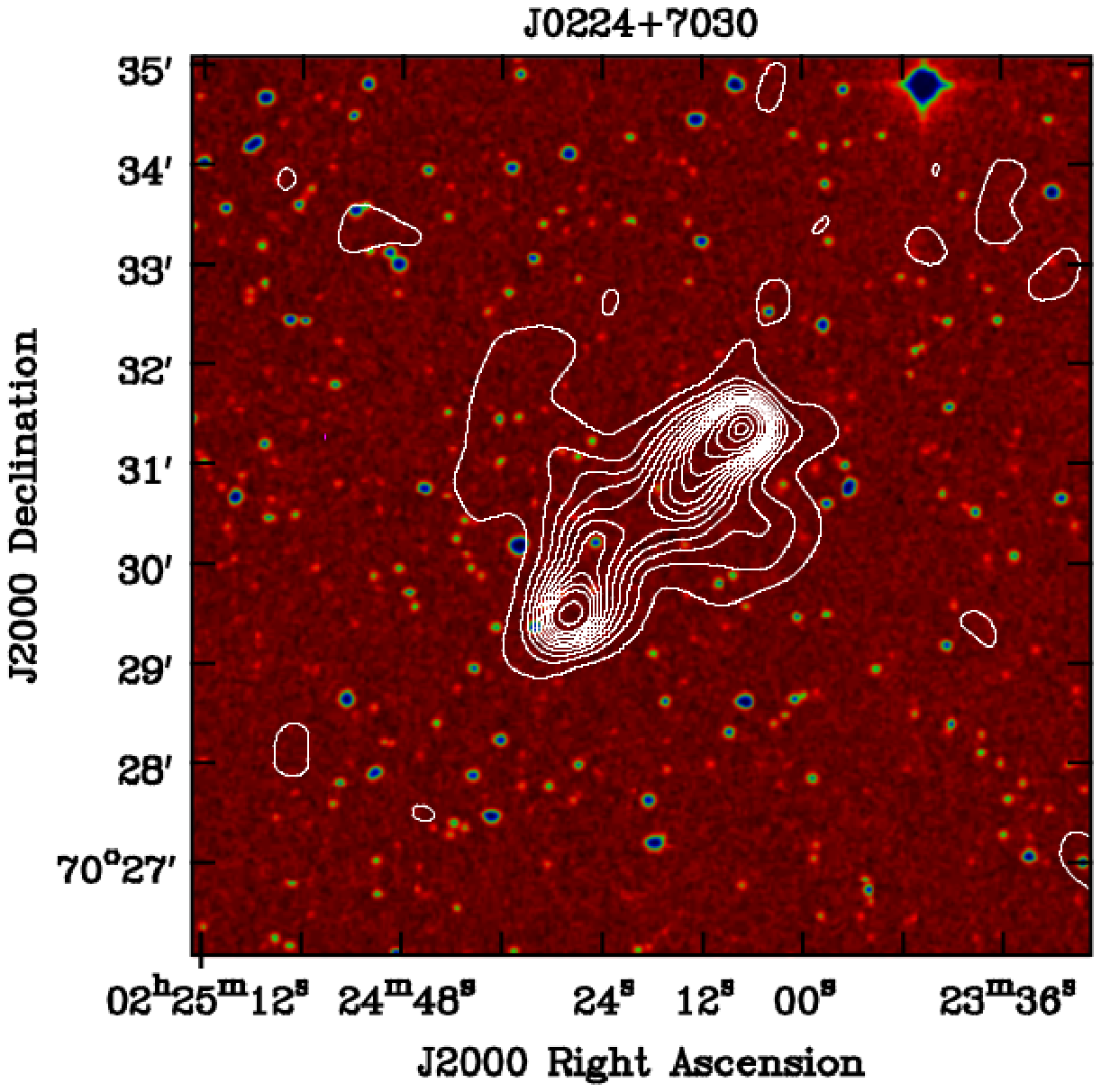,width=5.5cm,height=5cm}
\vskip 0.8cm         

\psfig{figure=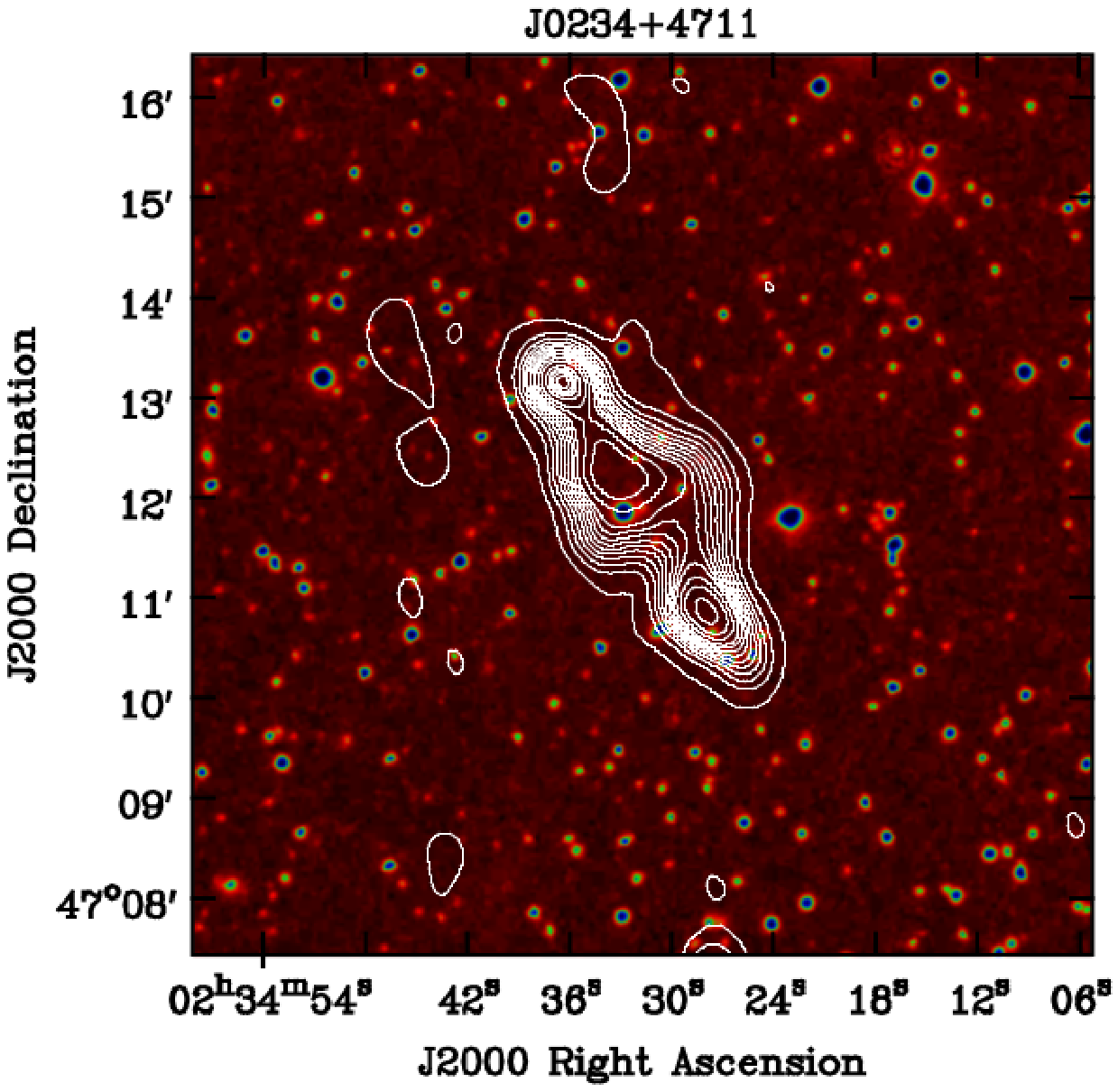,width=5.5cm,height=5cm}
\psfig{figure=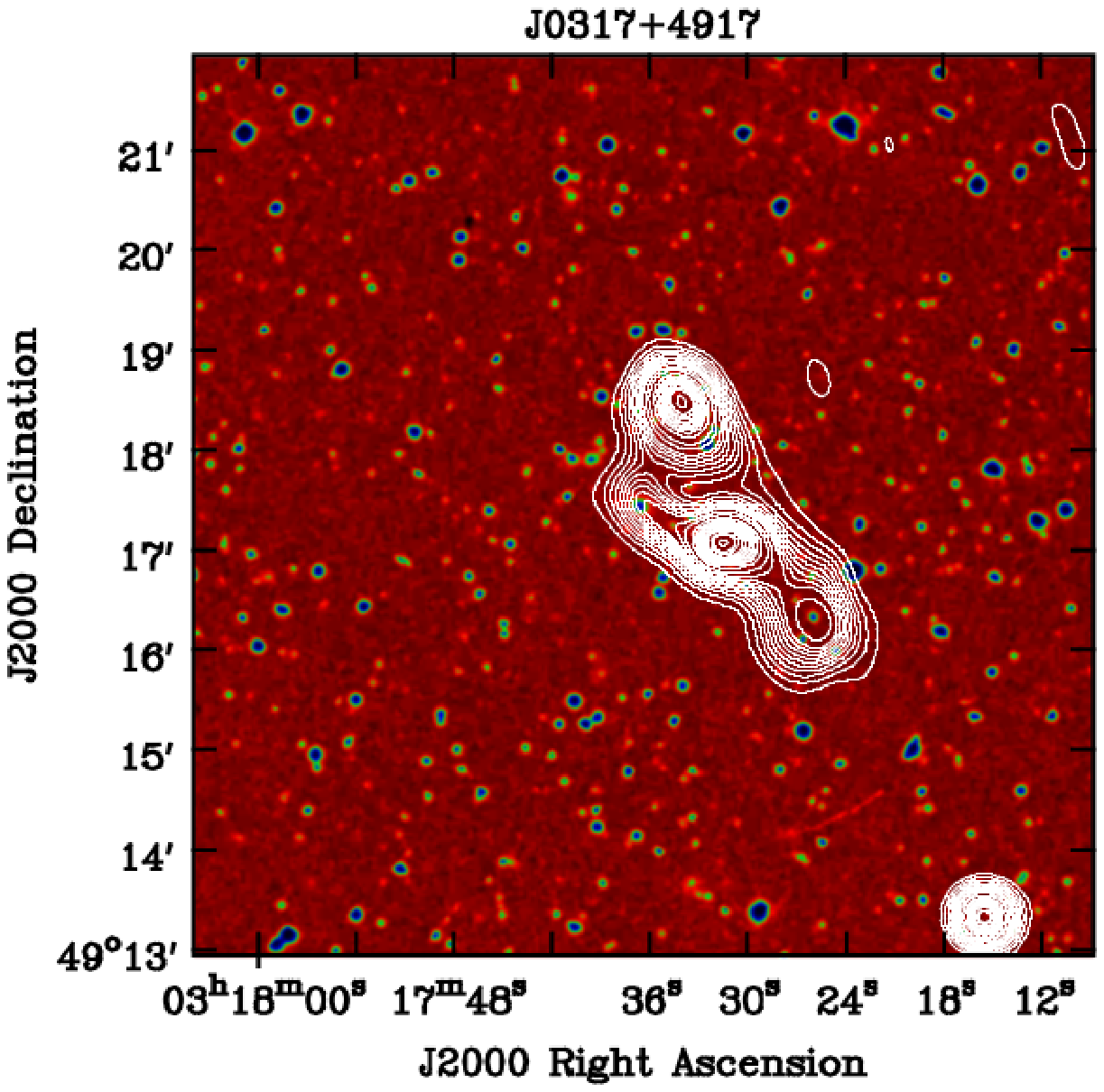,width=5.5cm,height=5cm}
\psfig{figure=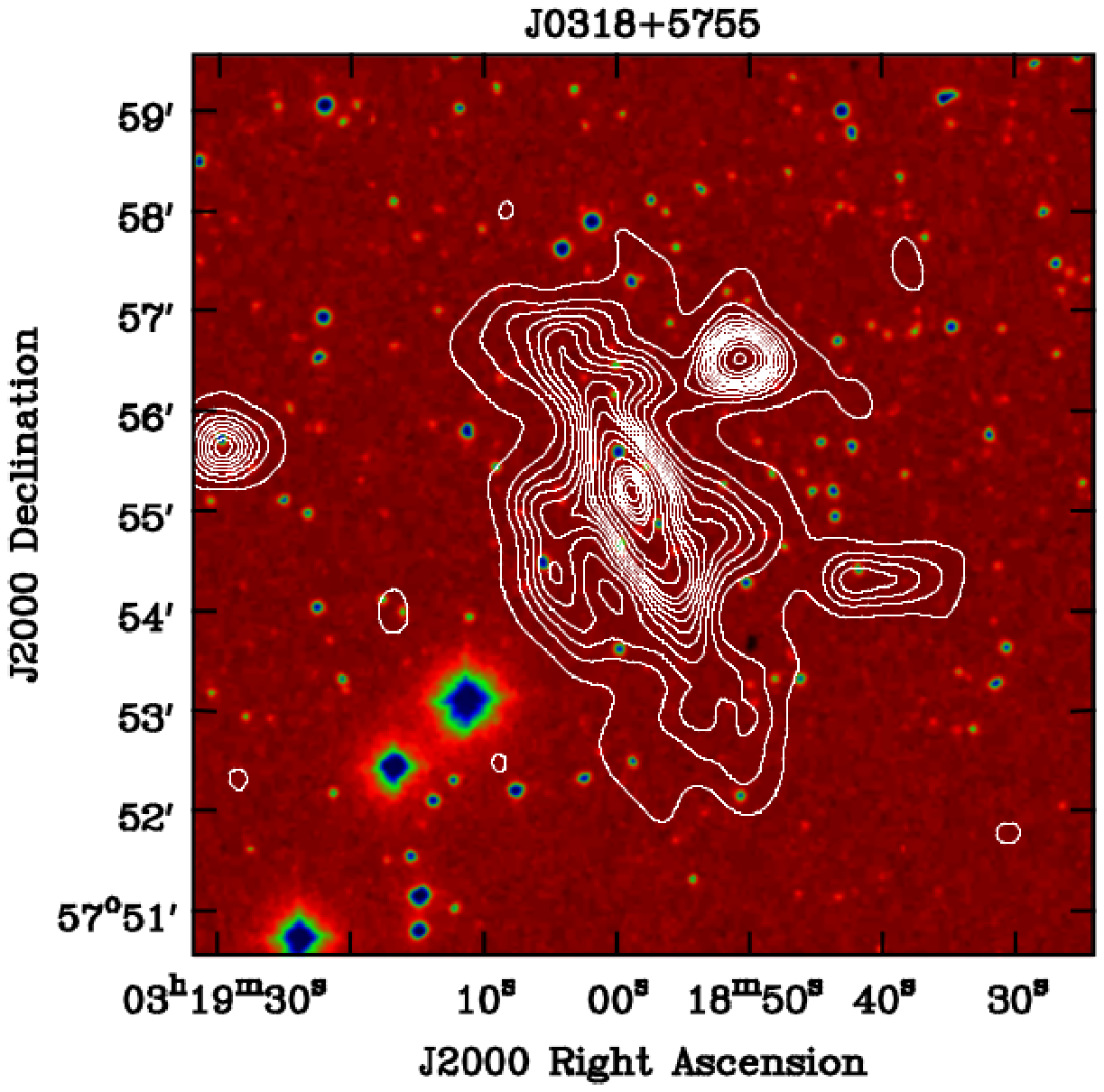,width=5.5cm,height=5cm}
\vskip 0.8cm 

\psfig{figure=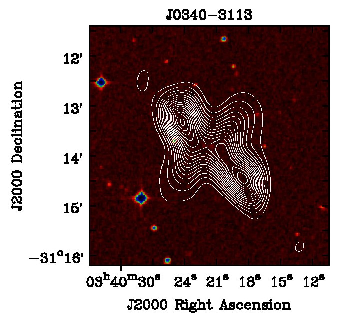,width=5.5cm,height=5cm} 
\psfig{figure=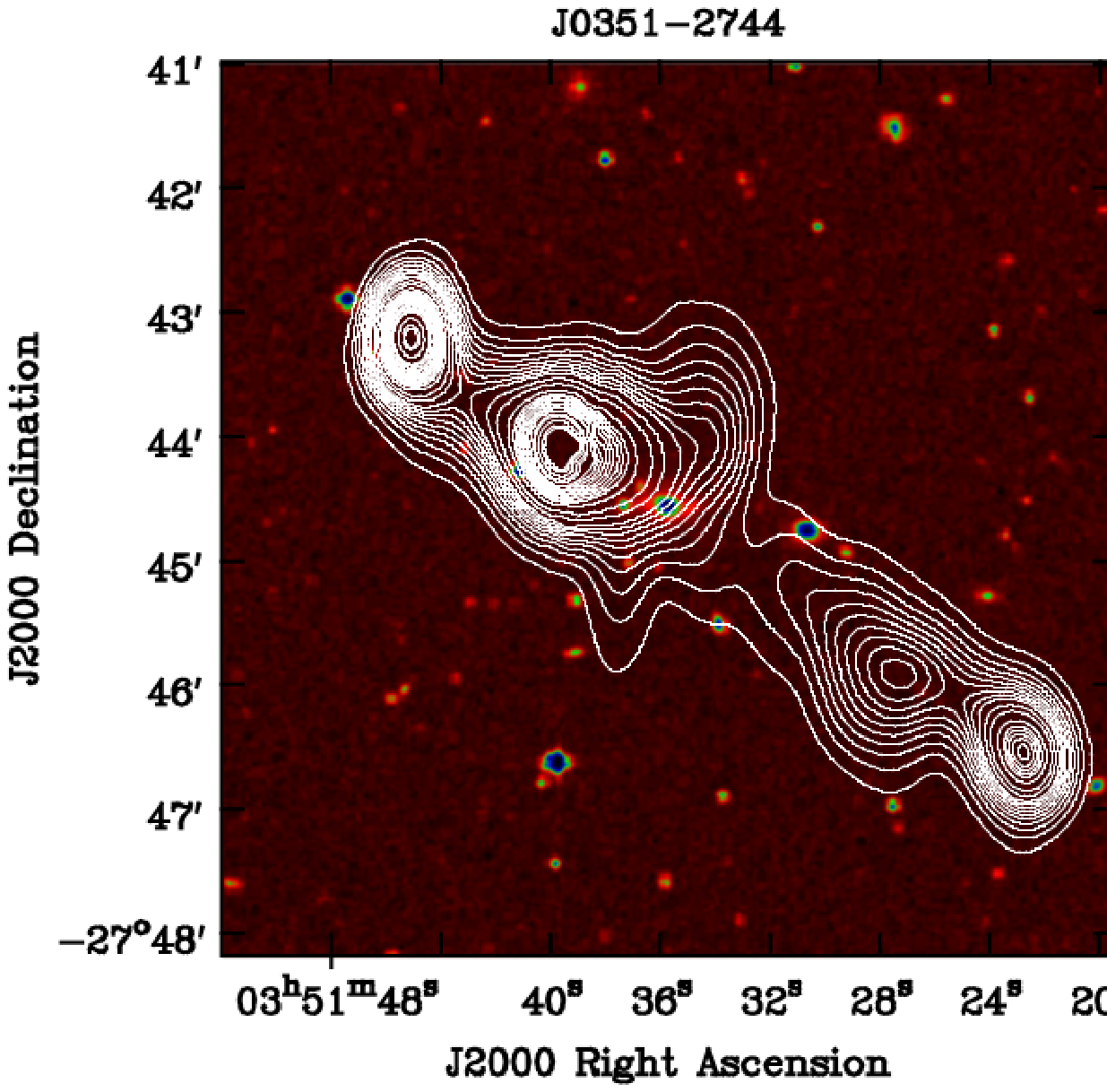,width=5.5cm,height=5cm} 
\psfig{figure=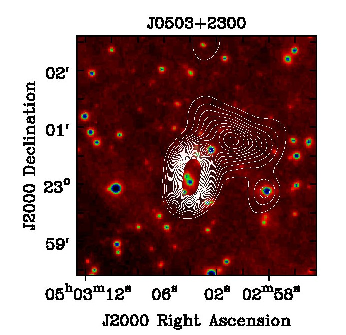,width=5.5cm,height=5cm}
\caption {Here we present radio maps of newly discovered XRGs. The white colour contour represents the radio emission detected in the TGSS ADR 1. The background colour images are from the DSS r-band. Contour levels are at 3$\sigma \times$[ 1, 1.41, 2, 2.83, 4, 5.66, 8, 11.31, 16, 22.63, 32, 45.25, 64, 90.51, 128, 181.02, 256], where $\sigma$ = 3.5 mJy beam$^{-1}$ is the local rms noise.}
\label{fig:XRG}
\end{figure*}

\begin{figure*}
\psfig{figure=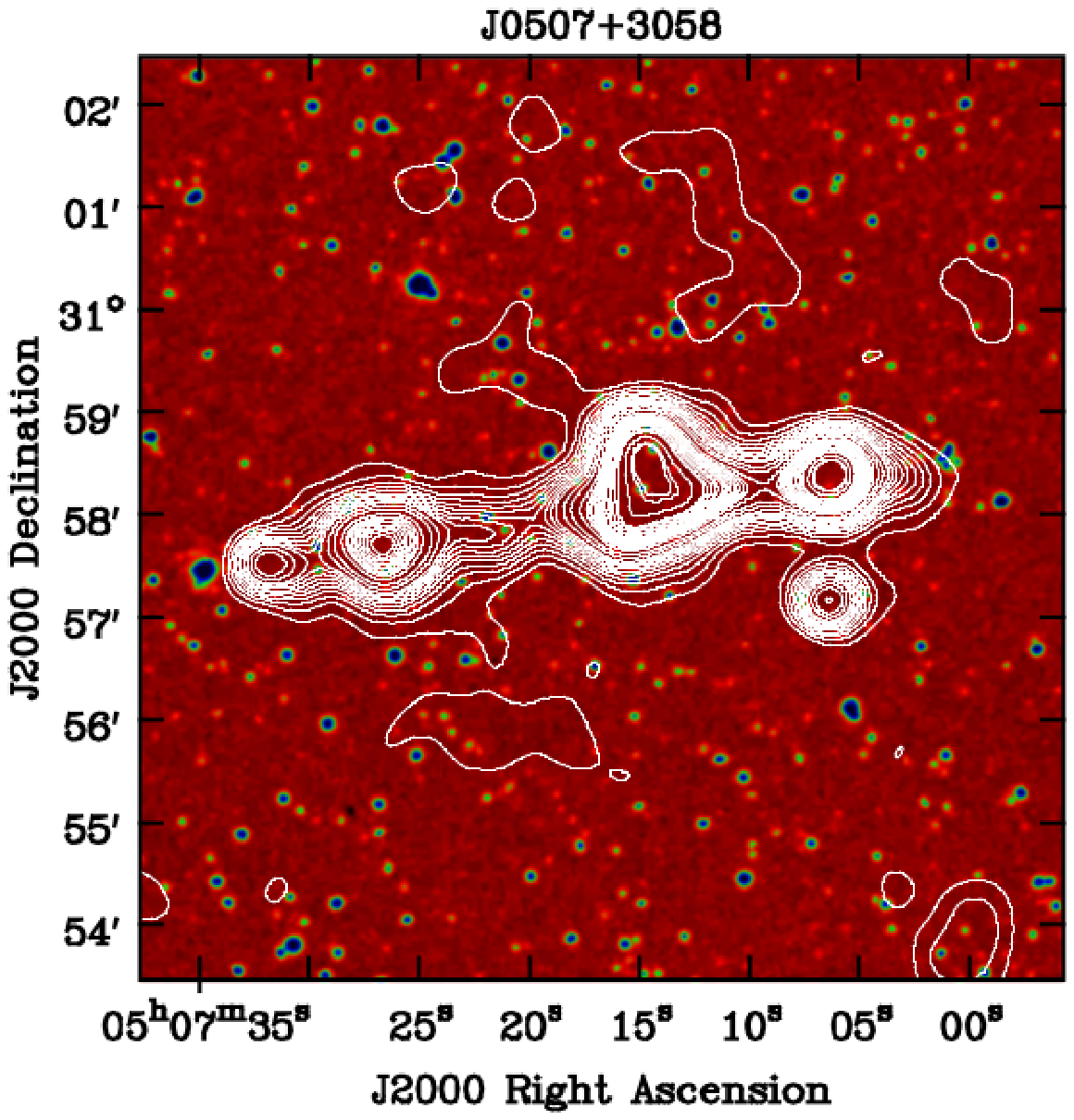,width=5.5cm,height=5cm}
\psfig{figure=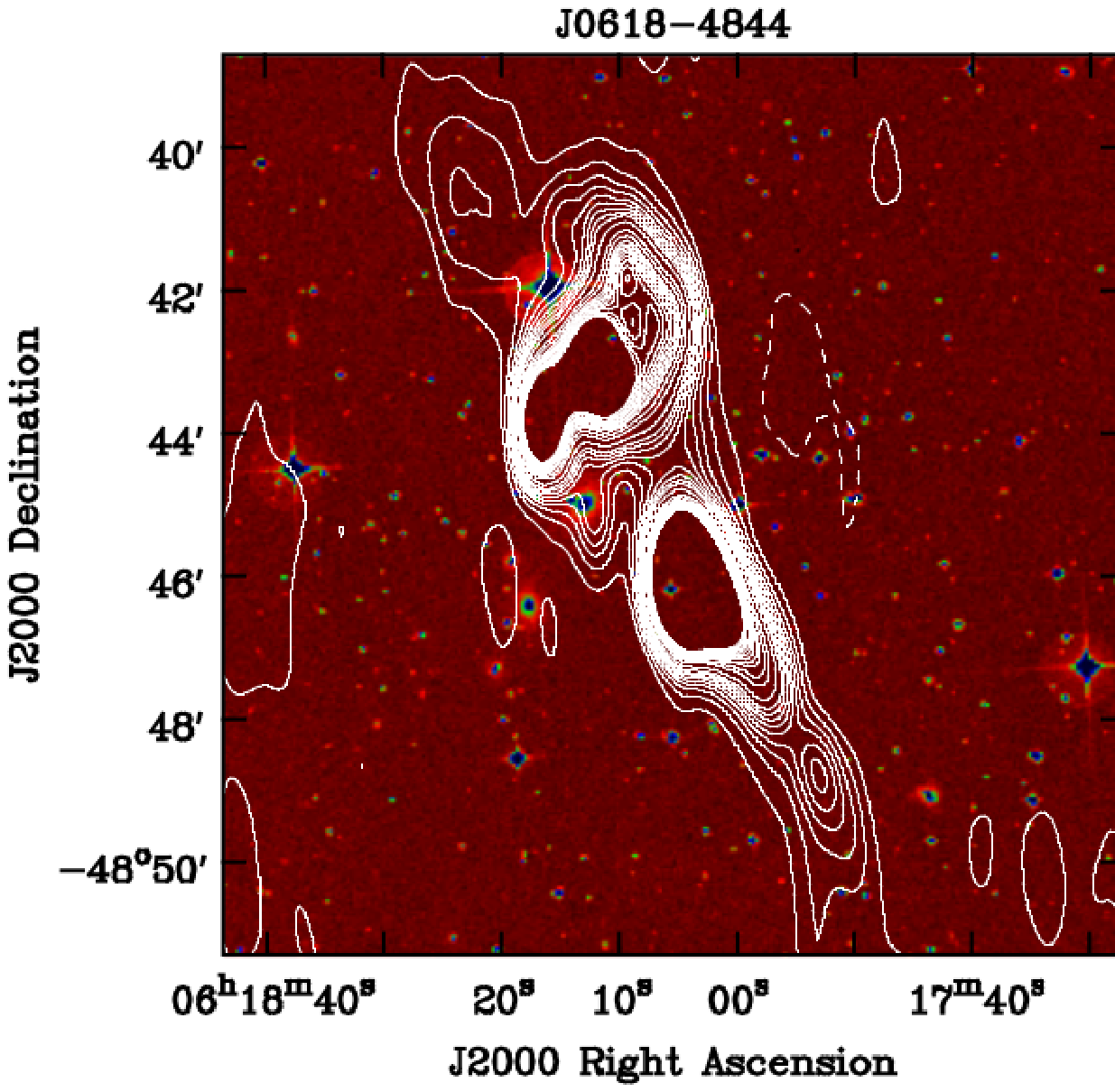,width=5.5cm,height=5cm} 
\psfig{figure=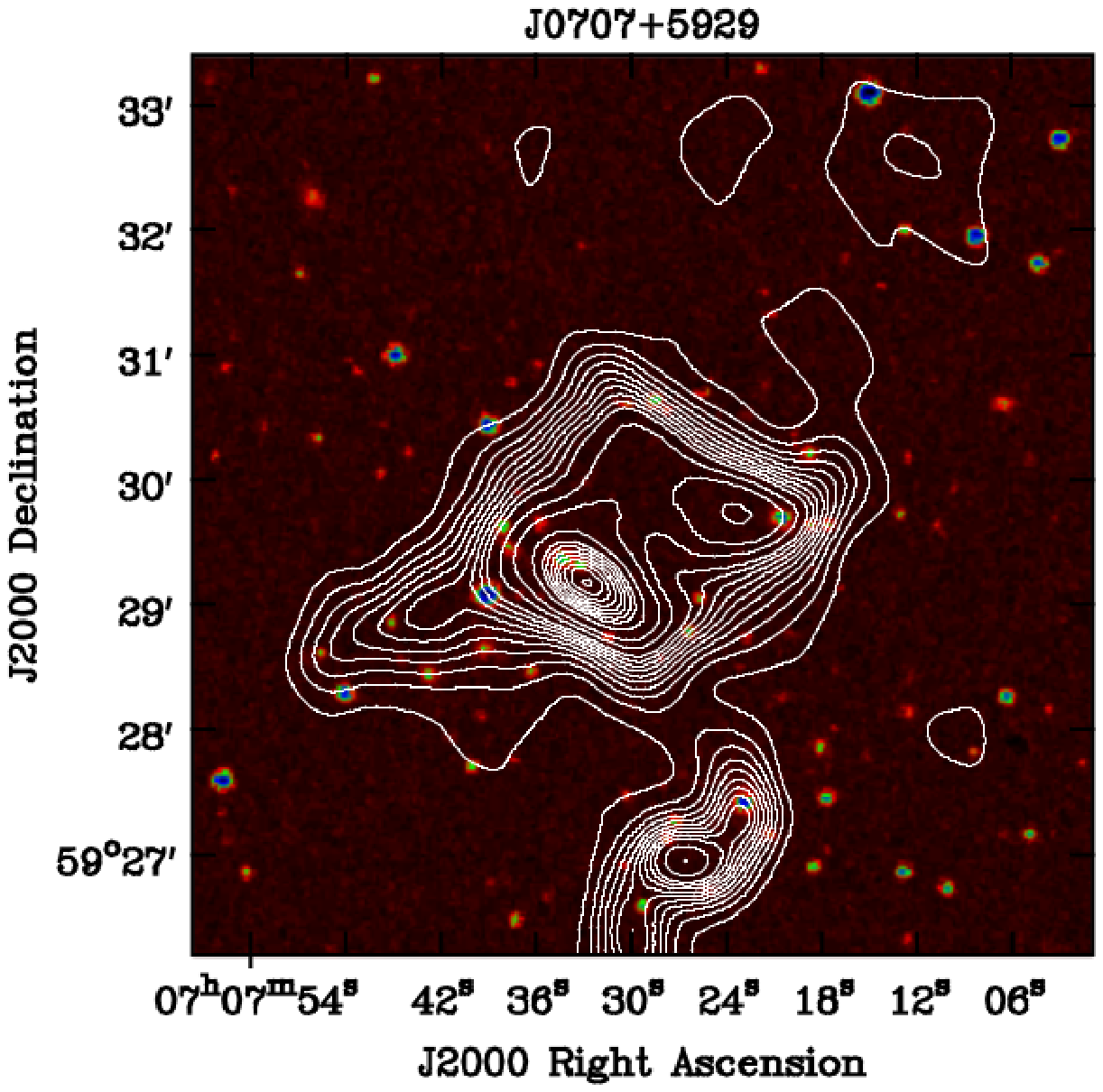,width=5.5cm,height=5cm}
\vskip 0.8cm         

\psfig{figure=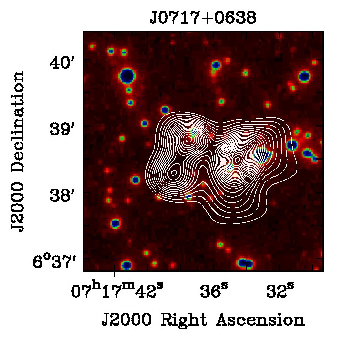,width=5.5cm,height=5cm} 
\psfig{figure=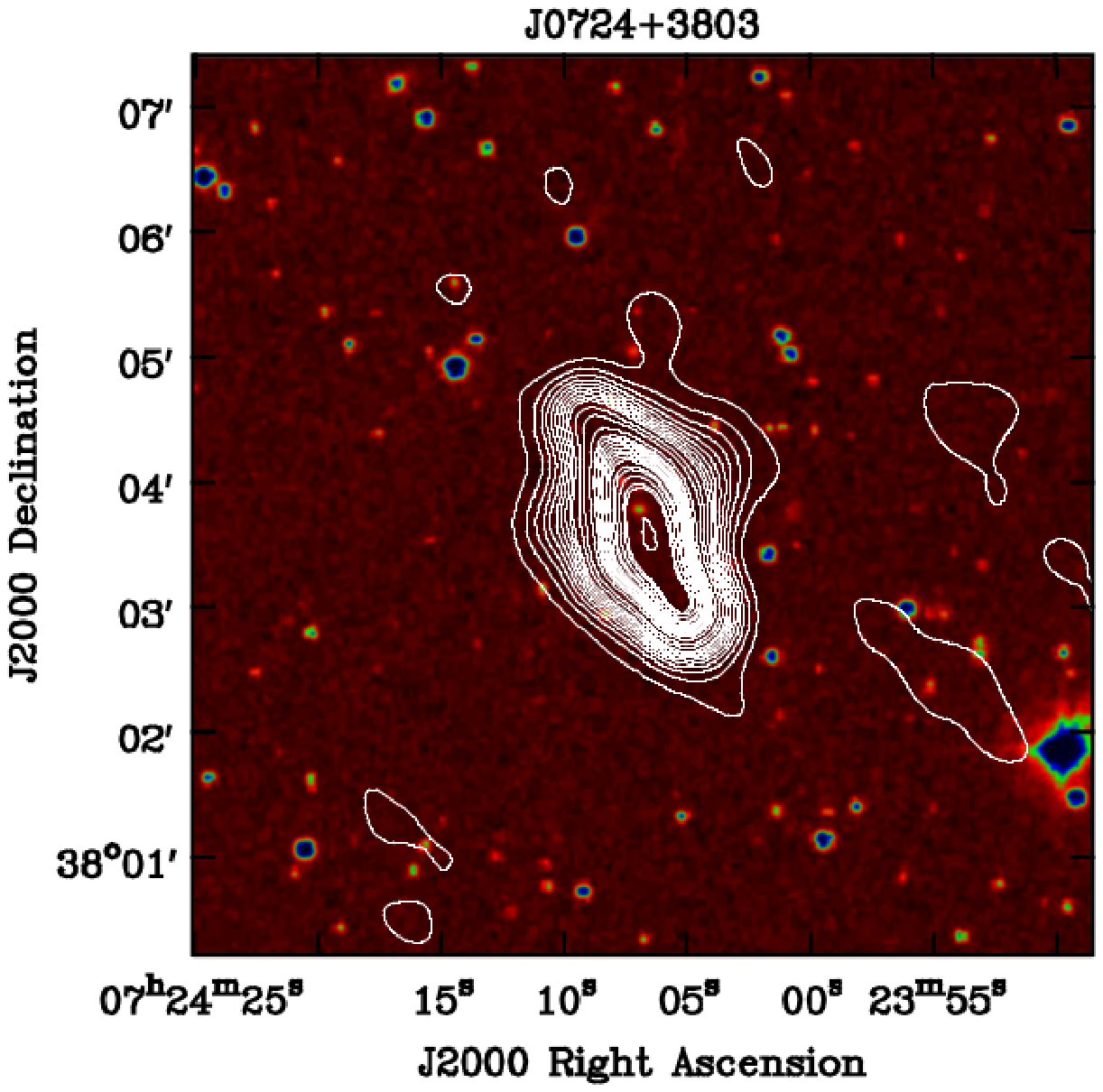,width=5.5cm,height=5cm}
\psfig{figure=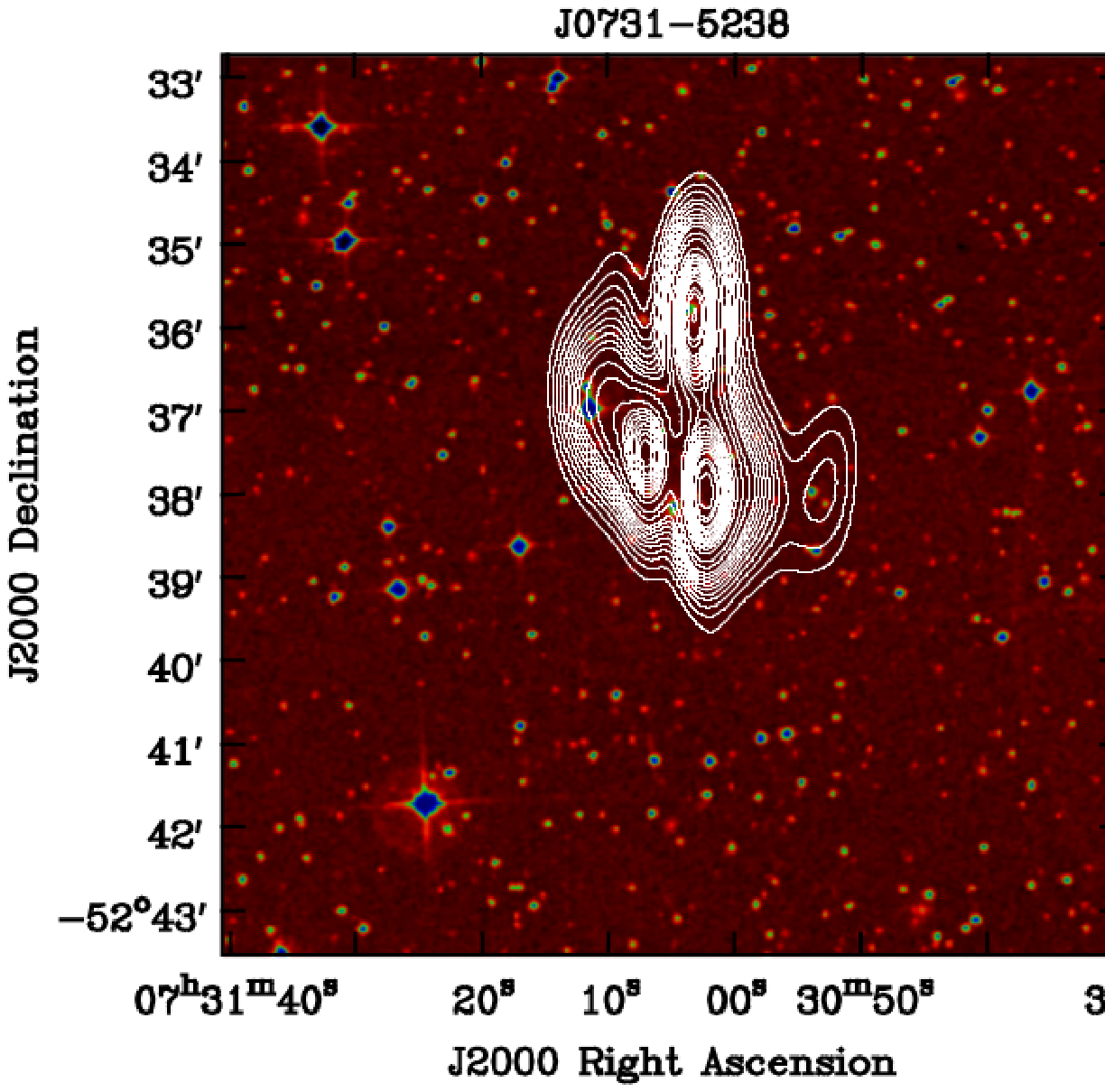,width=5.5cm,height=5cm}
\vskip 0.8cm

\psfig{figure=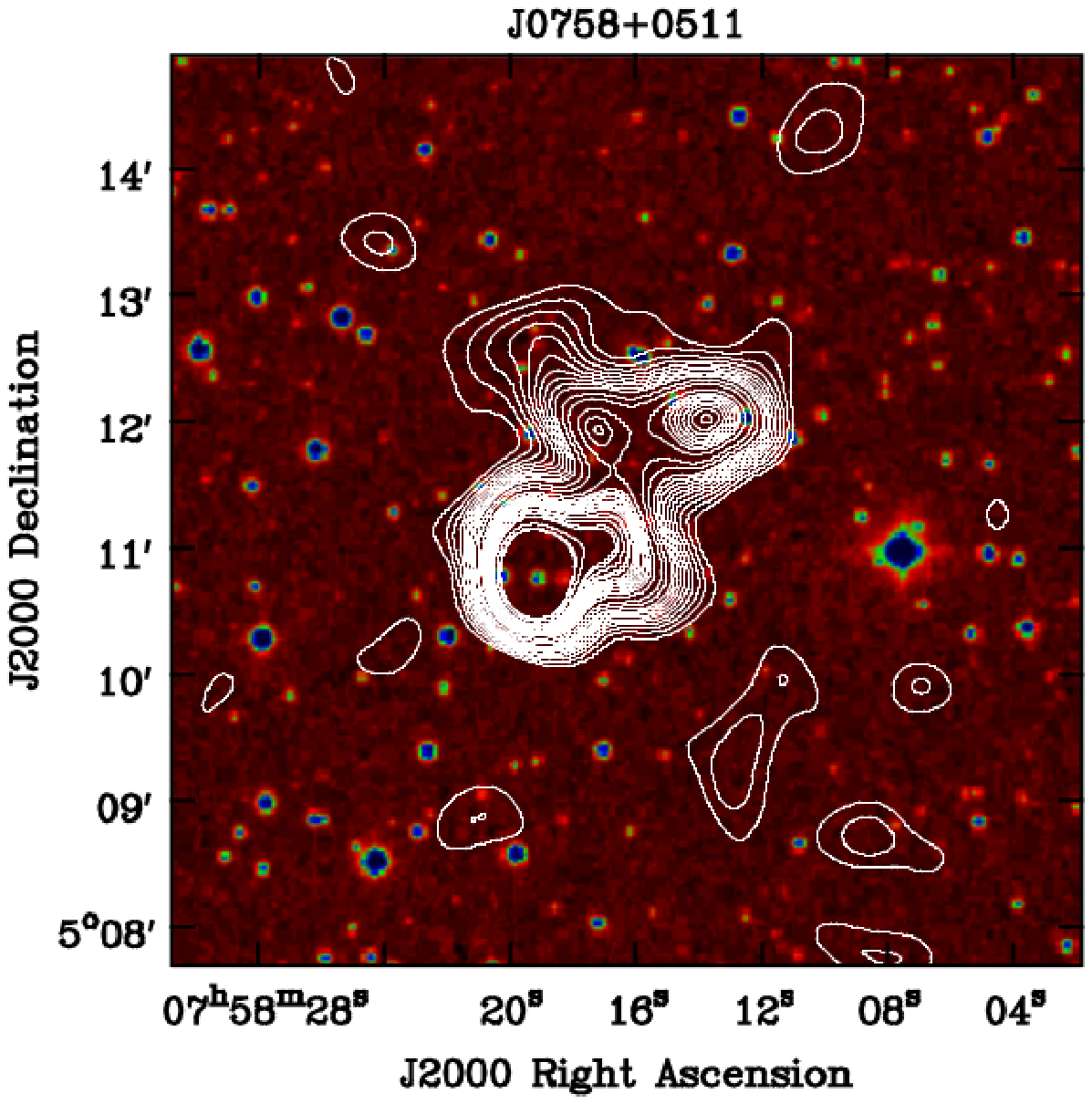,width=5.5cm,height=5cm}
\psfig{figure=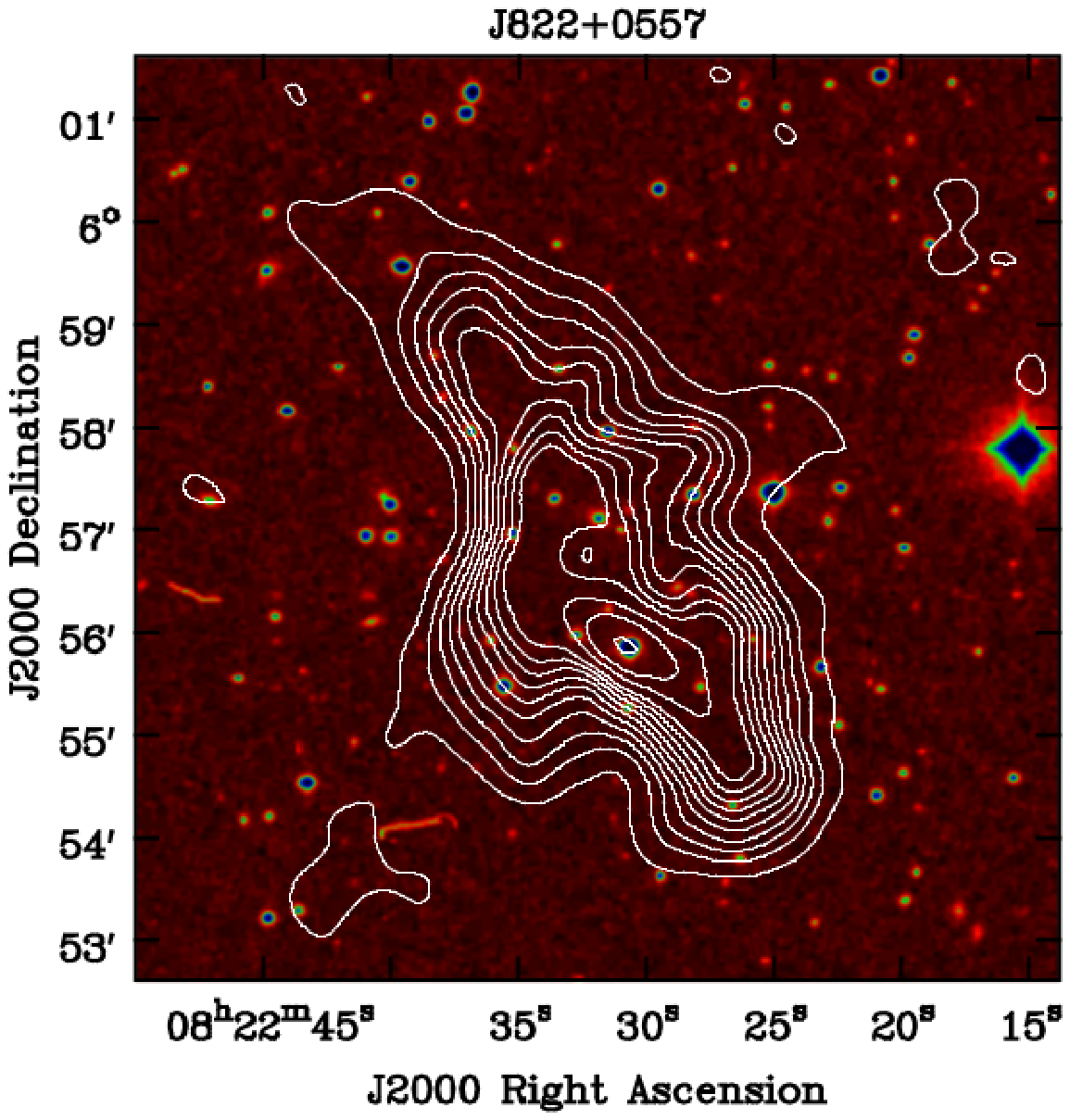,width=5.5cm,height=5cm}
\psfig{figure=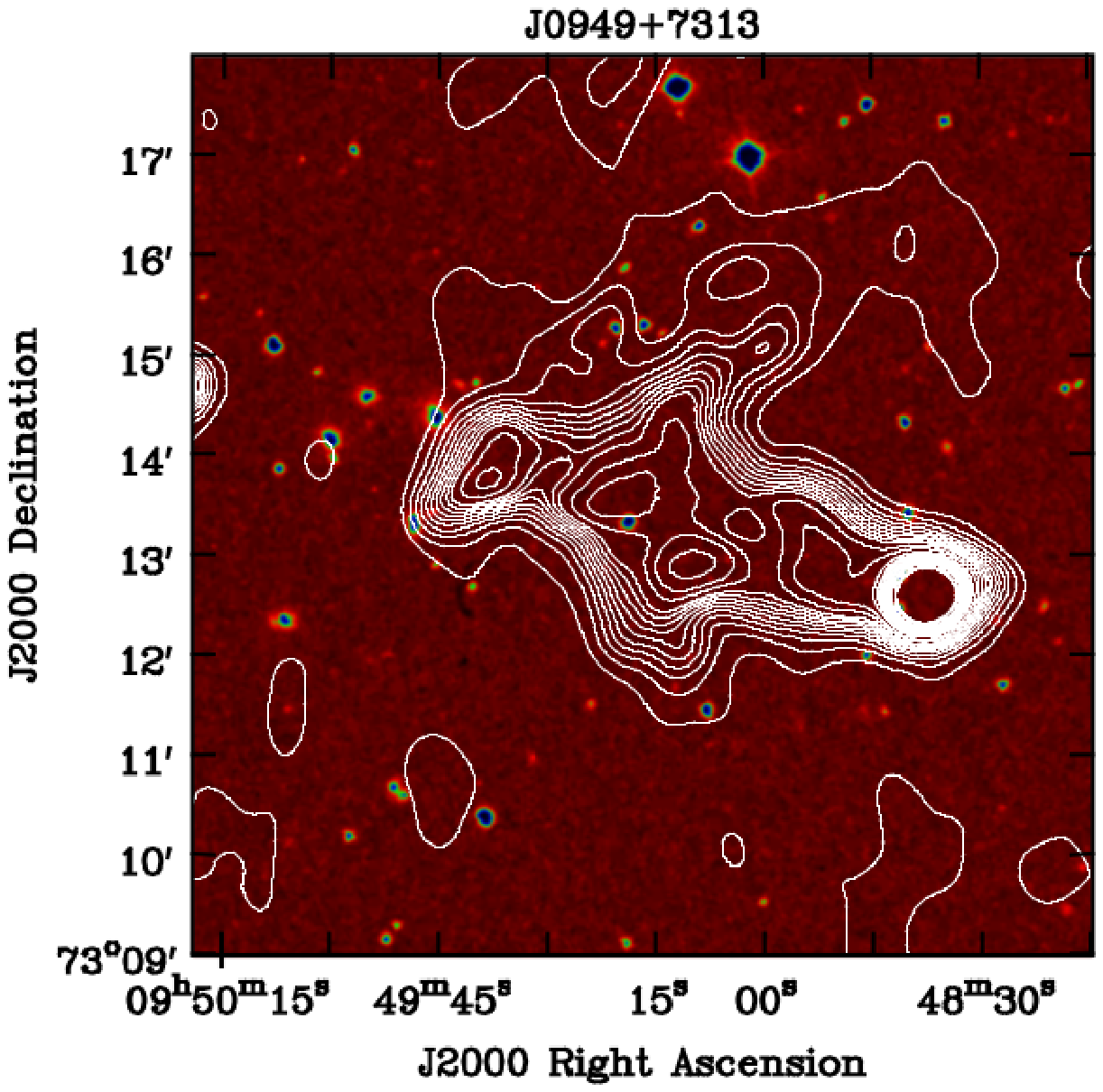,width=5.5cm,height=5cm}

\vskip 0.8cm
\psfig{figure=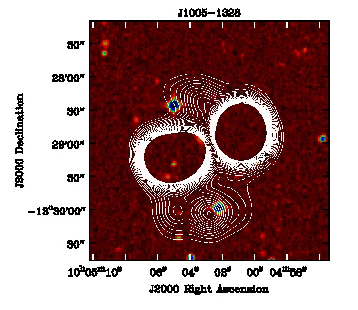,width=5.5cm,height=5cm}
\psfig{figure=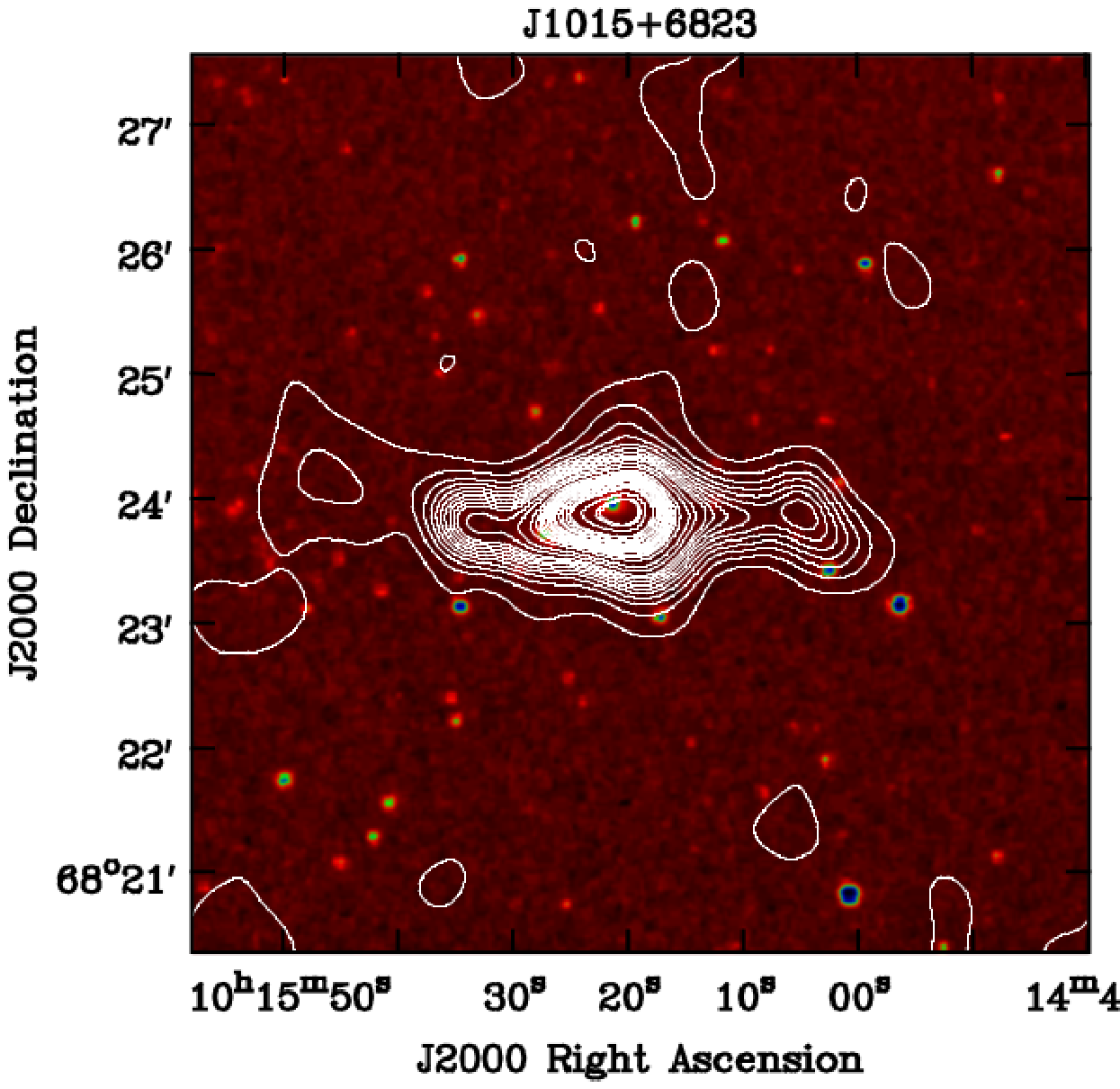,width=5.5cm,height=5cm}
\psfig{figure=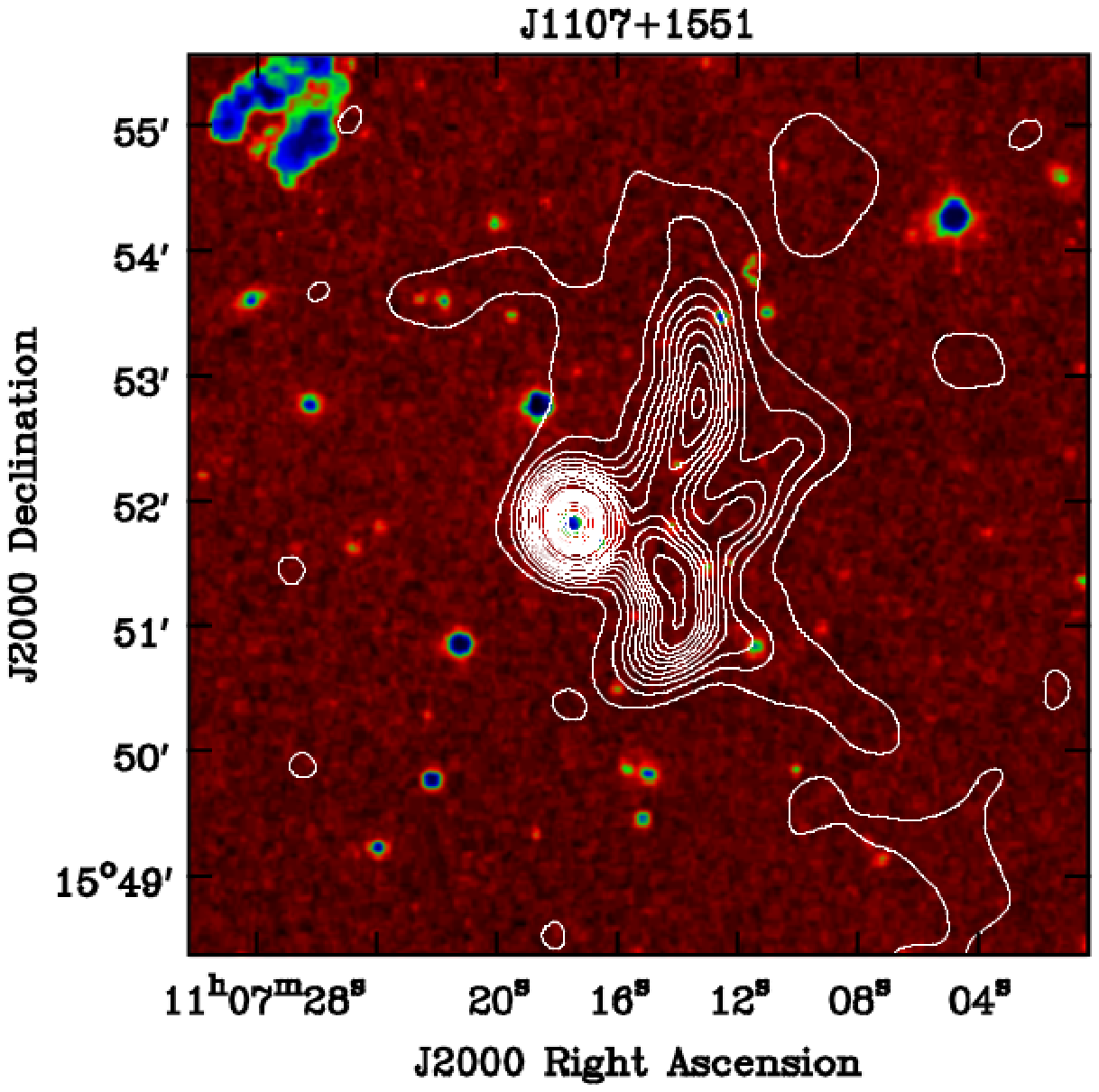,width=5.5cm,height=5cm}
\contcaption {Here we present radio maps of newly discovered XRGs. The white colour contour represents the radio emission detected in the TGSS ADR 1. The background colour images are from the DSS r-band. Contour levels are at 3$\sigma \times$[ 1, 1.41, 2, 2.83, 4, 5.66, 8, 11.31, 16, 22.63, 32, 45.25, 64, 90.51, 128, 181.02, 256], where $\sigma$ = 3.5 mJy beam$^{-1}$ is the local rms noise.}
\end{figure*}

\begin{figure*}

\psfig{figure=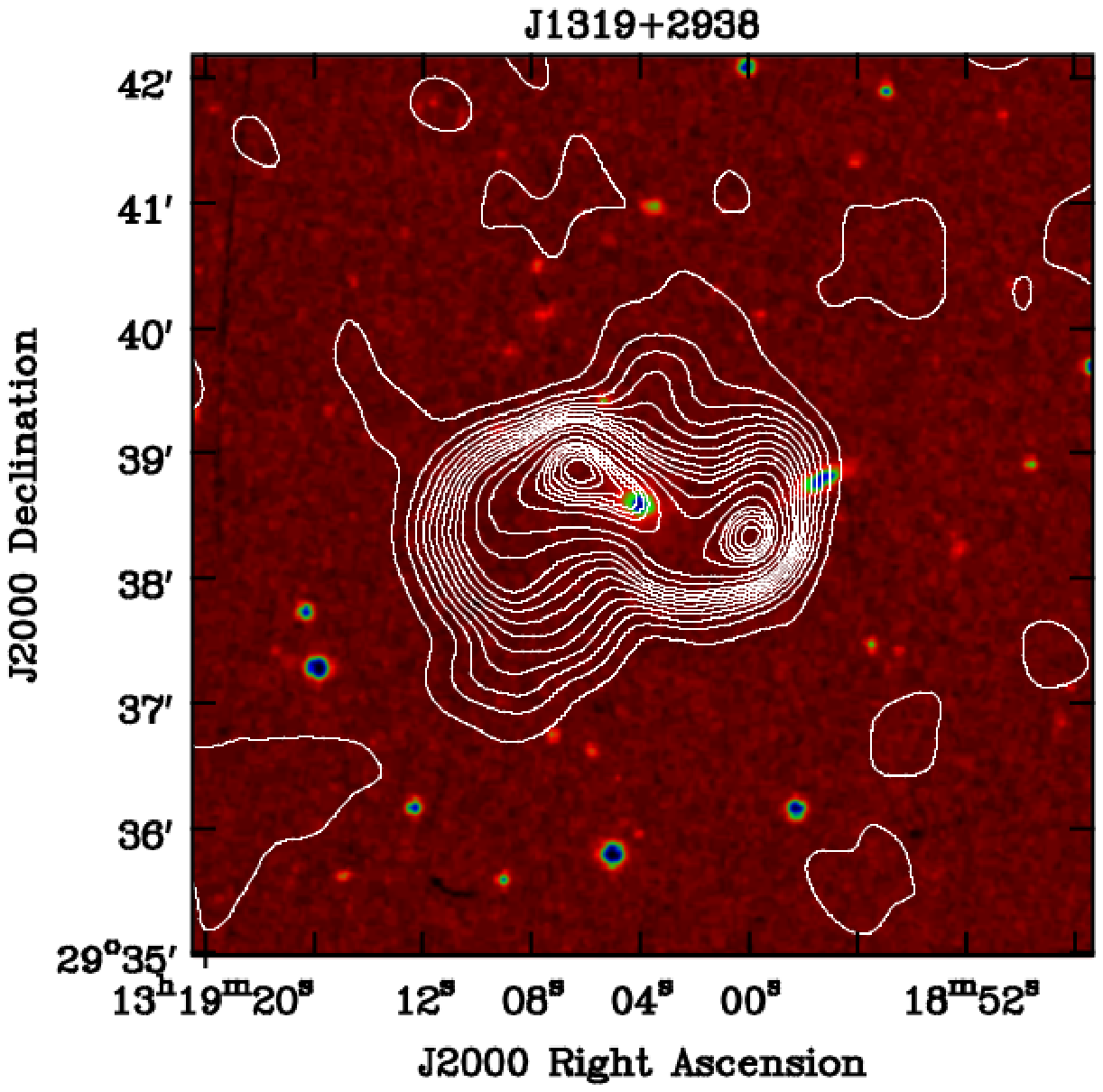,width=5.5cm,height=5cm}
\psfig{figure=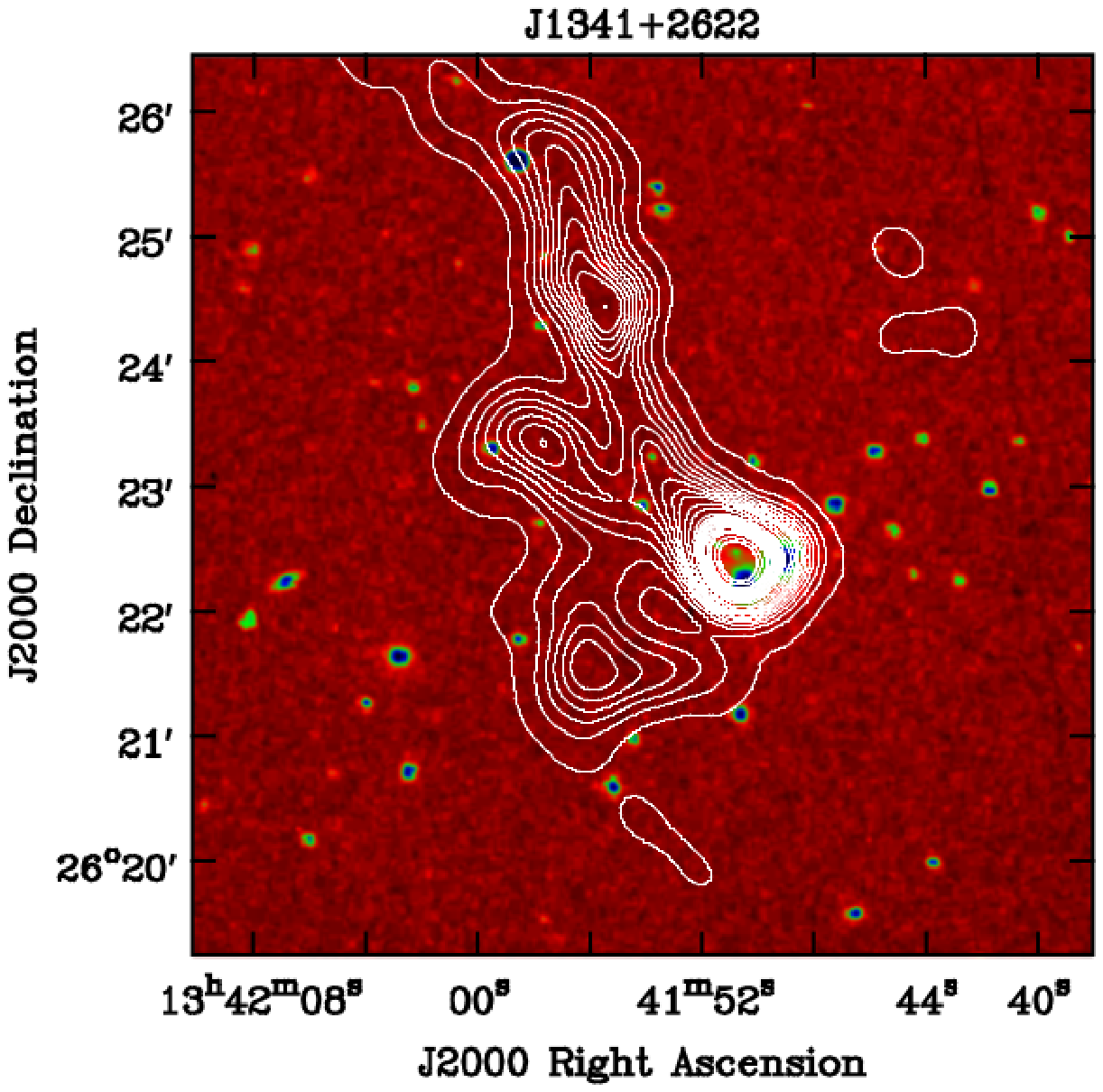,width=5.5cm,height=5cm} 
\psfig{figure=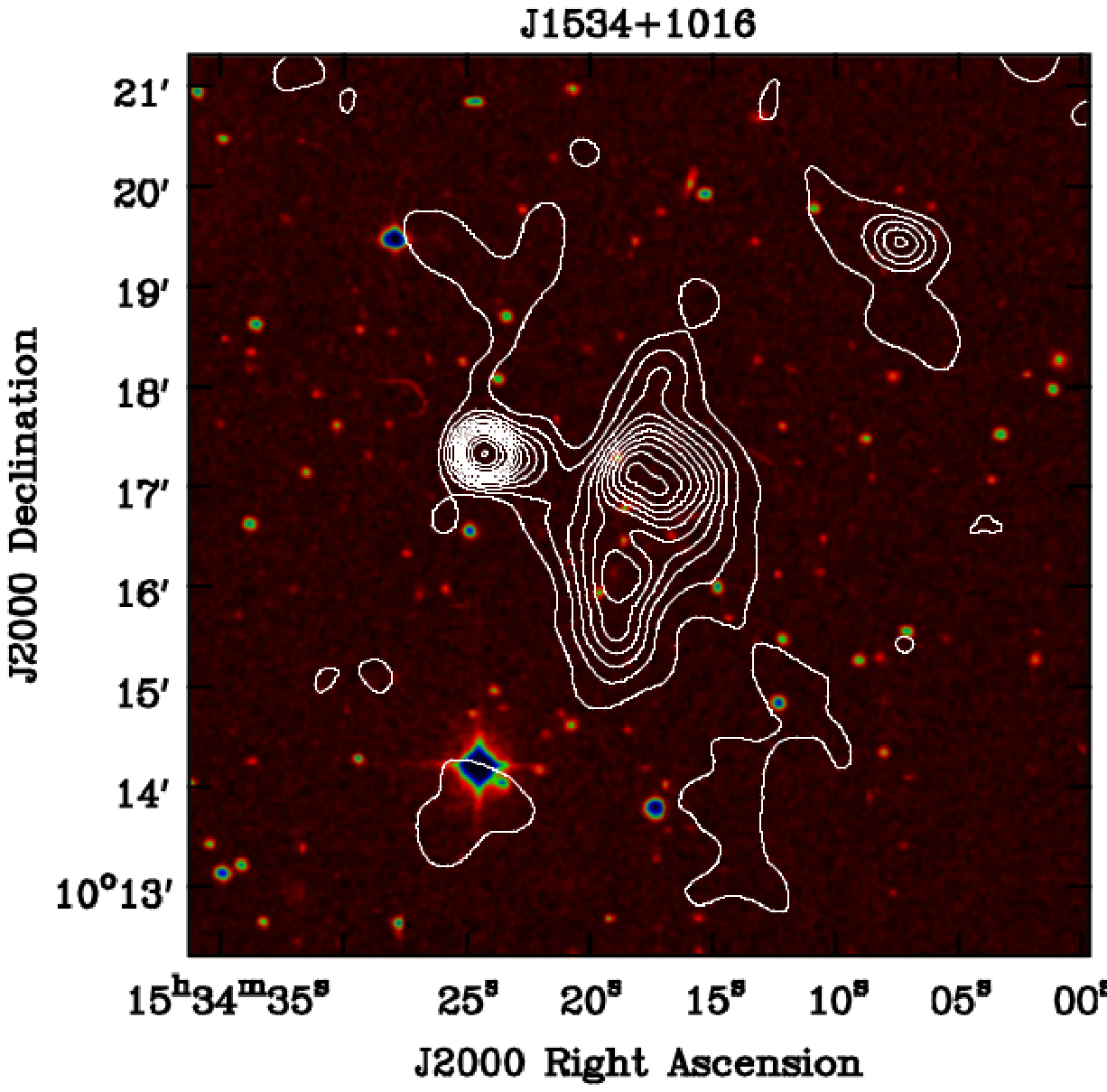,width=5.5cm,height=5cm}
\vskip 0.8cm

\psfig{figure=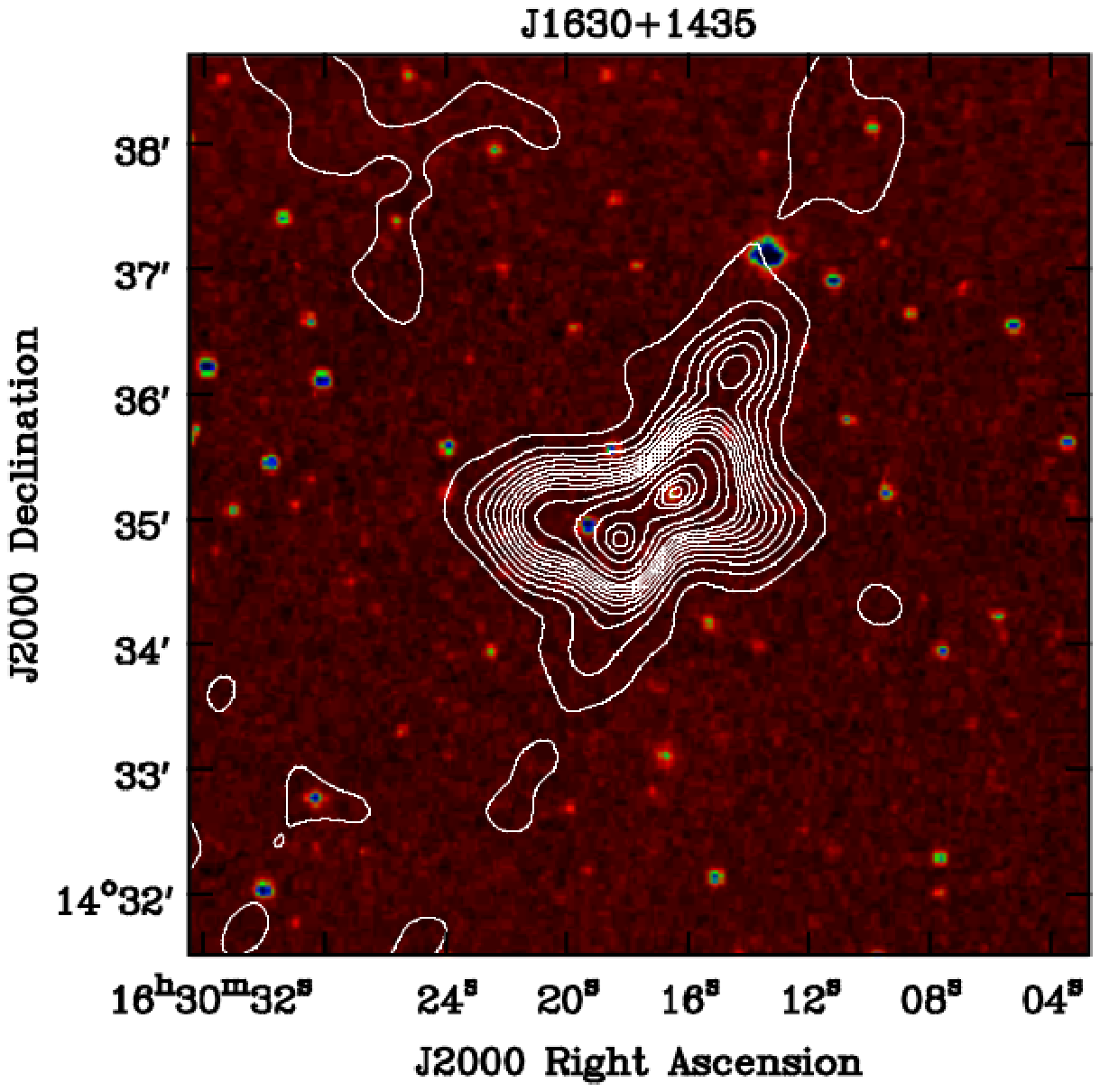,width=5.5cm,height=5cm} 
\psfig{figure=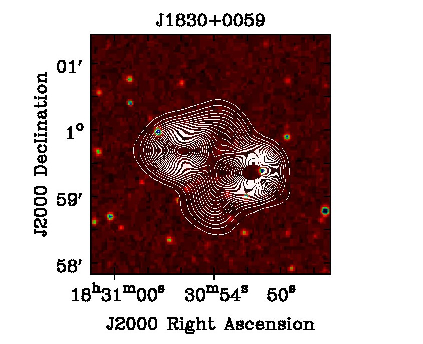,width=5.5cm,height=5cm}
\psfig{figure=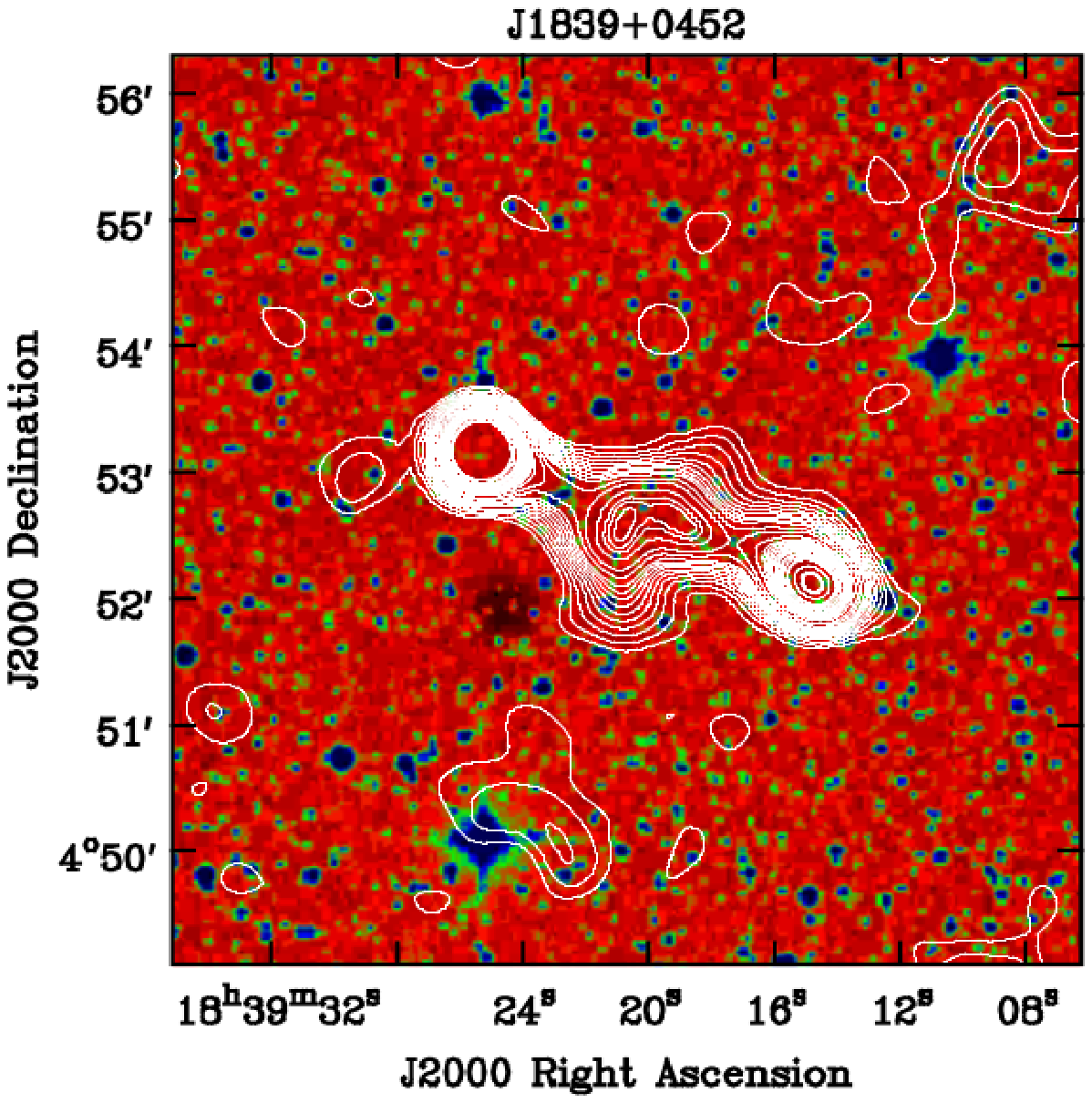,width=5.5cm,height=5cm}
\vskip 0.8cm

\psfig{figure=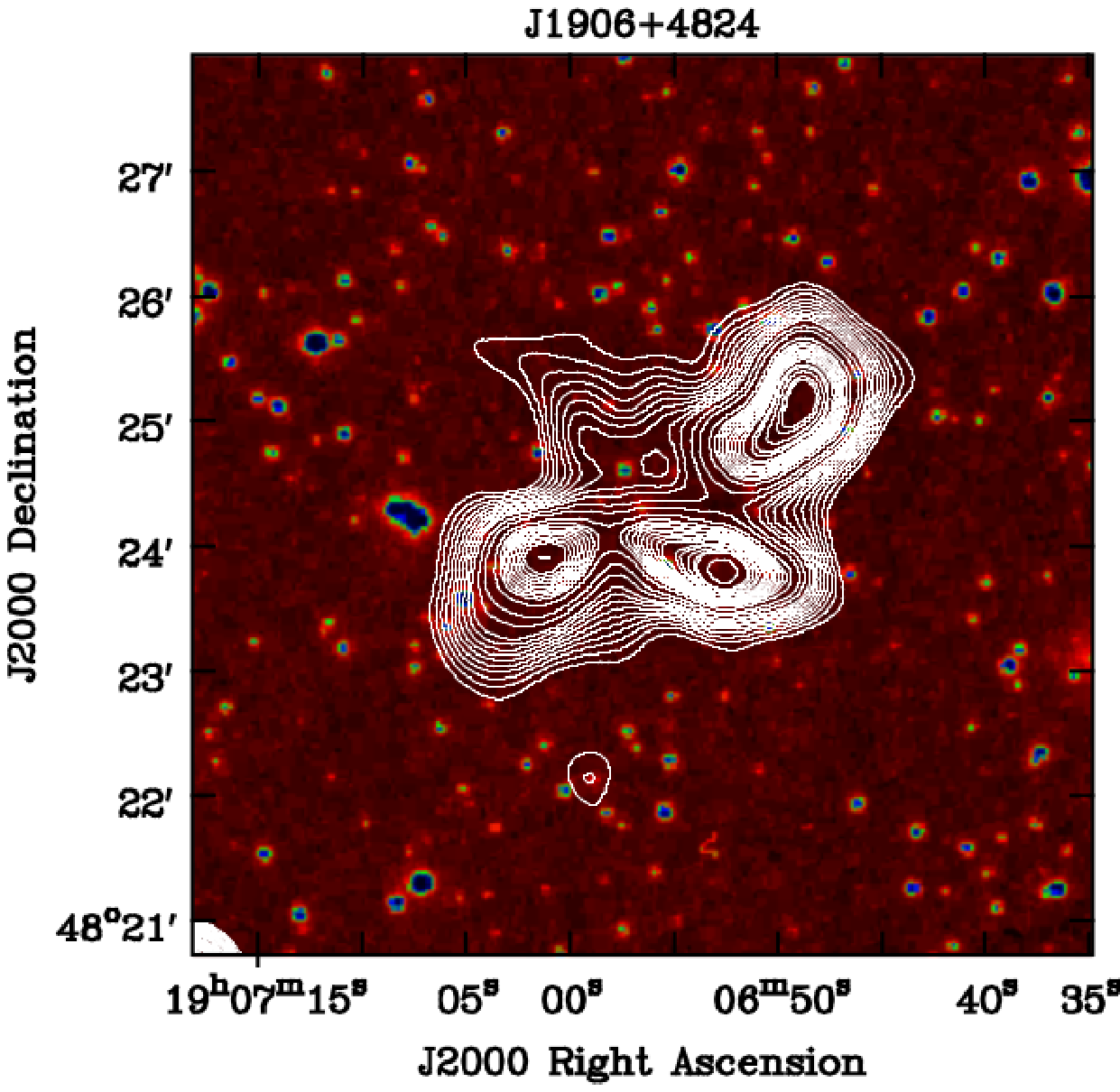,width=5.5cm,height=5cm}
\psfig{figure=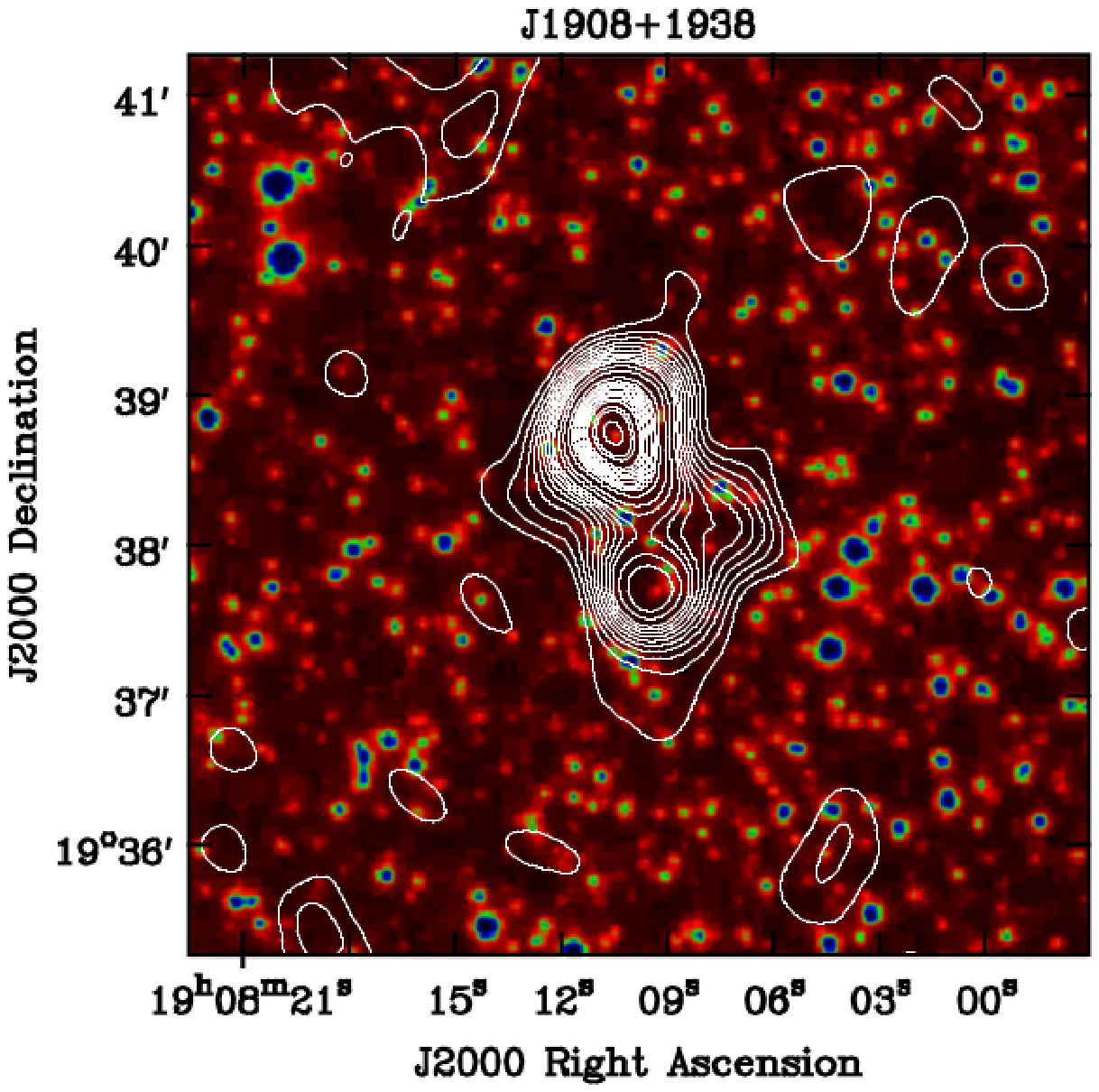,width=5.5cm,height=5cm}
\psfig{figure=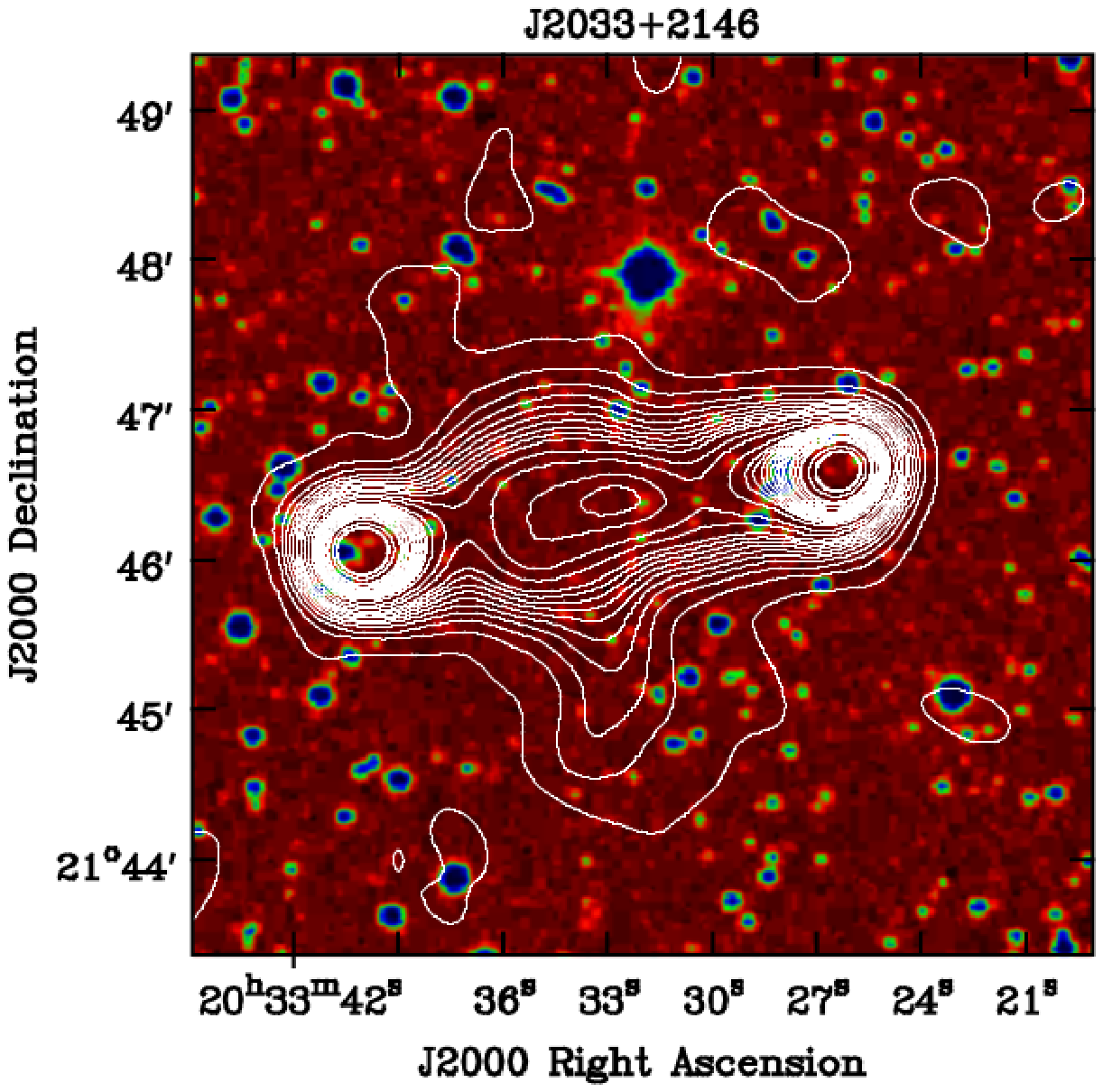,width=5.5cm,height=5cm}
\vskip 0.8cm

\psfig{figure=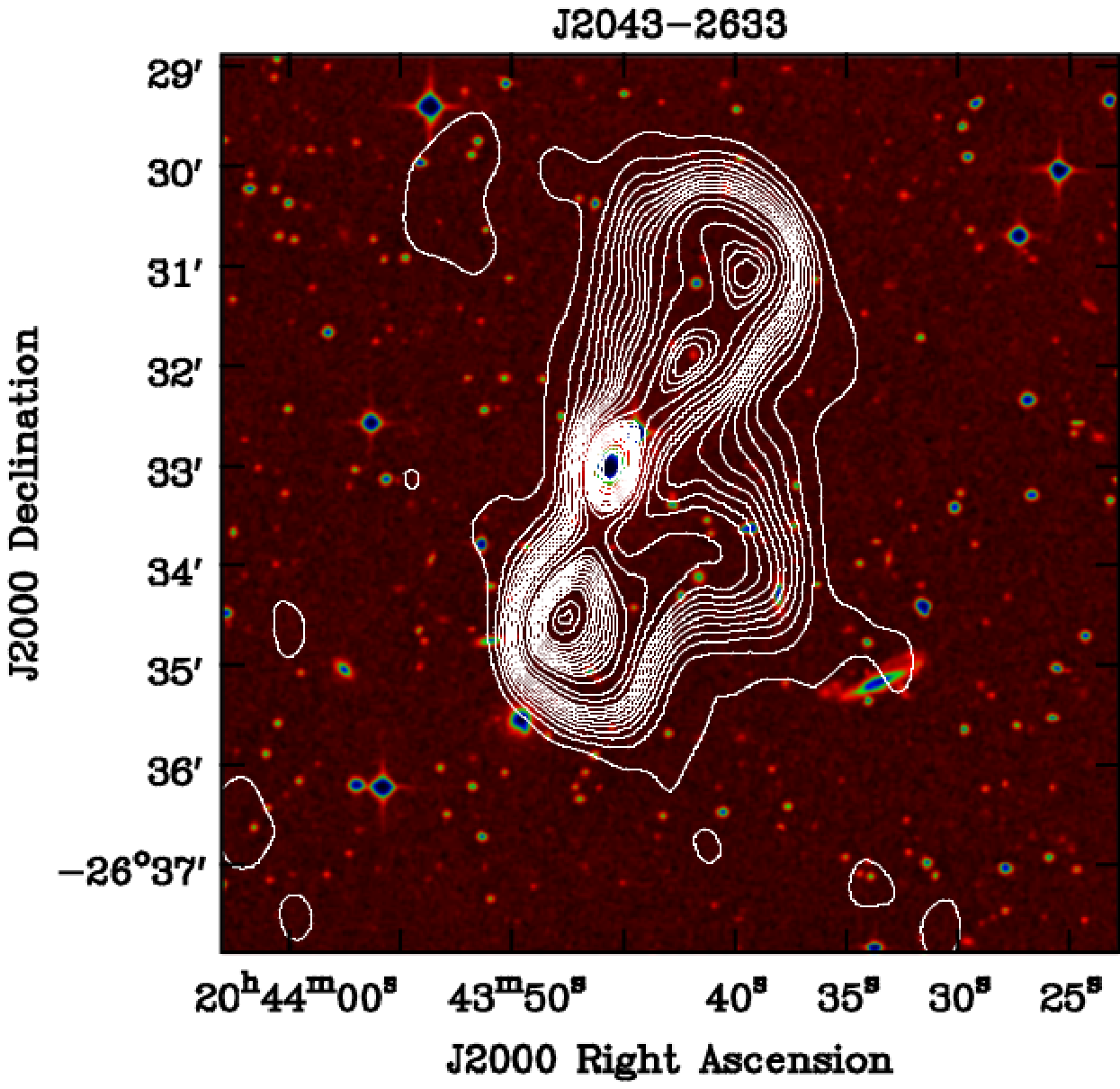,width=5.5cm,height=5cm}
\psfig{figure=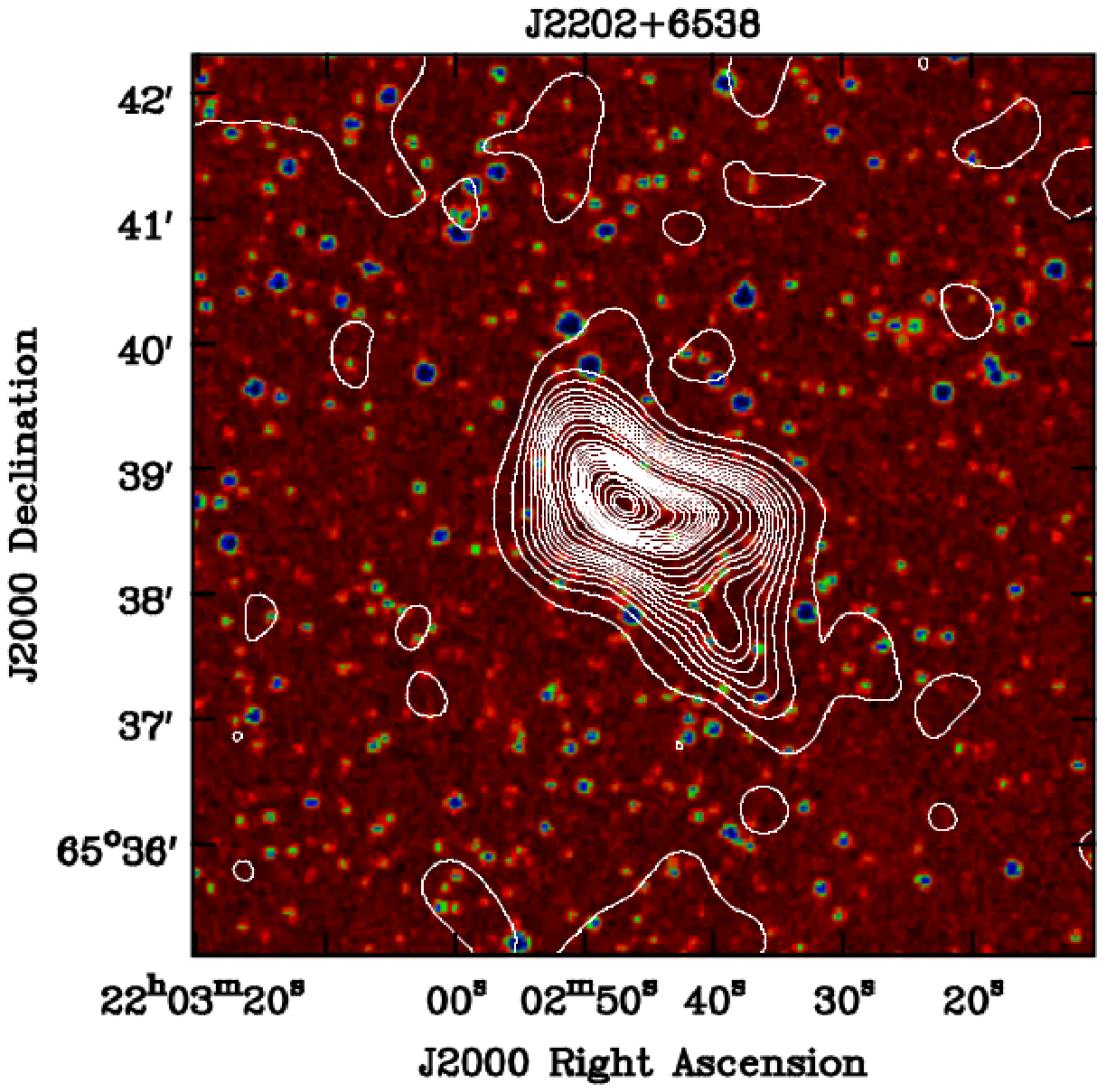,width=5.5cm,height=5cm}
\psfig{figure=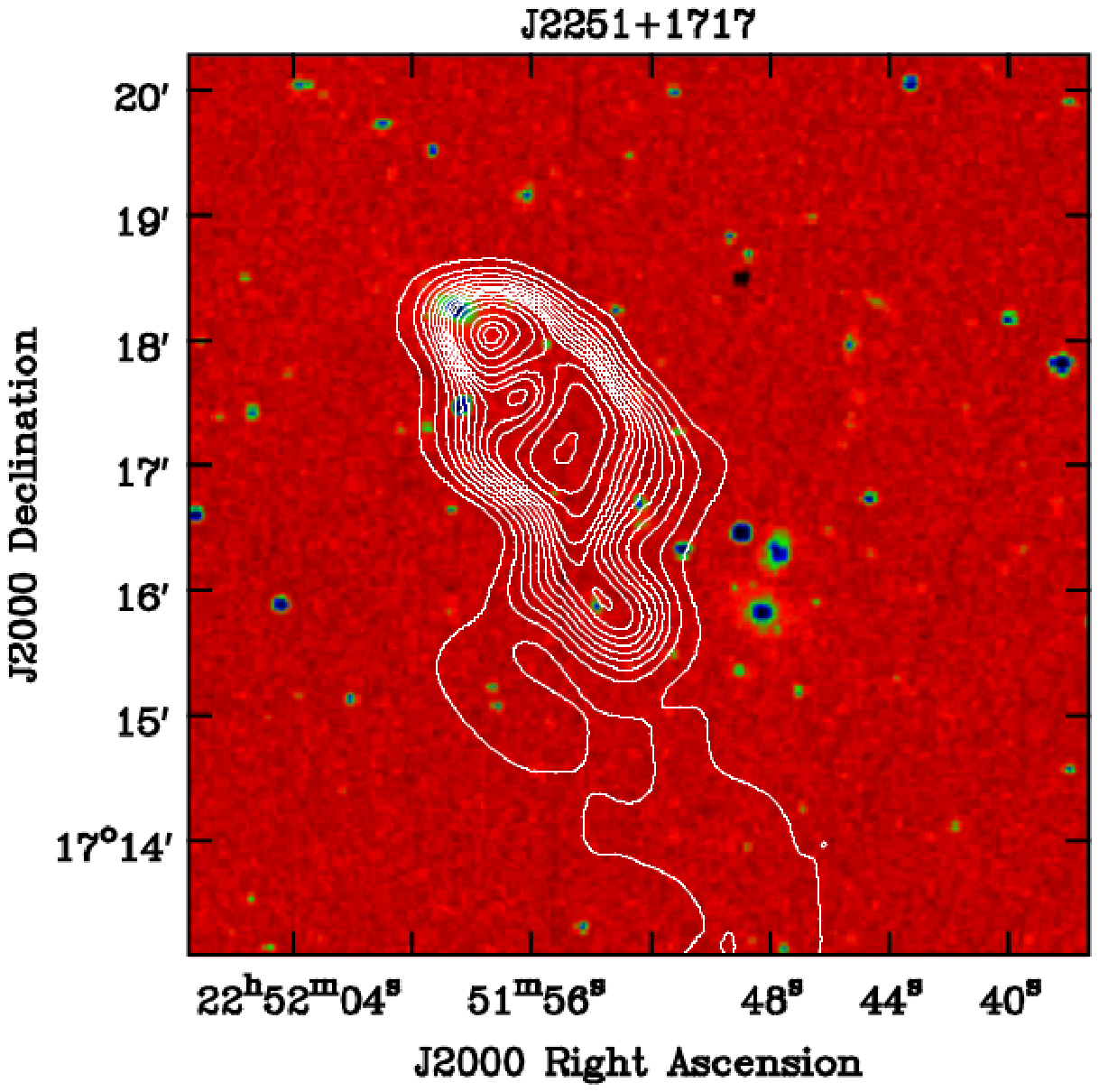,width=5.5cm,height=5cm}
\contcaption {Here we present radio maps of newly discovered XRGs. The white colour contour represents the radio emission detected in the TGSS ADR 1. The background colour images are from the DSS r-band. Contour levels are at 3$\sigma \times$[ 1, 1.41, 2, 2.83, 4, 5.66, 8, 11.31, 16, 22.63, 32, 45.25, 64, 90.51, 128, 181.02, 256], where $\sigma$ = 3.5 mJy beam$^{-1}$ is the local rms noise.}
\end{figure*}

\begin{figure*}
\psfig{figure=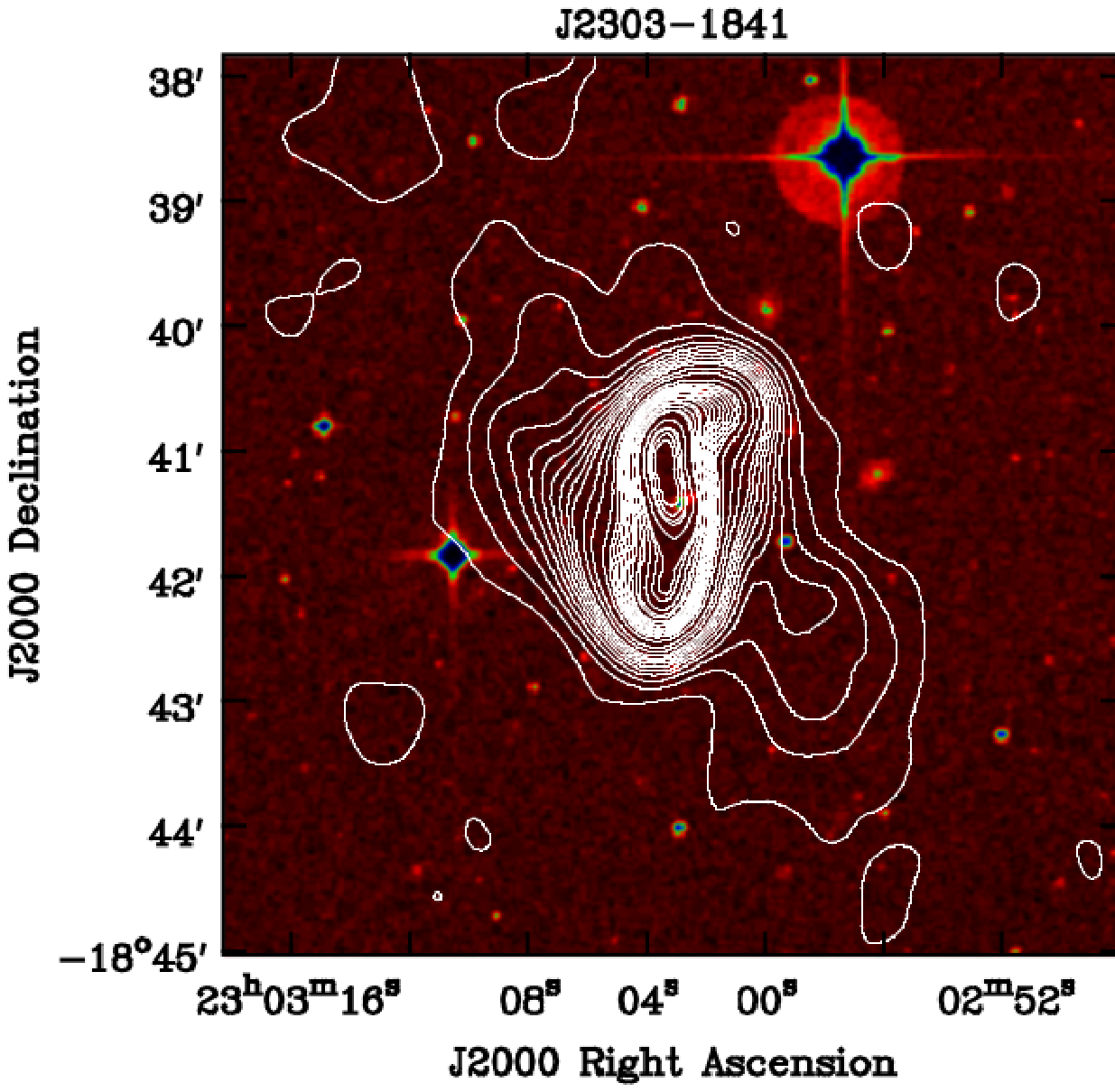,width=5.5cm,height=5cm} 
\psfig{figure=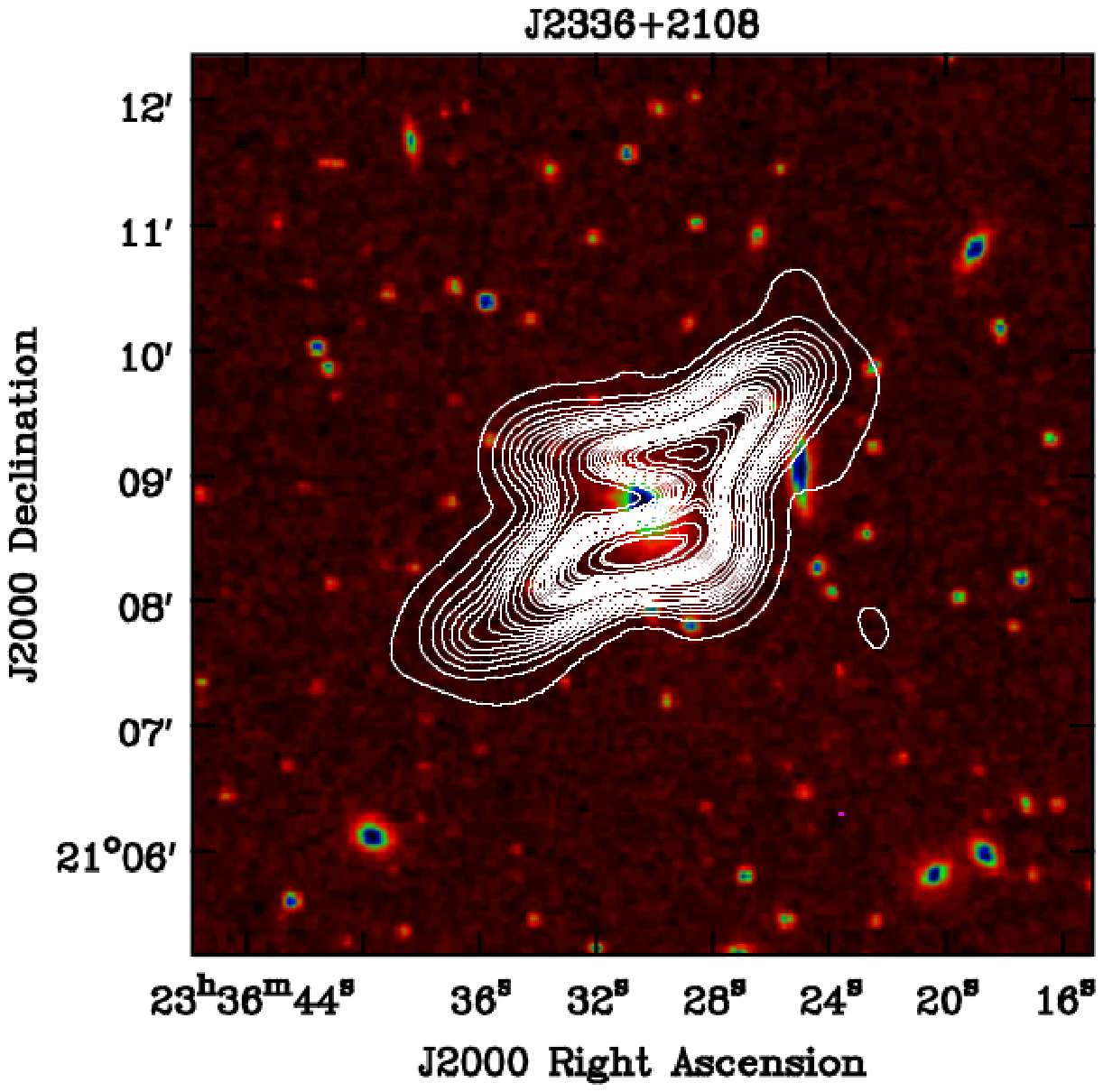,width=5.5cm,height=5cm}
\psfig{figure=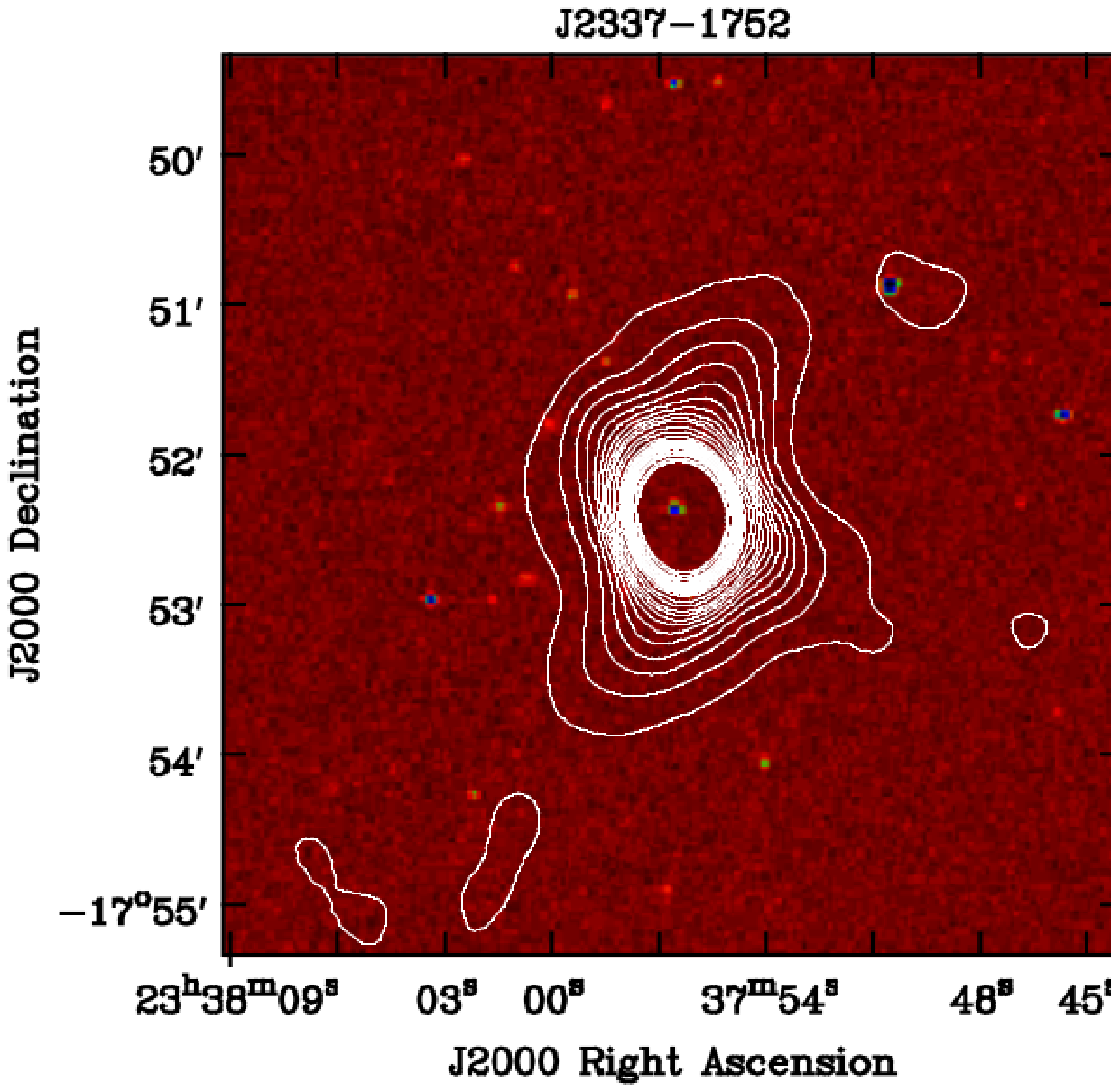,width=5.5cm,height=5cm}
\vskip 0.8cm
\psfig{figure=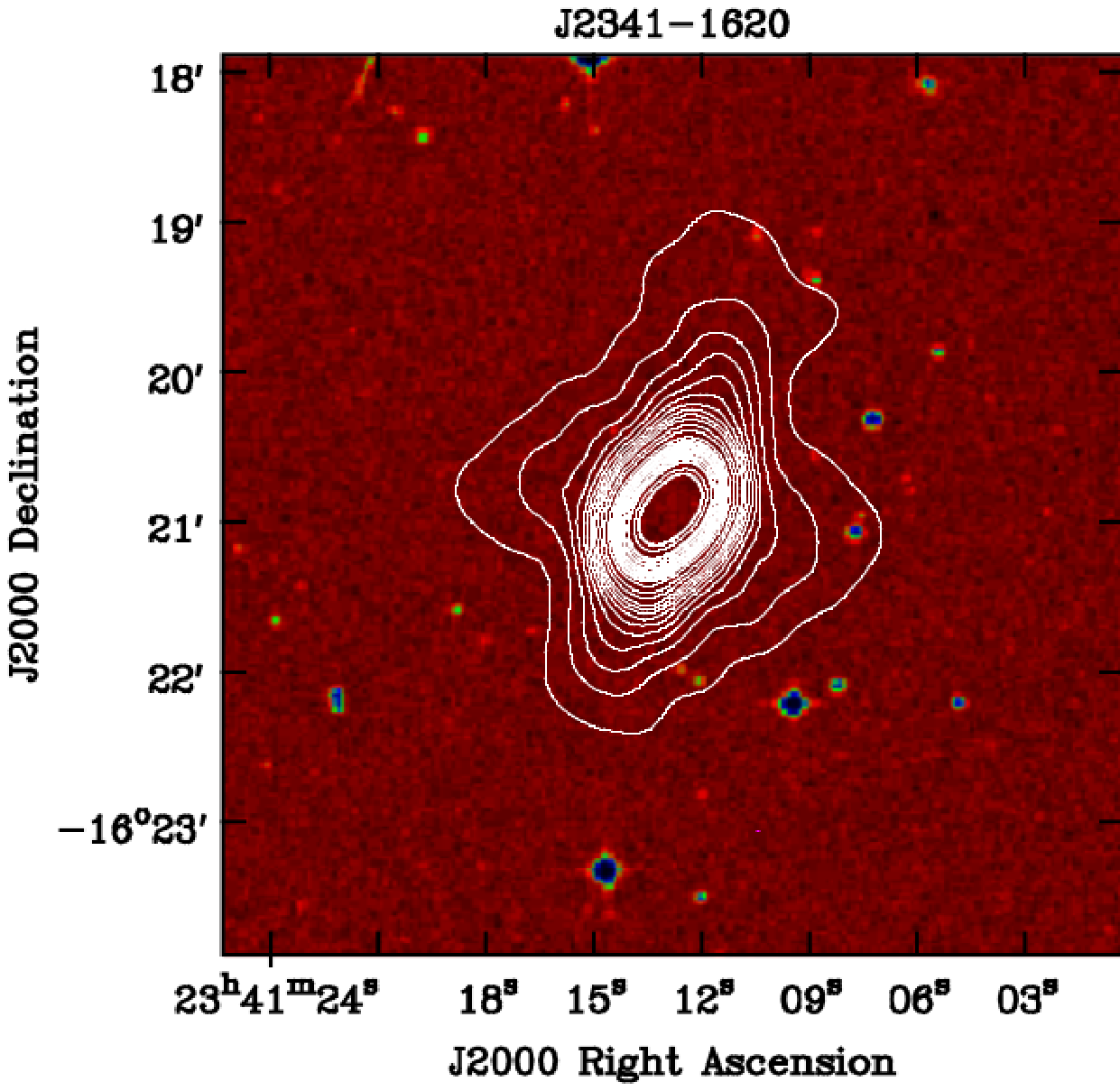,width=5.5cm,height=5cm} 
\contcaption {Here we present radio maps of newly discovered XRGs. The white colour contour represents the radio emission detected in the TGSS ADR 1. The background colour images are from the DSS r-band. Contour levels are at 3$\sigma \times$[ 1, 1.41, 2, 2.83, 4, 5.66, 8, 11.31, 16, 22.63, 32, 45.25, 64, 90.51, 128, 181.02, 256], where $\sigma$ = 3.5 mJy beam$^{-1}$ is the local rms noise.}
\end{figure*}
\begin{figure*}
\psfig{figure=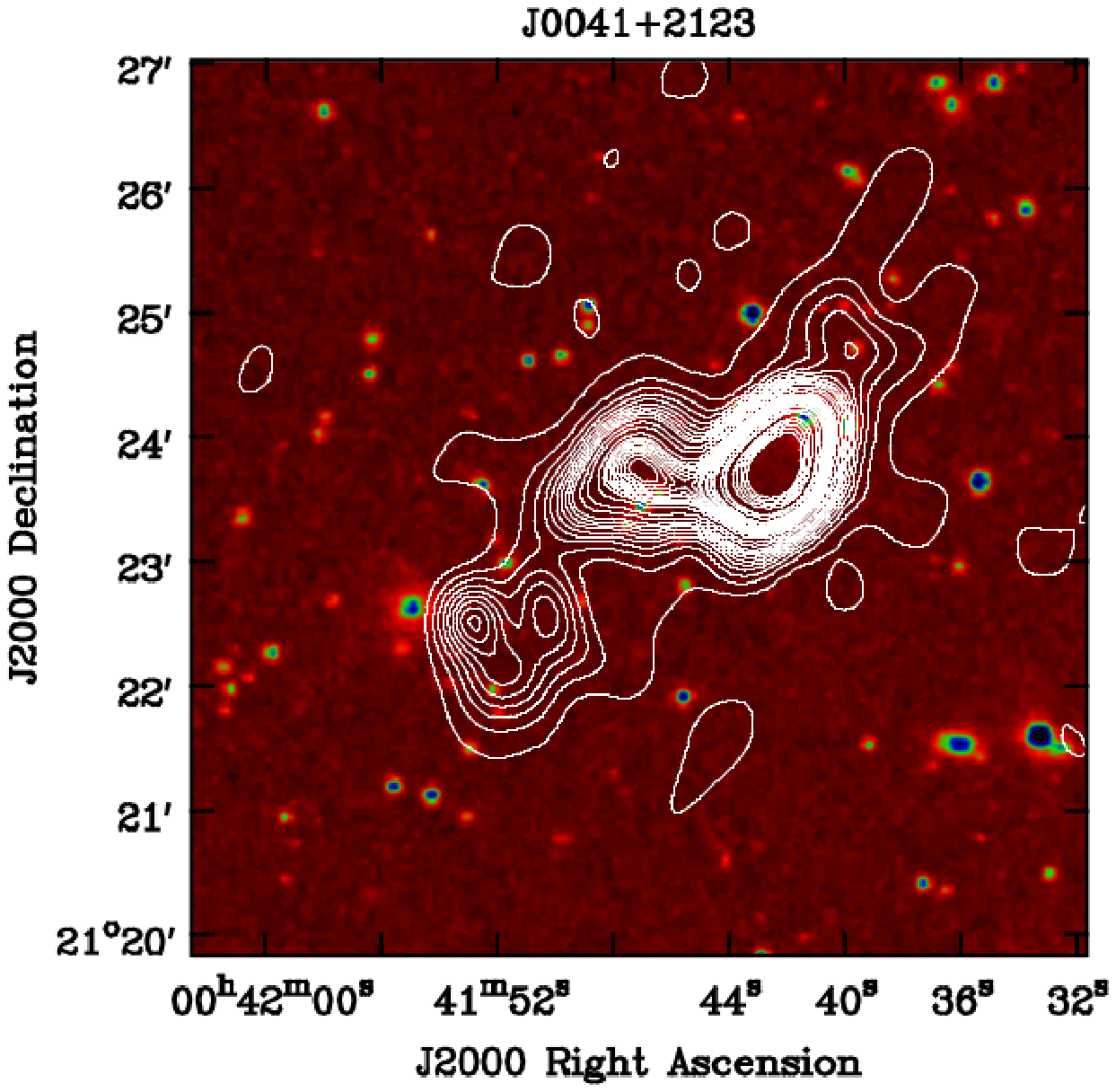,width=5.5cm,height=5cm} 
\psfig{figure=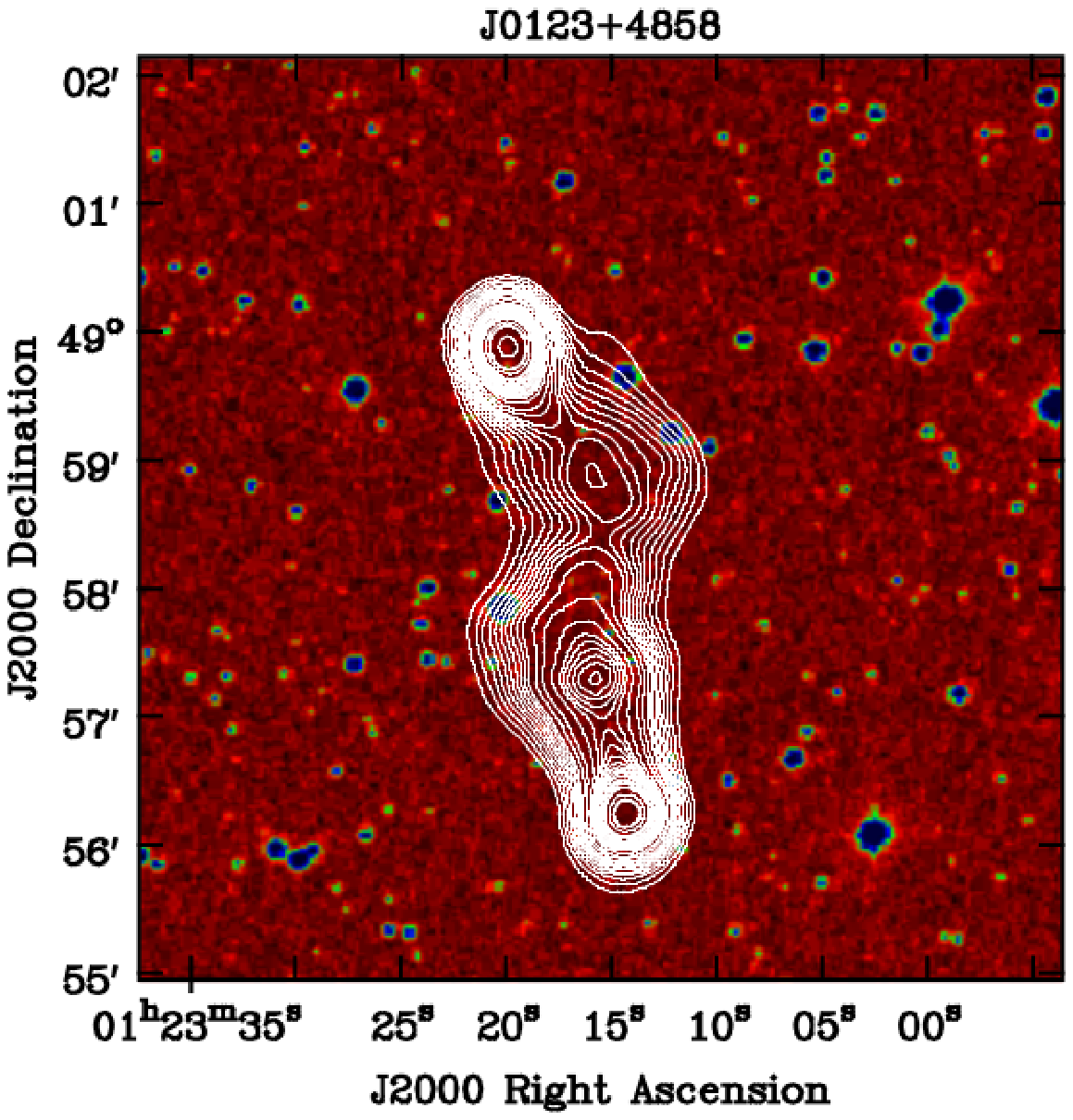,width=5.5cm,height=5cm}
\psfig{figure=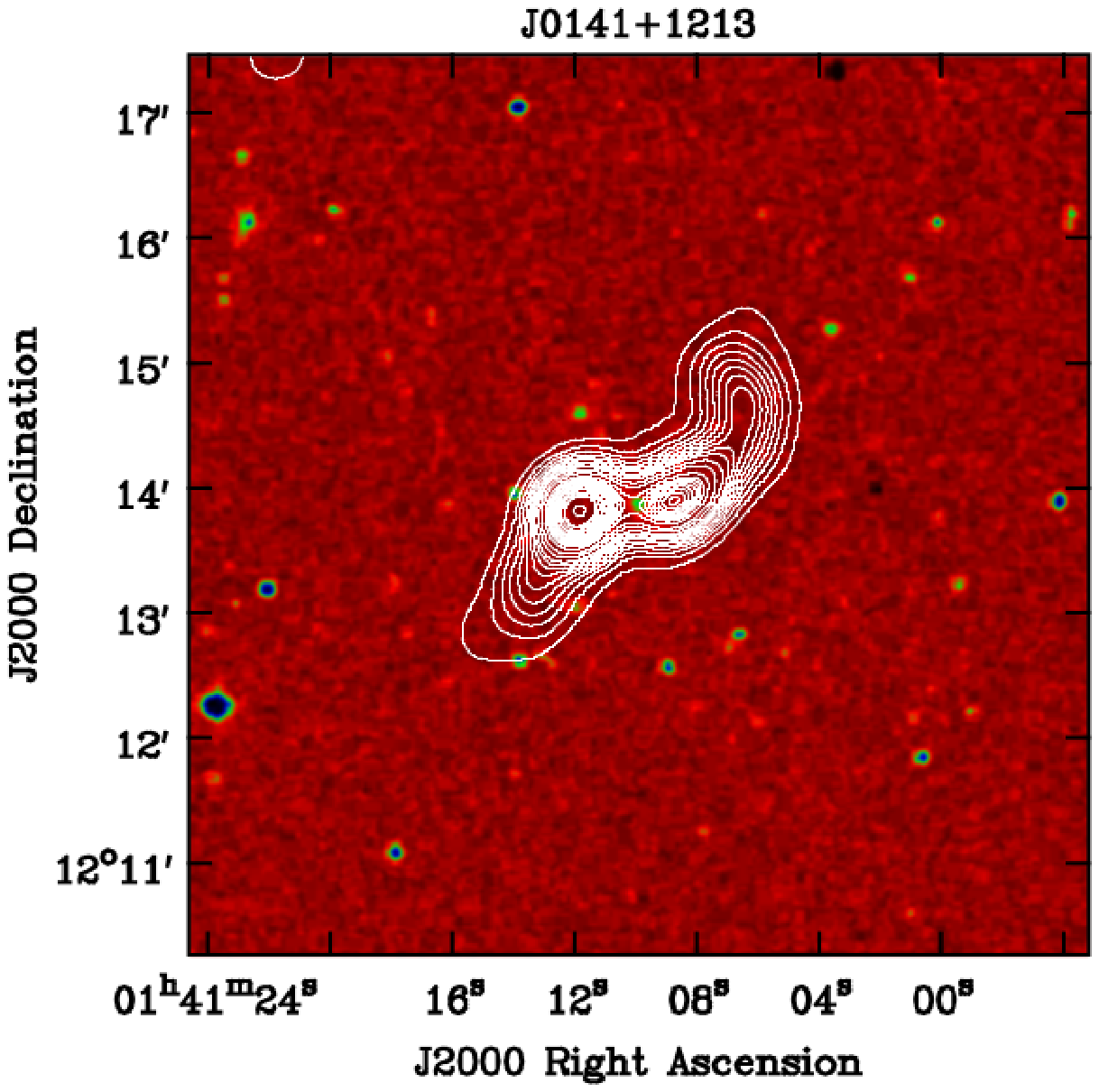,width=5.5cm,height=5cm}
\vskip 0.8cm         
 
\psfig{figure=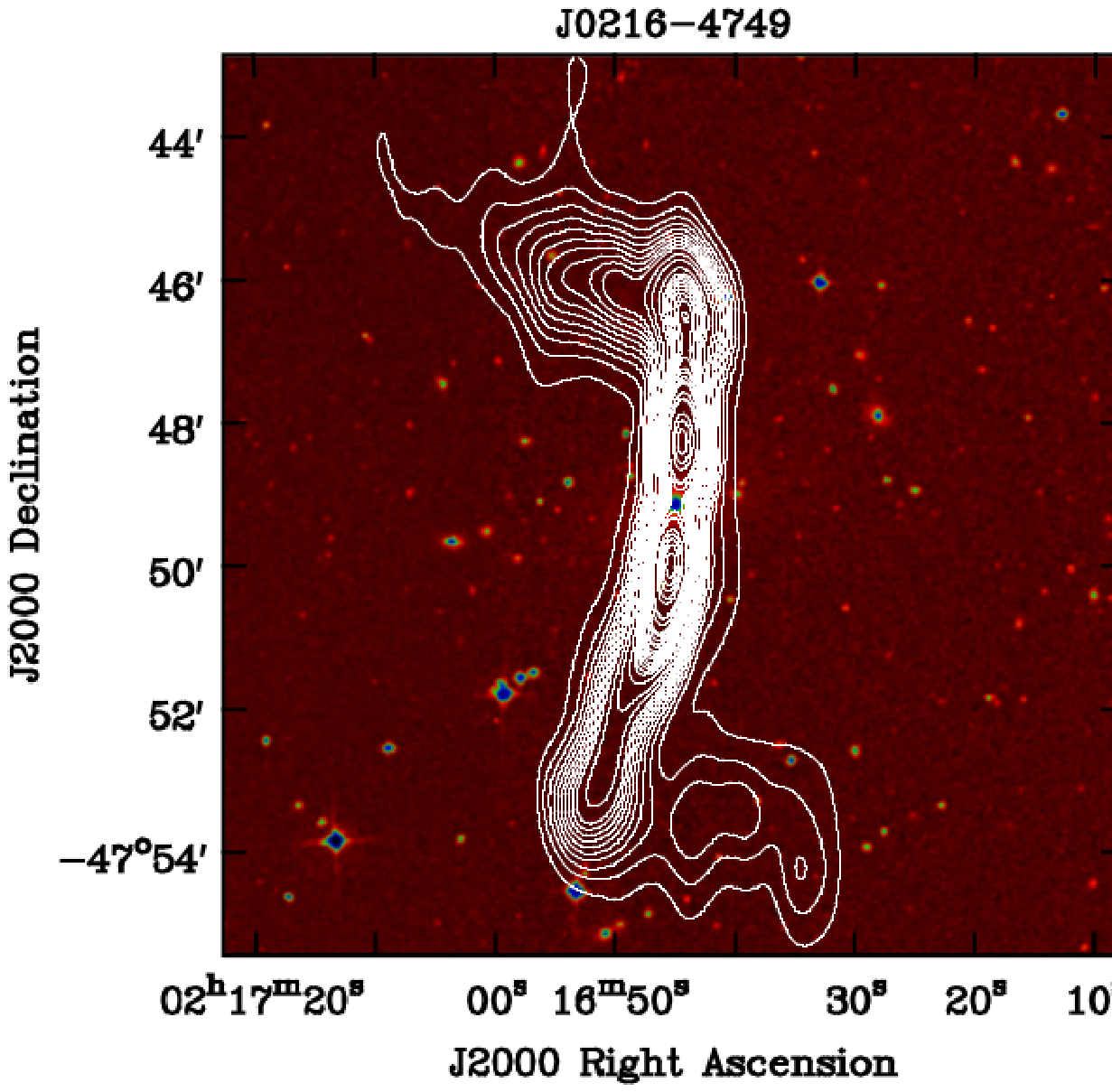,width=5.5cm,height=5cm}
\psfig{figure=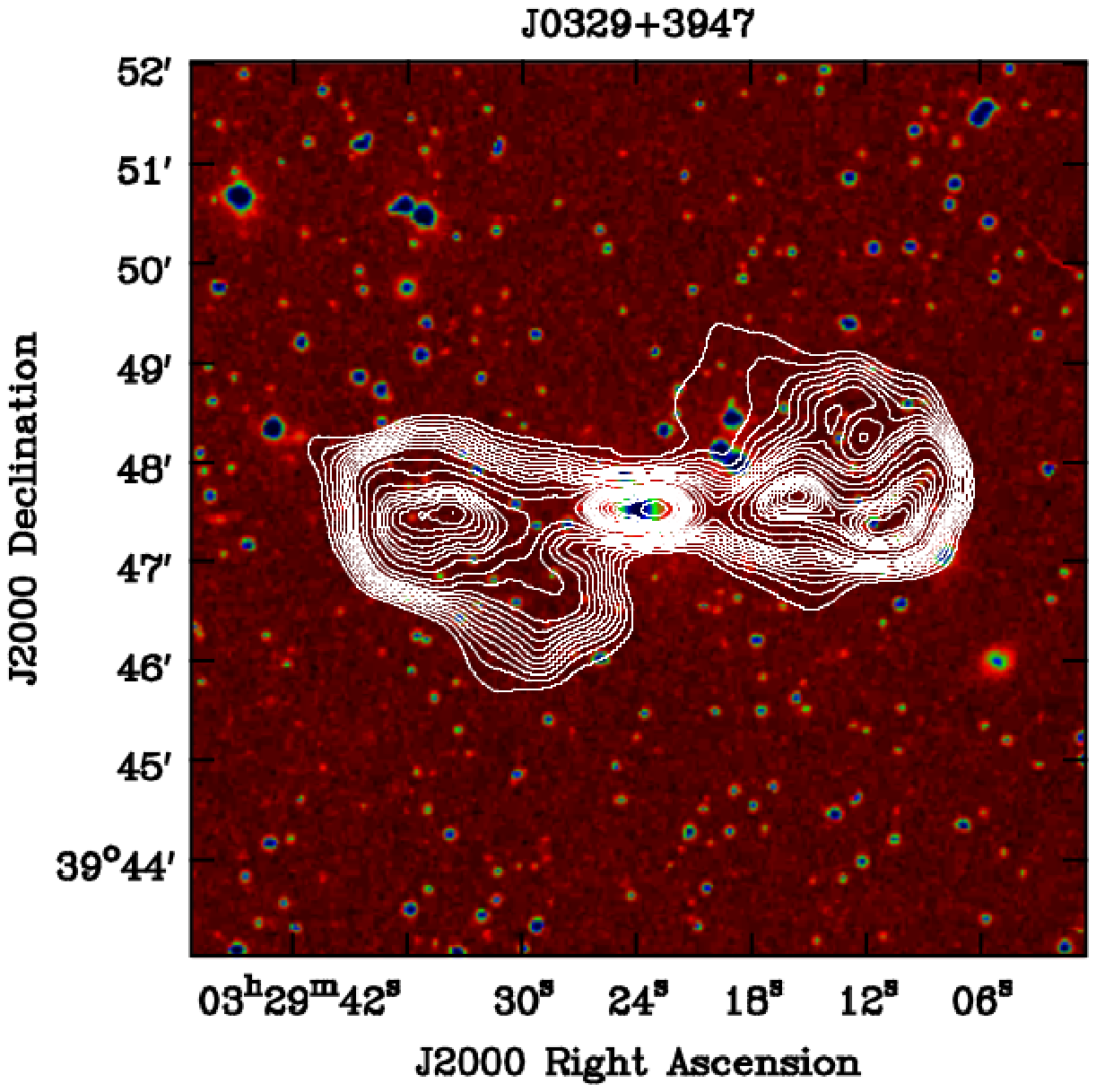,width=5.5cm,height=5cm}
\psfig{figure=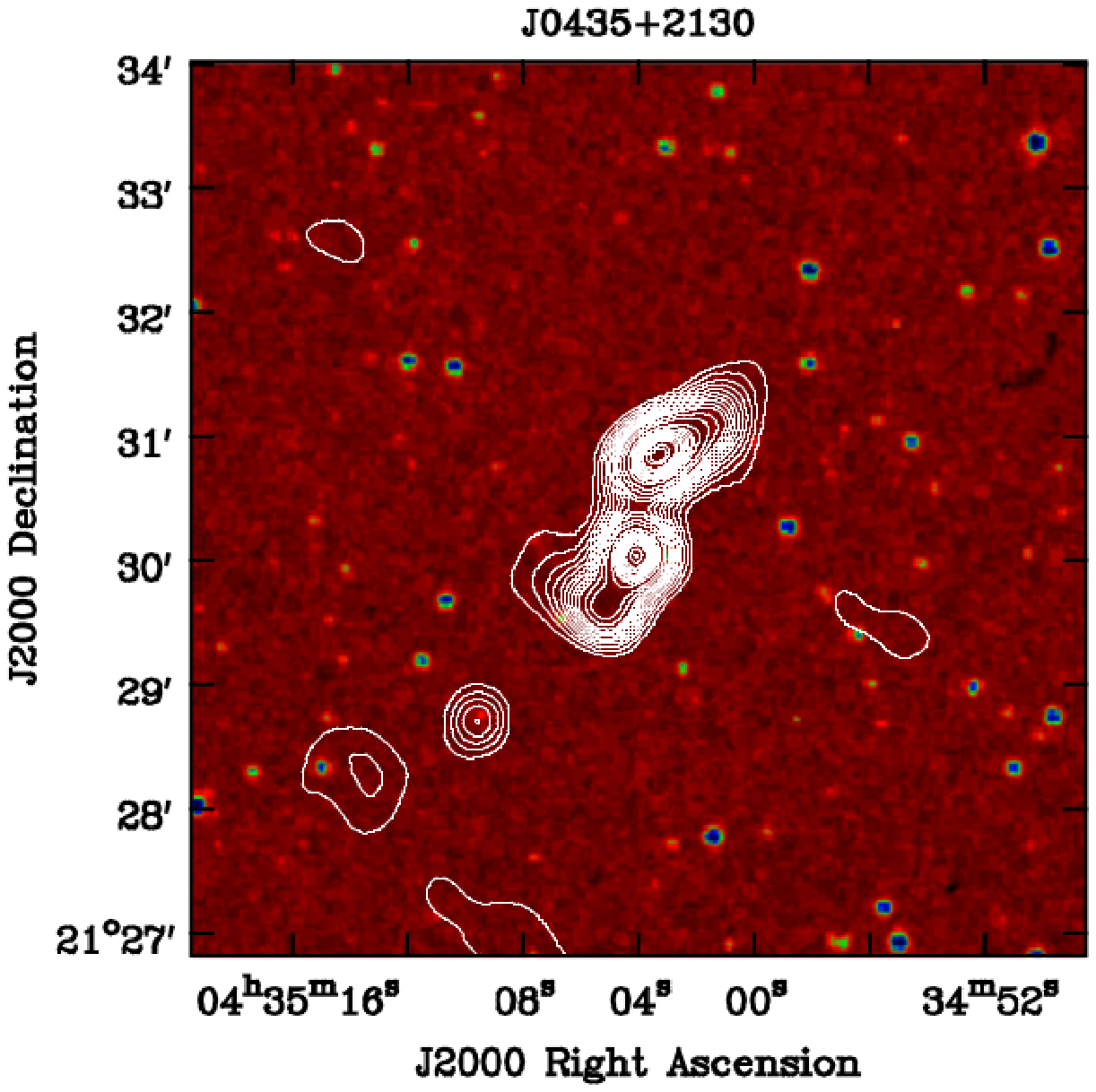,width=5.5cm,height=5cm}
\vskip 0.8cm         

\psfig{figure=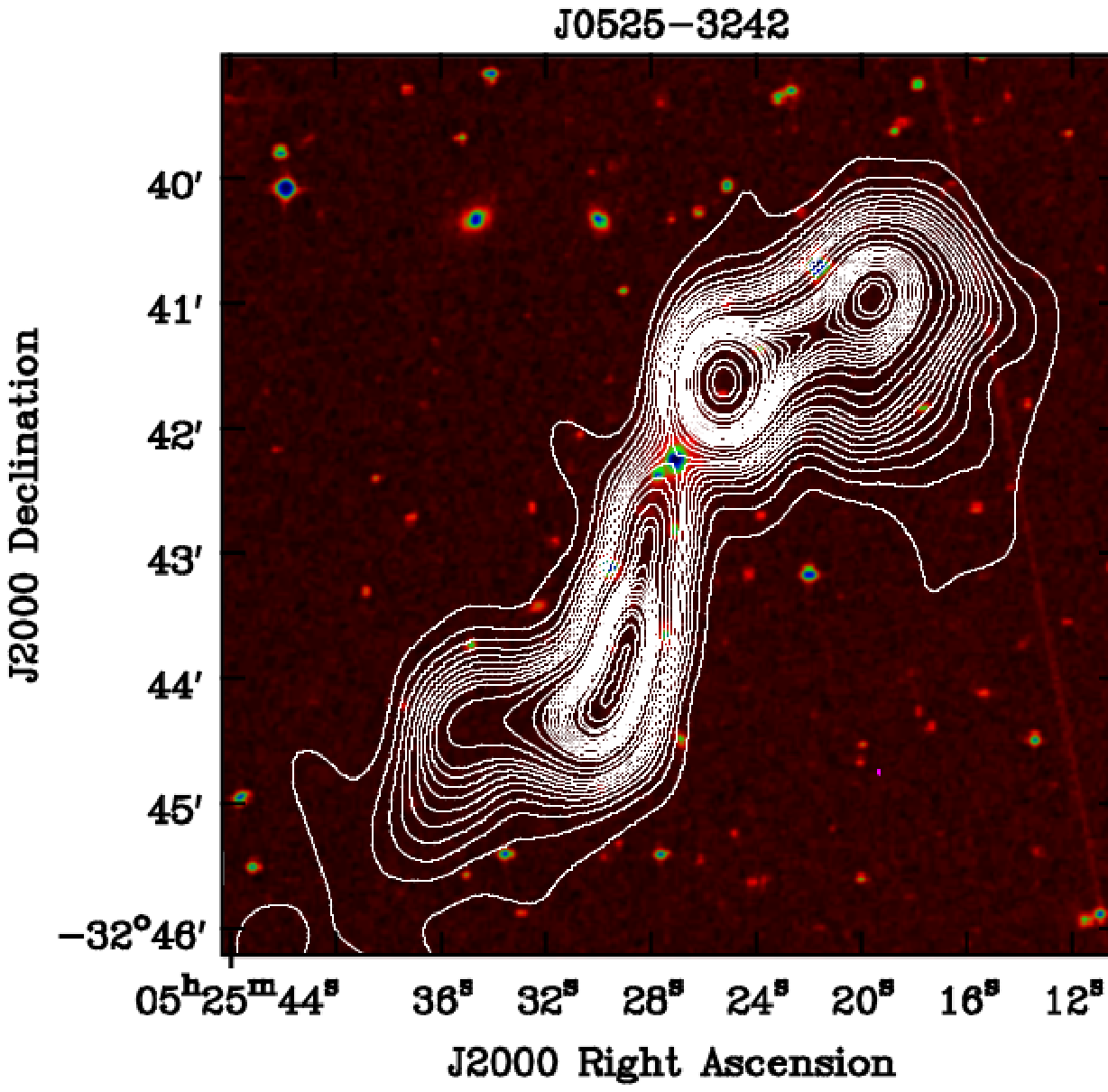,width=5.5cm,height=5cm}
\psfig{figure=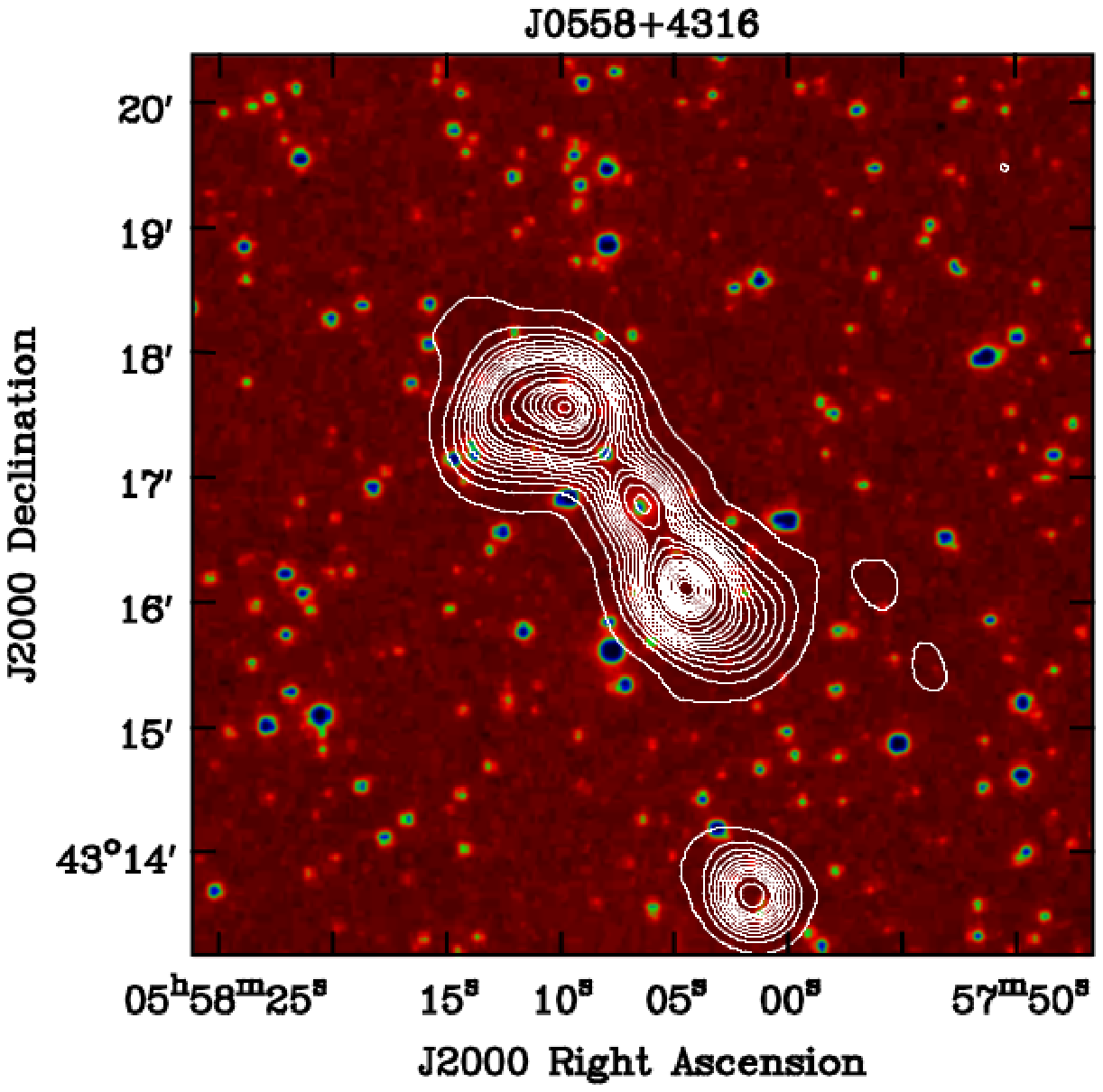,width=5.5cm,height=5cm}
\psfig{figure=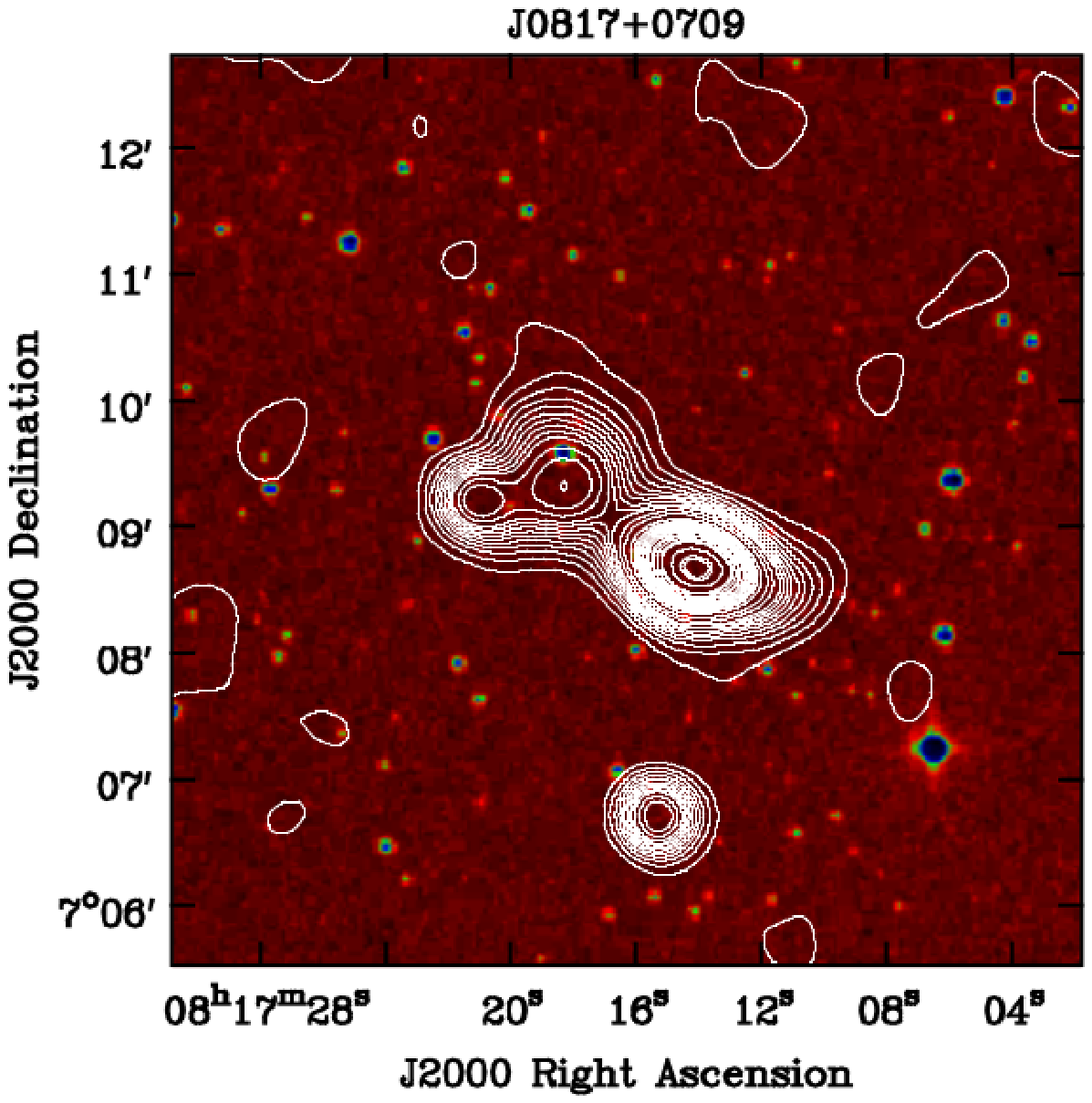,width=5.5cm,height=5cm}
\vskip 0.8cm         
 
\psfig{figure=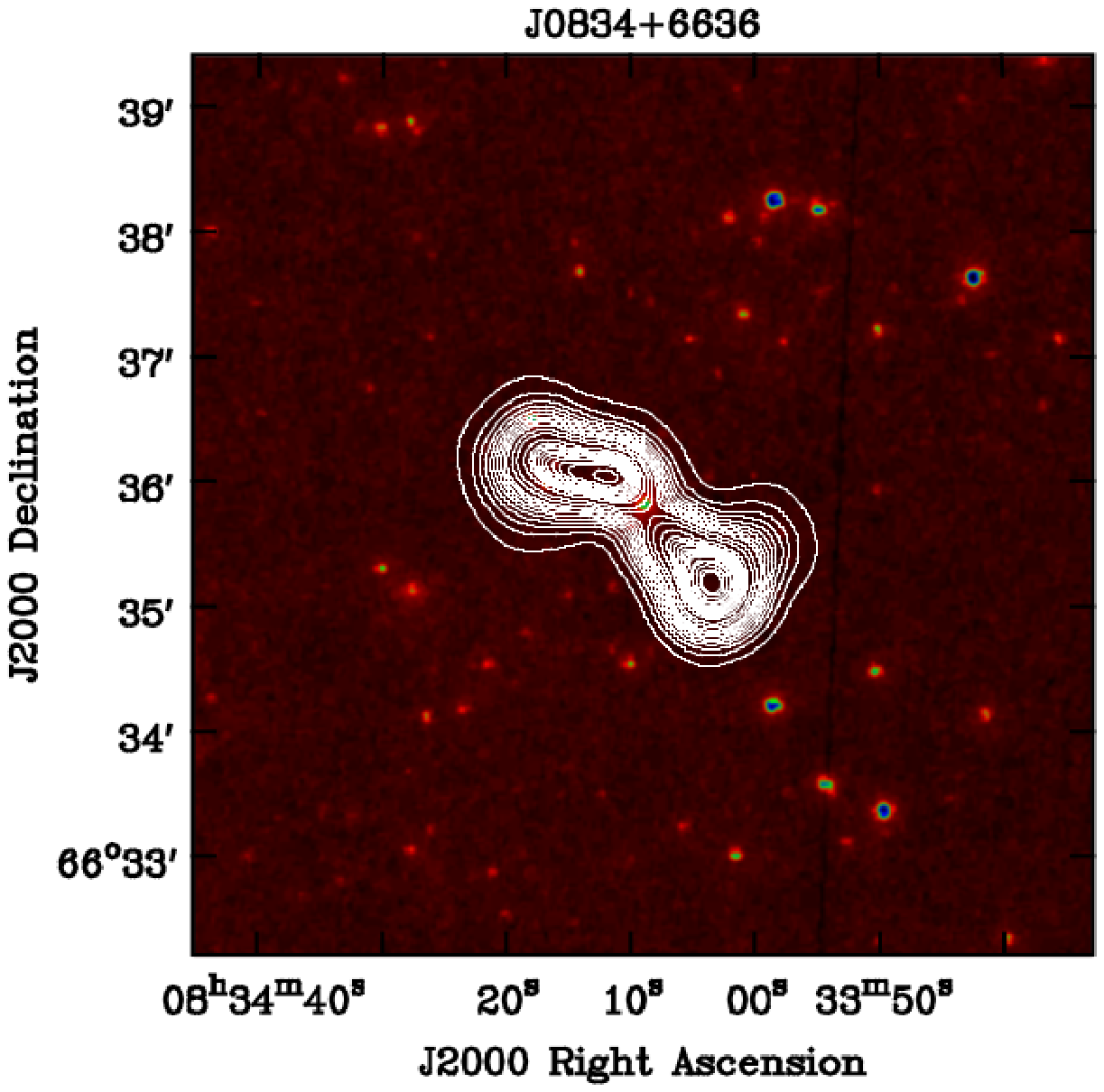,width=5.5cm,height=5cm}
\psfig{figure=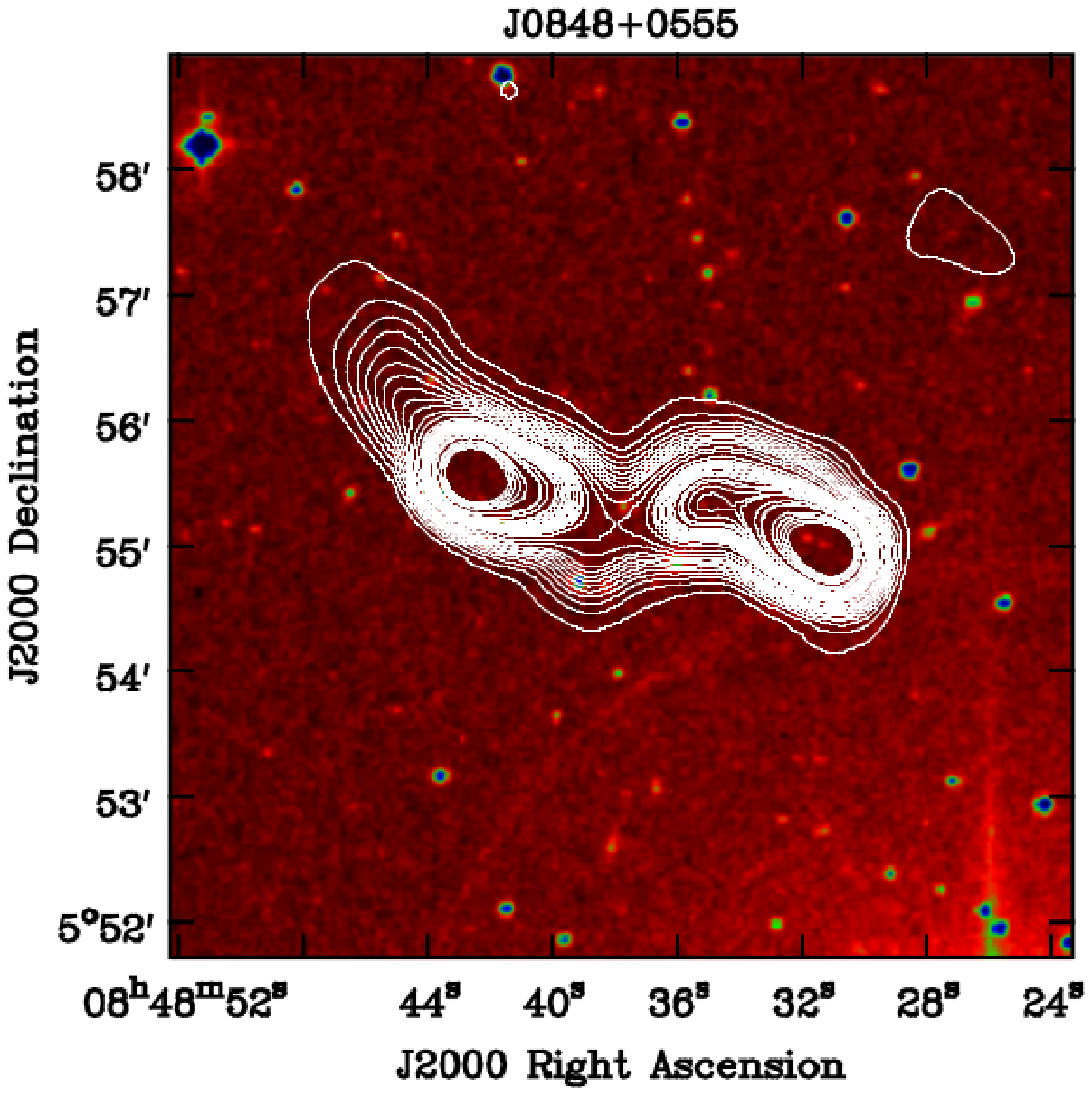,width=5.5cm,height=5cm}
\psfig{figure=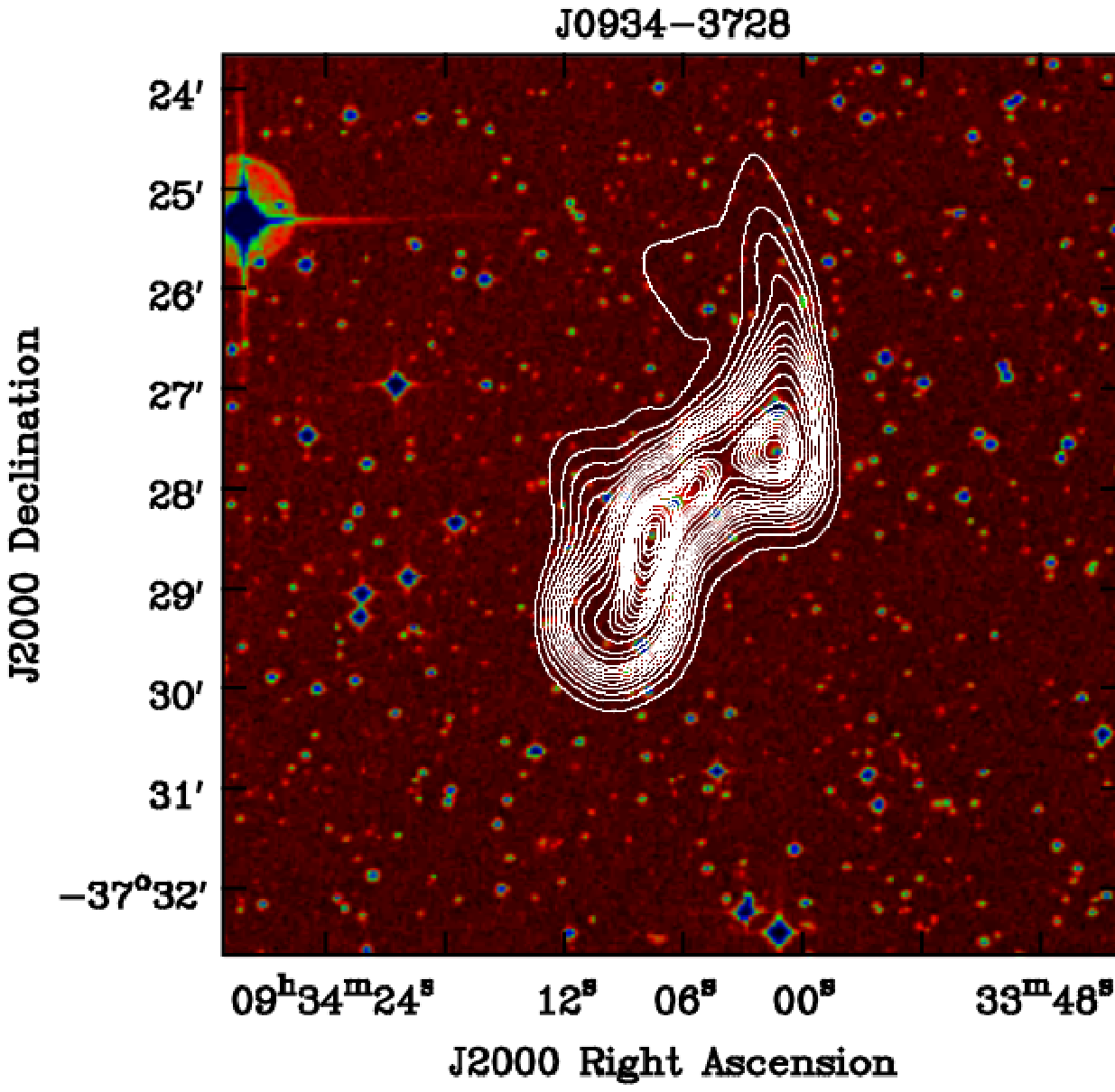,width=5.5cm,height=5cm} 
\caption {Here we present radio maps of newly discover ZRGs. The white colour contour represents the radio emission detected in the TGSS ADR 1. The background colour images are from the DSS r-band. Contour levels are at 3$\sigma \times$[ 1, 1.41, 2, 2.83, 4, 5.66, 8, 11.31, 16, 22.63, 32, 45.25, 64, 90.51, 128, 181.02, 256], where $\sigma$ = 3.5 mJy beam$^{-1}$ is the local rms noise.}
\label{fig:ZRG}
\end{figure*}

\begin{figure*}
\psfig{figure=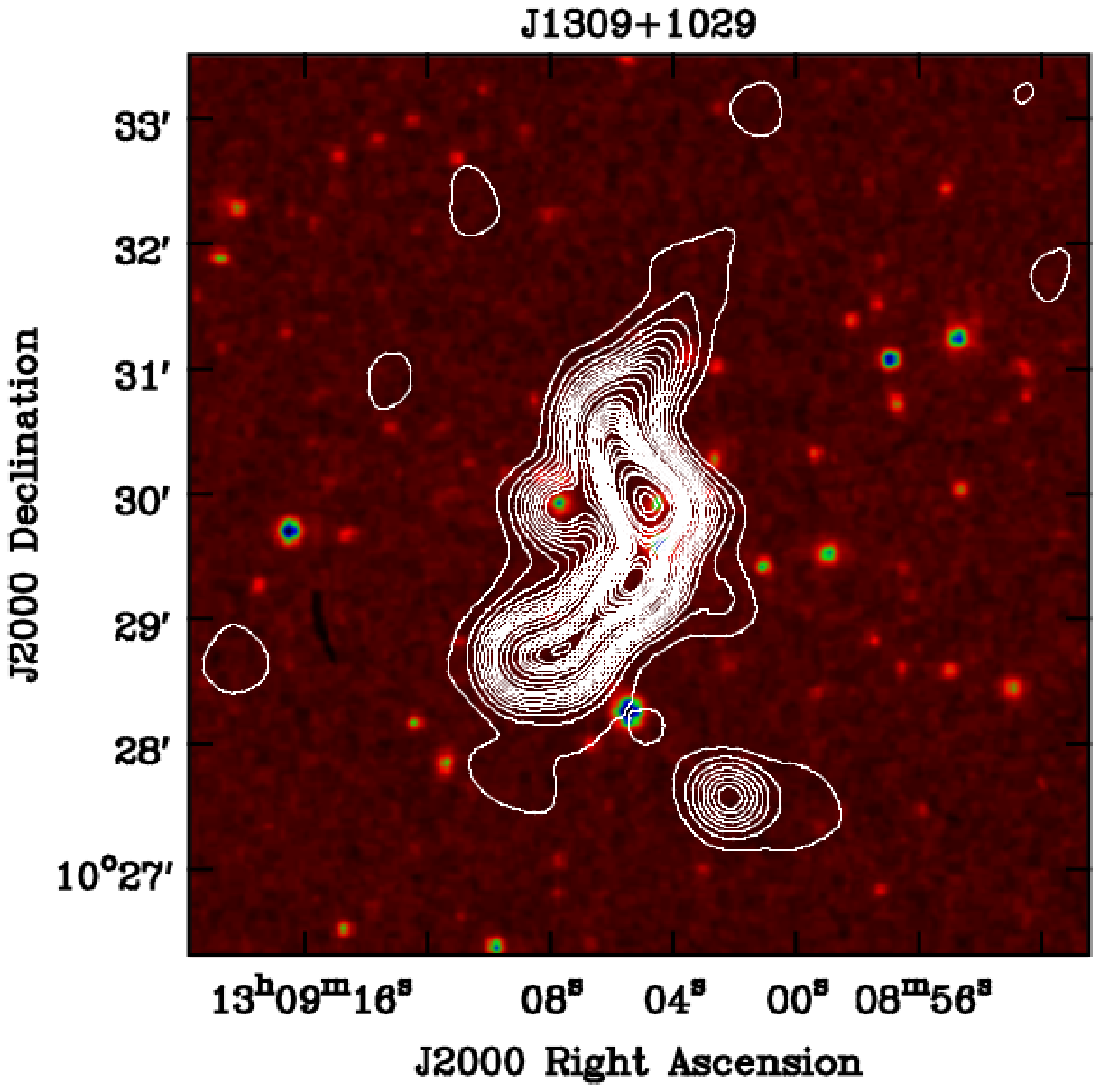,width=5.5cm,height=5cm} 
\psfig{figure=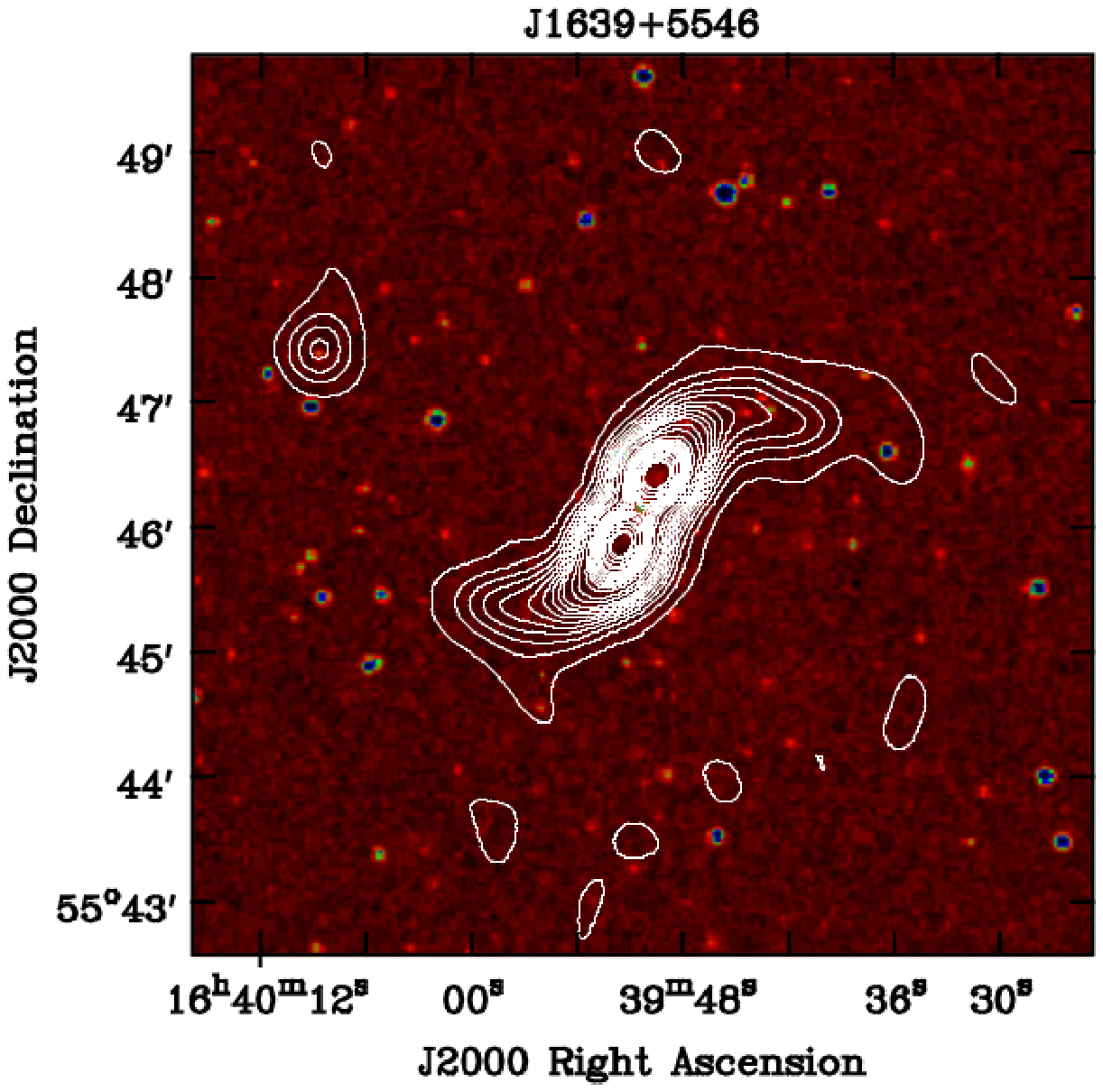,width=5.5cm,height=5cm}
\psfig{figure=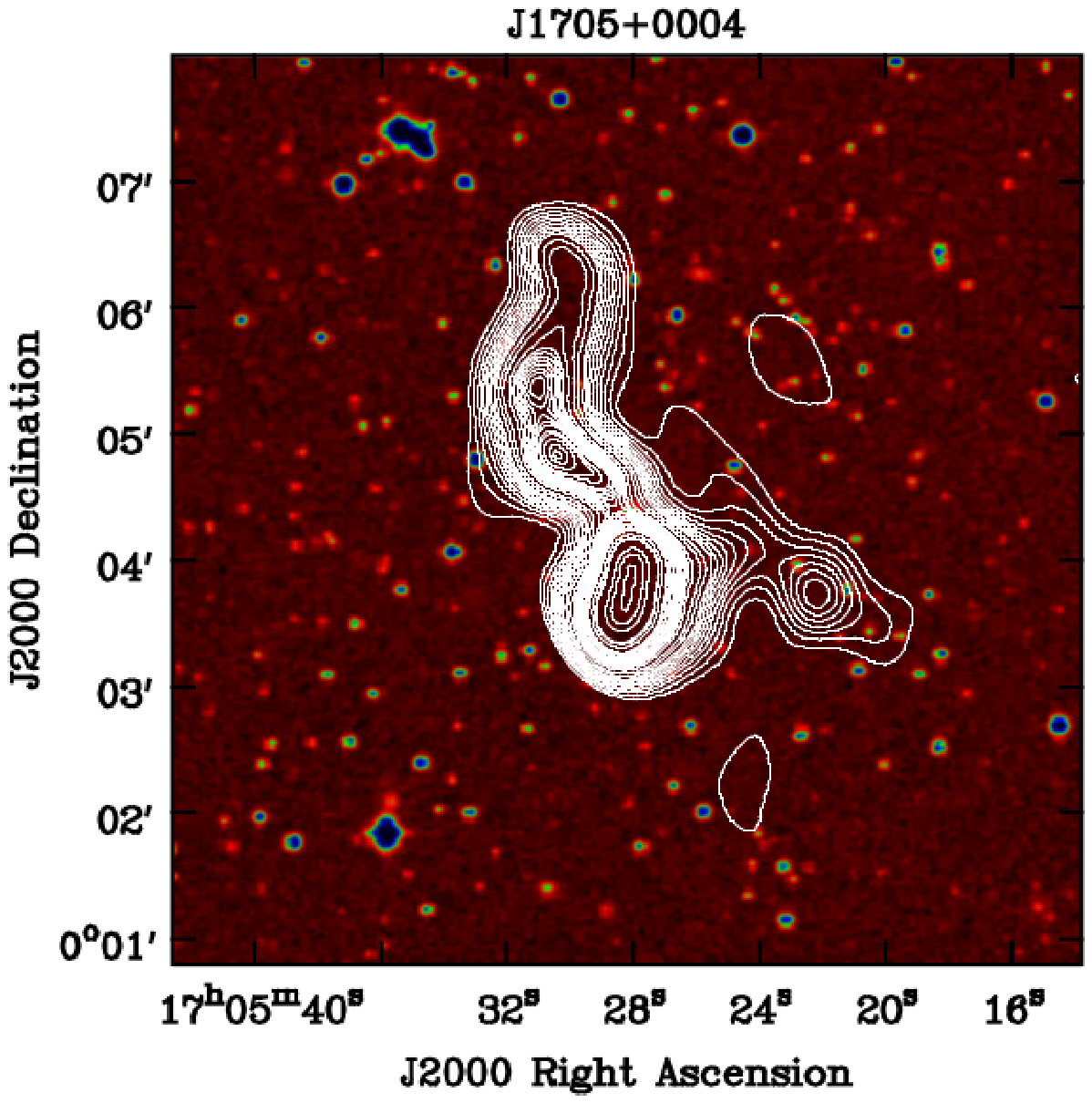,width=5.5cm,height=5cm}
\vskip 0.8cm         
 
\psfig{figure=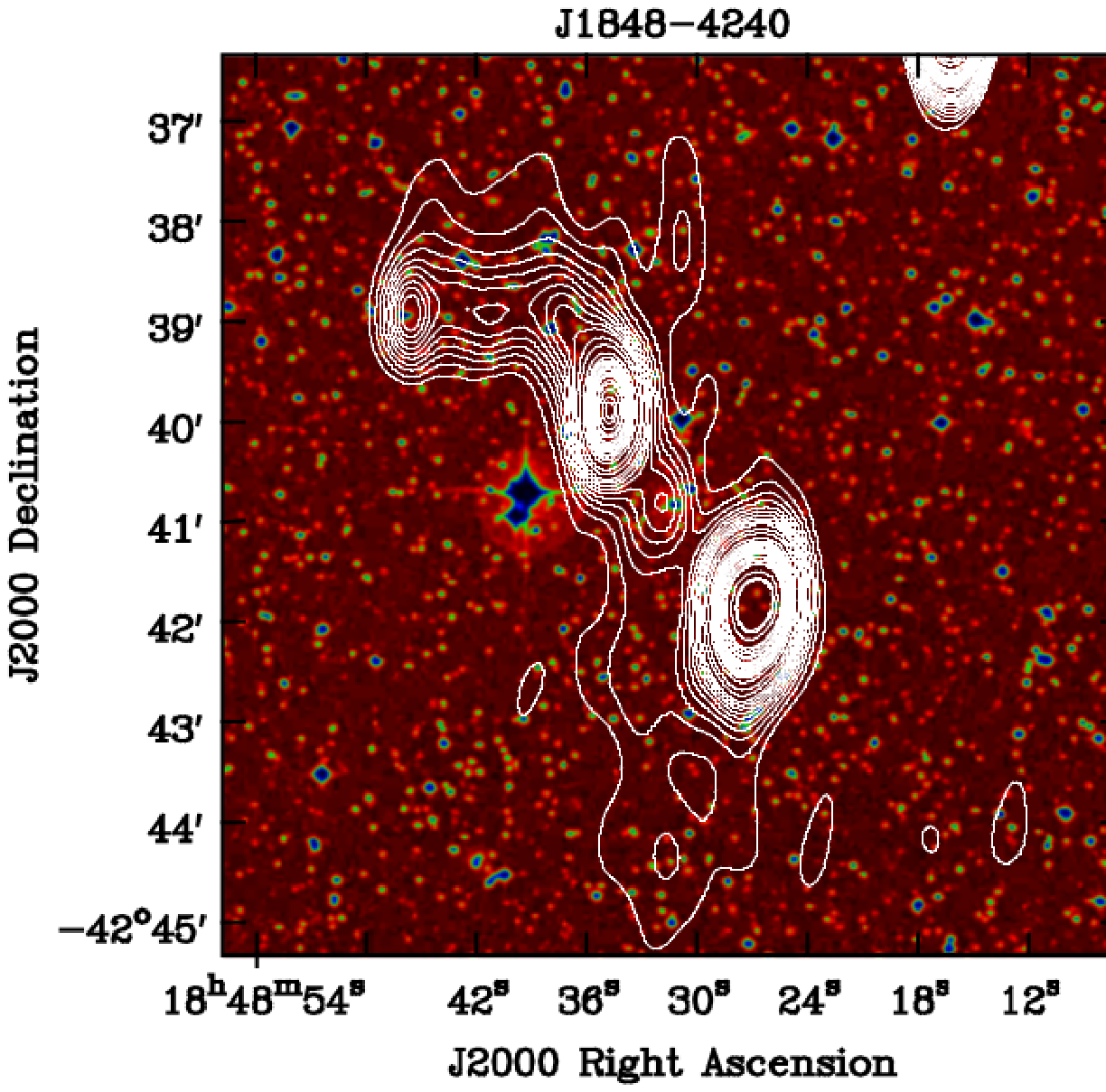,width=5.5cm,height=5cm}
\psfig{figure=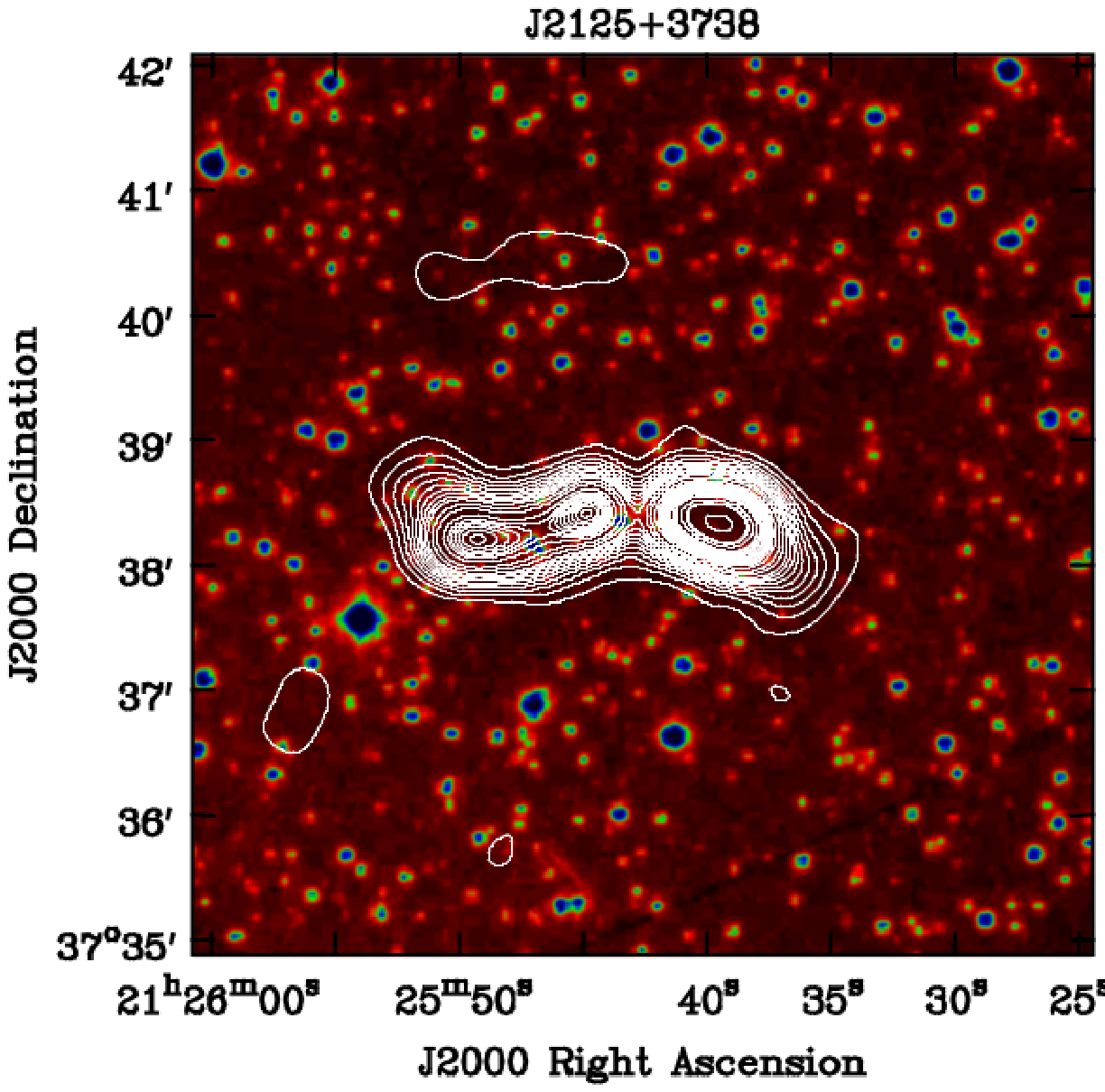,width=5.5cm,height=5cm}
\psfig{figure=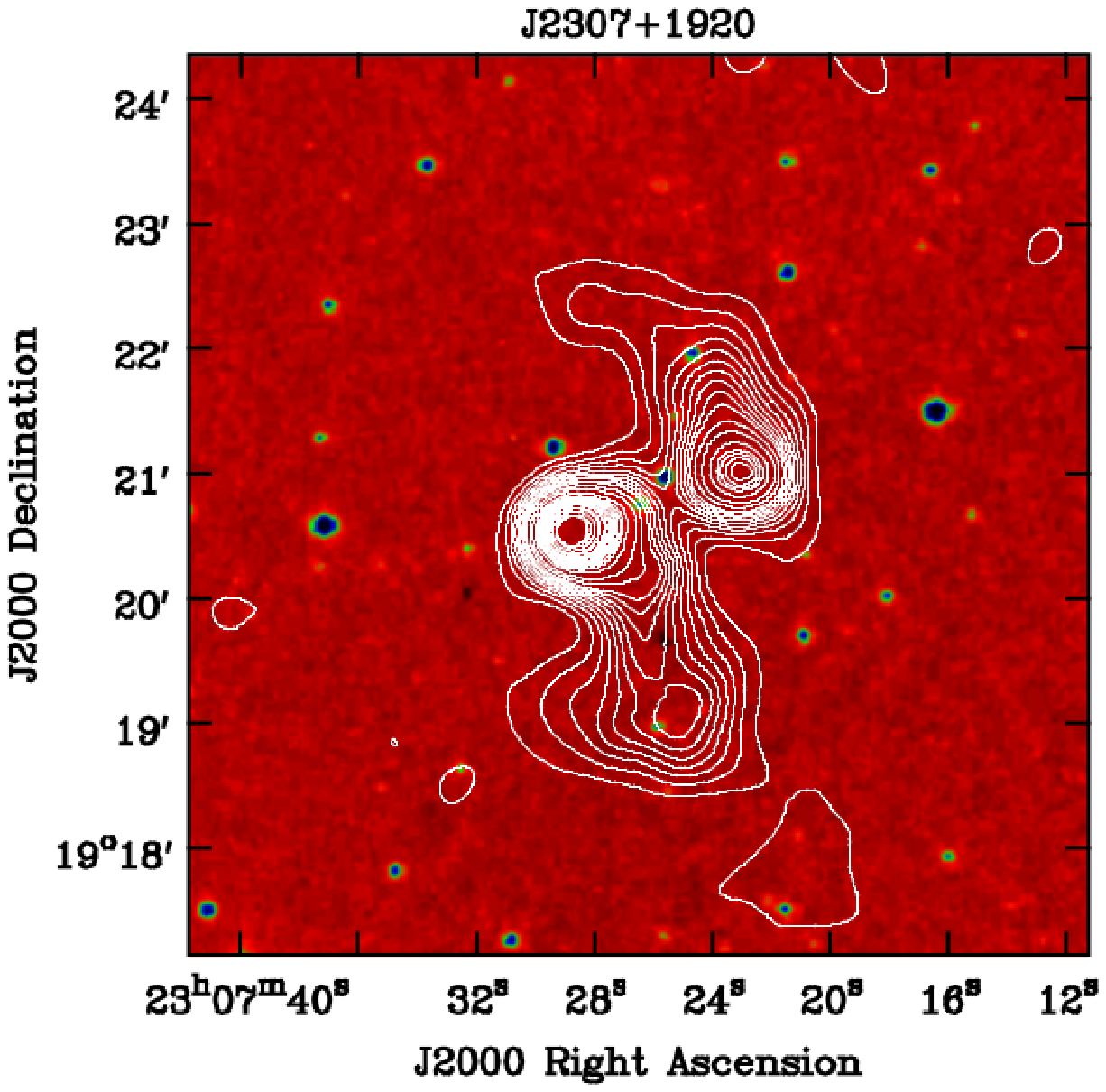,width=5.5cm,height=5cm}
\vskip 0.8cm         
\contcaption {Here we present radio maps of newly discover ZRGs. The white colour contour represents the radio emission detected in the TGSS ADR 1. The background colour images are from the DSS r-band. Contour levels are at 3$\sigma \times$[ 1, 1.41, 2, 2.83, 4, 5.66, 8, 11.31, 16, 22.63, 32, 45.25, 64, 90.51, 128, 181.02, 256], where $\sigma$ = 3.5 mJy beam$^{-1}$ is the local rms noise.}
\end{figure*}



\bsp	
\label{lastpage}
\end{document}